\documentclass{aa}  
\usepackage{graphicx}
\usepackage{eqnarray,amsmath}
\usepackage{multicol}
\usepackage[hypcap]{caption}
\usepackage{txfonts}
\usepackage{subcaption}
\usepackage{hyperref}
\usepackage{xcolor}
\graphicspath{{./figures/}} % Location of the graphics files
\begin{document} 
\titlerunning{Evolution of rotating massive stars with new hydrodynamic wind models}
\title{Evolution of rotating massive stars with new hydrodynamic wind models}
\authorrunning{Gormaz-Matamala et al.}
\author{A. C. Gormaz-Matamala\inst{1,2,3}
\and
J. Cuadra\inst{1}
\and
G. Meynet\inst{4}
\and
M. Curé\inst{5,6}
}
\institute{Departamento de Ciencias, Facultad de Artes Liberales, Universidad Adolfo Ib\'a\~nez, Av. Padre Hurtado 750, Vi\~na del Mar, Chile\\
\email{alex.gormaz@uv.cl}
\and
Instituto de Astrofísica, Facultad de Física, Pontificia Universidad Católica de Chile, 782-0436 Santiago, Chile
\and
Nicolaus Copernicus Astronomical Center, Polish Academy of Sciences, ul. Bartycka 18, 00-716 Warsaw, Poland
\and
Geneva Observatory, University of Geneva, Maillettes 51, 1290 Sauverny, Switzerland
\and
Instituto de Física y Astronomía, Universidad de Valparaíso. Av. Gran Breta\~na 1111, Casilla 5030, Valpara\'iso, Chile.
\and
Centro de Astrofísica, Universidad de Valparaíso. Av. Gran Breta\~na 1111, Casilla 5030, Valpara\'iso, Chile.}

\date{}

\abstract % 5 {} token are mandatory
{Mass loss due to radiatively line-driven winds is central to our understanding of the evolution of massive stars in both single and multiple systems.
This mass loss plays a key role modulating the stellar evolution at different metallicities, particularly in the case of massive stars with $M_*\ge25\,M_\odot$.}
{We extend the evolution models introduced in Paper I, where the mass loss recipe is based on the simultaneous calculation of the wind hydrodynamics and the line-acceleration, by incorporating the effects of stellar rotation.}
{As in Paper I, we introduce a grid of self-consistent line-force parameters $(k,\alpha,\delta)$ for a set of standard evolutionary tracks using \textsc{Genec}.
Based on this grid, we analyse the effects of stellar rotation, CNO abundances and He/H ratio on the wind solutions, to derive extra terms for the recipe predicting the self-consistent mass loss rate, $\dot M_\text{sc}$.
With that, we generate a new set of evolutionary tracks with rotation for $M_\text{ZAMS}=25,40,70,$ and $120\,M_\odot$, and metallicities $Z=0.014$ (Galactic) and $0.006$ (LMC).}
{Besides the expected correction factor due to rotation, the mass loss rate decreases when the surface become more helium-rich, especially at the later moments of the main sequence phase.
The self-consistent approach gives lower mass loss rates than the standard values adopted in previous \textsc{Genec} evolution models.
This decrease impacts strongly on the tracks of the most massive models.
Weaker winds allow the star to retain more mass, but also more angular momentum.
As a consequence at a given age, weaker wind models rotate faster and show a less efficient mixing in their inner stellar structure.}
{The self-consistent tracks predict an evolution of the rotational velocities through the main sequence in close agreement with the range of $\varv\sin i$ values found by recent surveys of Galactic O-type stars.
As subsequent implications, the weaker winds from self-consistent models also suggest a reduction of the contribution of the isotope $^{26}$Al to the interstellar medium due to stellar winds of massive stars during the MS phase.
Moreover, the higher luminosities found for the self-consistent evolutionary models suggest that some populations of massive stars might be less massive than previously thought, as in the case of Ofpe stars at the Galactic Centre.
Therefore, this study opens a wide range of consequences for further research based on the evolution of massive stars.}


\maketitle

%_____INTRODUCTION________________________________________________________________________________
%\begin{multicols}{1}
\section{Introduction}
	Evolution of massive stars is an important topic in stellar astrophysics.
	Indeed some of them are the progenitors of core-collapse supernova events, and give birth to neutron stars and black holes \citep{heger03}.
	Massive stars are also important for the study of nucleosynthesis, production of ionising flux, feedback due to wind momentum, studies of star formation history, and galaxy evolution.

	Massive stars with $M_*\gtrsim25\,M_\odot$ are characterised by strong stellar winds and large mass loss rates ($\dot M$), which determine the evolution of the star such as the phases involving Luminous Blue Variables (LBV) and Wolf-Rayet (WR) stars \citep{groh14}.
	The wind properties through all these stages constrain which kind of core-collapse process the star will experience at the end of its life, and what will be the final mass of the remnant \citep{belczynski10}.
	This last item is relevant, because theoretical masses of stellar black holes need to be in agreement with those measured by gravitational waves, in merger events detected by the Advanced-LIGO interferometers \citep{abbott16}.
	The study of the stellar winds has further implications that reach far beyond stellar astrophysics. 
	For instance, the stellar winds from numerous evolved massive stars collide to produce the plasma that fills up the interstellar medium in the Galactic Centre \citep[e.g.,][]{cuadra08,cuadra15,ressler18,ressler20,calderon20}.
	The properties of the stellar winds determine whether the plasma is able to cool and form clumps, which affects the accretion process on to the Galactic supermassive black hole, Sgr~A* \citep{cuadra05, calderon16, calderon20bin}.
	Therefore, a proper determination of the stellar winds is needed to correctly interpret the images of the black hole silhouette obtained by the \citet{EHT22V}.
	These examples highlight that theoretical calculations for the stellar winds can even influence the field of relativistic astrophysics.
	
	In the last decade, computational codes such as \textsc{Mesa} \citep{paxton19} and the Geneva evolution code \citep[][hereafter \textsc{Genec}]{eggenberger08} have been used to study stellar evolution.
	The recipe for theoretical mass loss rate varies, depending on the stellar evolution stage and the region of the HR diagram (HRD).
	For O-type main sequence (MS) stars, the recipe most commonly adopted has been the so-called Vink's formula \citep{vink01}.
	However, diagnostics of mass loss rates performed in recent years consider that values from Vink's formula are overestimated by a factor of $\sim3$ \citep{bouret12,surlan13,vink21b}, and therefore the quest has been the development of evolution models adopting more updated and accurate recipes for mass loss rate.
	In that direction, we mention the studies from \citet{bjorklund22} using \textsc{Mesa}, and \citet[][hereafter Paper I]{alex22b} using \textsc{Genec}.

	In Paper I, we developed new evolution models for stars born with masses $M_\text{zams}\ge25$ $M_\odot$, introducing a new recipe for the theoretical mass loss rate based on the self-consistent m-CAK wind solutions from \citet{alex19,alex22a}.
	These new self-consistent tracks show that stars retain more mass through their evolution, therefore remaining larger and more luminous compared with models for massive stars using Vink's formula for $\dot M$ \citep{ekstrom12,georgy13,eggenberger21}.
	A similar result is obtained by \citet{bjorklund22}, where evolution models adopting new mass loss rates are shifted to higher luminosities in HRD.
	It is important to remark that, to our knowledge, both \citet{bjorklund22} and \citet{alex22b} are the first studies on developing tracks for stellar evolution of massive stars adopting a self-consistent treatment for the stellar wind.

        In this work, we extend the self-consistent evolutionary tracks introduced in Paper I, including rotation to our models.
	Rotation is important, because it modifies the mass loss rate and the surface abundances due to the rotational mixing \citep{maeder10}, and because it makes the star not only mass but also losing angular momentum through the stellar wind \citep{georgy11,keszthelyi17}.
	Because of the extra details to be considered into the analysis of our results, this work includes only Galactic ($Z=0.014$) and LMC ($Z=0.006$) metallicities, leaving the supra-solar ($Z>0.014$) and SMC and lower-metallicity ($Z\le0.002$) cases to forthcoming studies.

        This paper is organised as follows.
        Section~\ref{selfconsistent} summarises the most relevant results and conclusions obtained from our self-consistent evolutionary tracks in Paper I.
        Section~\ref{rotationmdot} introduces the updates on $\dot M_\text{sc}$ to account for the effects of rotation.
        Section~\ref{rotationtracks} presents the new evolution models adopting self-consistent winds, whereas their comparison with observational diagnostics are analysed in Section~\ref{discussion}.
        In Section~\ref{implications}, we deliberate about the implications of these new evolution models in a Galactic scale.
        Finally, conclusions of this work are summarised in Section~\ref{conclusion}.

%_____Table Galactic case__________________________________________________________________________
	\begin{table*}[t!]
		\centering
		\caption{\small{Self-consistent line-force parameters $(k,\alpha,\delta)$ calculated for the stellar parameters corresponding to the positions indicated by circles in Fig.~\ref{hrd_initial}, together with their resulting terminal velocities and mass loss rates ($\dot M_\text{sc}$).
		Ratios showing the enhancement of mass loss rate due to rotation, adopting either our self-consistent procedure or MM00, are also shown in the last columns.}}
		\resizebox{\textwidth}{!}{
		\begin{tabular}{ccccccccccc|ccc|cccc}
			\hline
			\hline
			& \multicolumn{9}{c}{Input stellar parameters} & & & & &\\
			Name & $T_\text{eff}$ & $\log g$ & $R_*$ & $M_*$ & $\log L_*$ & He/H & C/C$_\odot$ & N/N$_\odot$ & O/O$_\odot$ & $\Omega$ & $k$ & $\alpha$ & $\delta$ & $\varv_\infty$ & $\log\dot M_\text{sc}$ & $\left(\frac{\dot M(\Omega)}{\dot M(0)}\right)_\text{sc}$ & $\left(\frac{\dot M(\Omega)}{\dot M(0)}\right)_\text{MM00}$\\
			& $[\text{kK}]$ & & $[R_\odot]$ & $[M_\odot]$ & $[L_\odot]$ & & & & & & & & & [km s$^{-1}$] & [$M_\odot\,\text{yr}^{-1}$]\\
			\hline
			\texttt{P120z10-01} & 52.0 & 4.12 & 15.8 & 120.0 & 6.22 & 0.085 & 1.0 & 1.0 & 1.0 & 0.0 & 0.125 & 0.566 & 0.026 & $3\,520\pm110$ & $-5.455\pm.100$\\%$^{90}$
			& & & & & & & & & & 0.4 & 0.113 & 0.578 & 0.025 & $3\,320\pm110$ & $-5.417\pm.100$ & $1.090$ & $1.092$\\%$^{90}$
			\texttt{P120z10-02} & 50.5 & 4.03 & 17.2 & 115.9 & 6.23 & 0.087 & 0.73 & 3.12 & 0.87 & 0.0 & 0.105 & 0.584 & 0.024 & $3\,360\pm110$ & $-5.407\pm.063$\\ %$^{5776}$
			& & & & & & & & & & 0.32 & 0.105 & 0.586 & 0.025 & $3\,230\pm110$ & $-5.387\pm.063$ & $1.057$ & $1.056$\\ %$^{5776}$
			\texttt{P120z10-03} & 52.0 & 4.01 & 16.9 & 106.5 & 6.27 & 0.155 & 0.08 & 11.66 & 0.09 & 0.0 & 0.106 & 0.584 & 0.028 & $3\,190\pm100$ & $-5.312\pm.065$\\
			& & & & & & & & & & 0.18 & 0.104 & 0.586 & 0.028 & $3\,150\pm100$ & $-5.307\pm.067$ & 1.013 & 1.013\\%$^{21151}
			\texttt{P120z10-04} & 54.0 & 4.00 & 16.3 & 96.1 & 6.31 & 0.308 & 0.05 & 12.35 & 0.02 & 0.0 & 0.101 & 0.573 & 0.025 & $2\,900\pm110$ & $-5.314\pm.073$\\%$^{42436}$
			& & & & & & & & & & 0.07 & 0.100 & 0.573 & 0.025 & $2\,890\pm110$ & $-5.312\pm.065$ & $1.004$ & $1.002$\\%$^{42436}$
			\hline
			\texttt{P070z10-01} & 49.0 & 4.18 & 11.3 & 70 & 5.82 & 0.085 & 1.0 & 1.0 & 1.0 & 0.0 & 0.159 & 0.537 & 0.027 & $3\,100\pm100$ & $-5.999\pm.081$\\
			& & & & & & & & & & 0.4 & 0.143 & 0.547 & 0.026 & $2\,930\pm100$ & $-5.969\pm.073$ & 1.072 & 1.078\\%$^{100}$
			\texttt{P070z10-02} & 47.0 & 4.06 & 12.7 & 67.8 & 5.85 & 0.085 & 0.94 & 1.45 & 0.97 & 0.0 & 0.154 & 0.557 & 0.023 & $3\,030\pm100$ & $-5.826\pm.070$\\
			& & & & & & & & & & 0.34 & 0.144 & 0.565 & 0.022 & $2\,910\pm100$ & $-5.803\pm.069$ & 1.054 & 1.050\\%$^{4225}
			\texttt{P070z10-03} & 47.0 & 3.98 & 13.6 & 63.7 & 5.91 & 0.116 & 0.40 & 7.51 & 0.47 & 0.0 & 0.135 & 0.576 & 0.026 & $2\,960\pm100$ & $-5.686\pm.079$\\
			& & & & & & & & & & 0.22 & 0.131 & 0.580 & 0.025 & $2\,900\pm100$ & $-5.672\pm.070$ & 1.029 & 1.022\\%$^{13924}
			\texttt{P070z10-04} & 49.0 & 3.97 & 13.3 & 59.8 & 5.96 & 0.199 & 0.15 & 10.86 & 0.16 & 0.0 & 0.112 & 0.574 & 0.029 & $2\,700\pm100$ & $-5.711\pm.068$\\
			& & & & & & & & & & 0.13 & 0.111 & 0.575 & 0.029 & $2\,710\pm100$ & $-5.708\pm.068$ & 1.007 & 1.006\\%$^{25301}
			\texttt{P070z10-05} & 52.5 & 3.97 & 12.5 & 53.3 & 6.03 & 0.481 & 0.05 & 12.24 & 0.03 & 0.0 & 0.105 & 0.560 & 0.027 & $2\,440\pm100$ & $-5.659\pm.073$\\	
			& & & & & & & & & & 0.05 & 0.104 & 0.561 & 0.027 & $2\,440\pm100$ & $-5.658\pm.067$ & 1.002 & 1.001\\%$^{49284}
			\hline
			\texttt{P040z10-01} & 43.5 & 4.22 & 8.1 & 40.0 & 5.33 & 0.085 & 1.0 & 1.0 & 1.0 & 0.0 & 0.247 & 0.468 & 0.019 & $2\,410\pm110$ & $-6.874\pm.110$\\%$^{90}$
			& & & & & & & & & & 0.4 & 0.214 & 0.479 & 0.018 & $2\,280\pm150$ & $-6.849\pm.146$ & $1.059$ & $1.069$\\%$^{90}$
			\texttt{P040z10-02} & 42.0 & 3.99 & 10.4 & 38.6 & 5.43 & 0.086 & 0.91 & 1.75 & 0.95 & 0.0 & 0.175 & 0.488 & 0.013 & $2\,180\pm100$ & $-6.742\pm.103$\\
			& & & & & & & & & & 0.34 & 0.160 & 0.498 & 0.014 & $2\,090\pm100$ & $-6.709\pm.093$ & 1.076 & 1.049\\%$^{4401}
			\texttt{P040z10-03} & 39.5 & 3.72 & 13.6 & 35.5 & 5.60 & 0.157 & 0.41 & 7.51 & 0.46 & 0.0 & 0.108 & 0.560 & 0.028 & $2\,050\pm110$ & $-6.268\pm.075$\\
			& & & & & & & & & & 0.16 & 0.106 & 0.562 & 0.029 & $2\,030\pm110$ & $-6.263\pm.073$ & 1.010 & 1.008\\%$^{20449}
			\texttt{P040z10-04} & 36.0 & 3.44 & 18.1 & 33.0 & 5.69 & 0.264 & 0.26 & 9.54 & 0.28 & 0.0 & 0.072 & 0.648 & 0.068 & $1\,990\pm160$ & $-5.855\pm.050$\\
			& & & & & & & & & & 0.04 & 0.072 & 0.649 & 0.069 & $1\,990\pm160$ & $-5.850\pm.050$ & 1.010& 1.001\\%$^{44581}$
			\texttt{P040z10-05} & 32.0 & 3.19 & 23.9 & 32.1 & 5.73 & 0.289 & 0.24 & 9.79 & 0.26 & 0.0 & 0.050 & 0.700 & 0.105 & $1\,770\pm440$ & $-5.761\pm.149$\\
			& & & & & & & & & & 0.02 & 0.050 & 0.700 & 0.105 & $1\,770\pm440$ & $-5.761\pm.149$ & $1.000$ & $1.000$\\%$^{62001}$
			\hline
			\texttt{P025z10-01} & 38.0 & 4.25 & 6.2 & 25.0 & 4.85 & 0.085 & 1.0 & 1.0 & 1.0 & 0.0 & 0.306 & 0.462 & 0.033 & $2\,120\pm100$ & $-7.510\pm.070$\\
			& & & & & & & & & & 0.4 & 0.287 & 0.462 & 0.032 & $1\,940\pm100$ & $-7.503\pm.088$ & $1.016$ & $1.064$\\%$^{100}$
			\texttt{P025z10-02} & 36.0 & 4.03 & 8.0 & 24.7 & 4.98 & 0.085 & 0.95 & 1.28 & 0.98 & 0.0 & 0.185 & 0.515 & 0.048 & $2\,070\pm110$ & $-7.237\pm.073$\\
			& & & & & & & & & & 0.39 & 0.171 & 0.532 & 0.056 & $1\,970\pm100$ & $-7.123\pm.075$ & 1.300 & 1.054\\%$^{1631}
			\texttt{P025z10-03} & 34.0 & 3.81 & 10.1 & 24.4 & 5.09 & 0.088 & 0.80 & 2.44 & 0.92 & 0.0 & 0.097 & 0.644 & 0.098 & $2\,370\pm110$ & $-6.615\pm.064$\\
			& & & & & & & & & & 0.38 & 0.090 & 0.653 & 0.099 & $2\,250\pm110$ & $-6.591\pm.059$ & 1.059 & 1.061\\%$^{3398}
			\texttt{P025z10-04} & 32.0 & 3.62 & 12.6 & 24.1 & 5.17 & 0.095 & 0.68 & 3.49 & 0.84 & 0.0 & 0.060 & 0.689 & 0.109 & $2\,390\pm120$ & $-6.558\pm.047$\\
			& & & & & & & & & & 0.37 & 0.058 & 0.691 & 0.108 & $2\,240\pm120$ & $-6.550\pm.038$ & 1.018 & 1.056\\%$^{5531}
			\texttt{P025z10-05} & 30.0 & 3.44 & 15.3 & 23.9 & 5.23 & 0.105 & 0.61 & 4.28 & 0.78 & 0.0 & 0.038 & 0.697 & 0.099 & $2\,230\pm230$ & $-6.740\pm.044$\\
			& & & & & & & & & & 0.35 & 0.036 & 0.705 & 0.099 & $2\,170\pm230$ & $-6.710\pm.053$ & $1.050$ & $1.051$\\%$^{7569}$
			\hline
		\end{tabular}}
		\label{standardtable}
	\end{table*}

%_____SELF-CONSISTENT EVOLUTIONARY MODELS______________________________________________________
\section{Self-consistent evolution models}\label{selfconsistent}
	\begin{figure}[t!]
		\centering
		\includegraphics[width=0.9\linewidth]{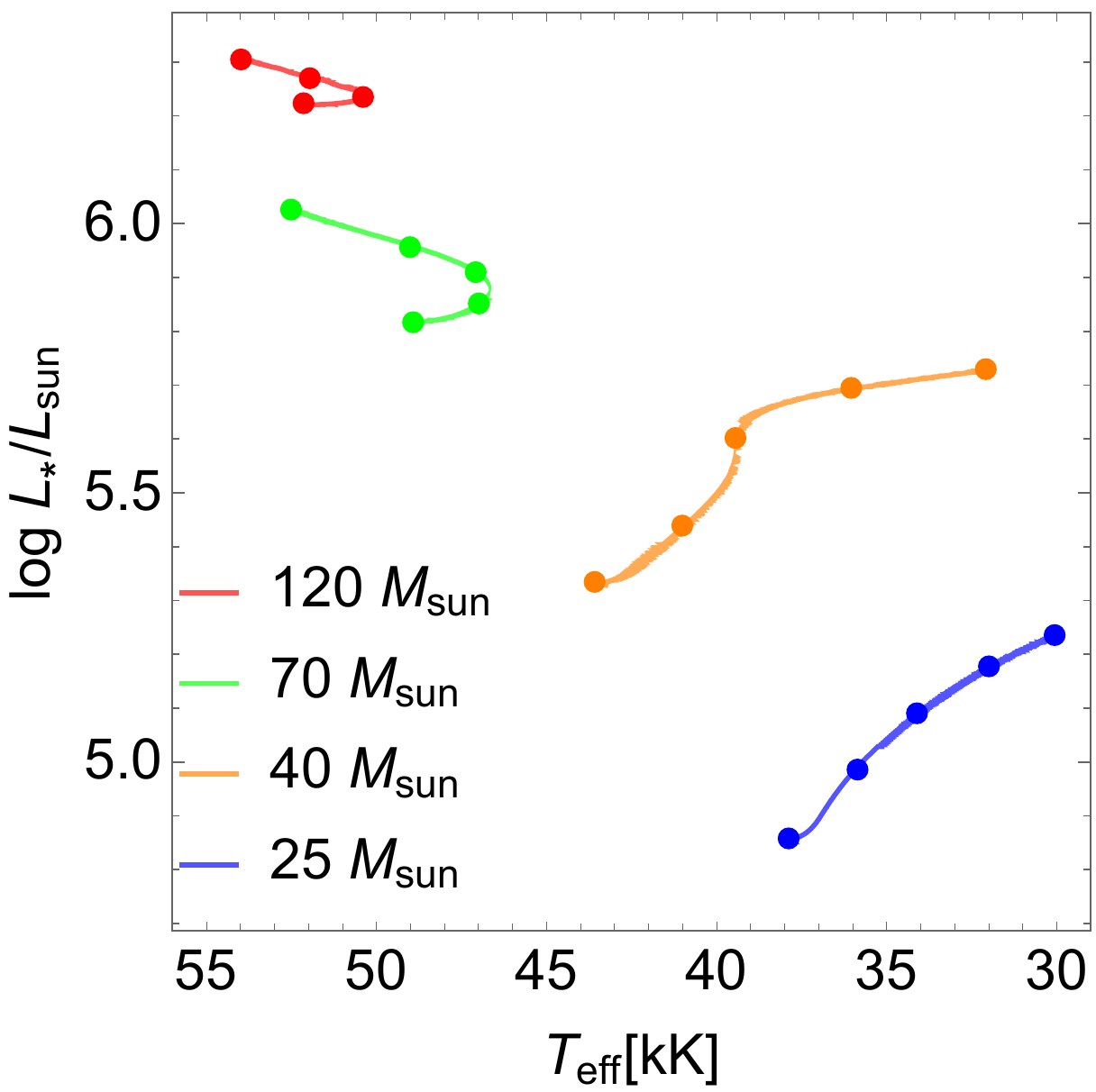}
		\caption{\small{Evolutionary tracks across the HRD for models with rotation using $\dot M_\text{Vink}$ and $Z/Z_\odot=1.0$ covering the main sequence stage, hereafter \textit{original tracks}.
		Circles represent the location of the selected stellar models to be tabulated in Table~\ref{standardtable} of Section~\ref{rotationmdot}.}}
		\label{hrd_initial}
	\end{figure}

	We call \textit{self-consistent evolutionary tracks} the evolution models that adopt a self-consistent recipe for the mass loss rate.
	Self-consistency implies that the wind hydrodynamics (i.e., velocity and density profiles for the wind) are calculated in agreement with the radiative acceleration (i.e., the acceleration over the wind particles due to the stellar radiation).
	As a consequence, mass loss rate is a direct solution for the wind under a specific set of stellar parameters (temperature, radius, mass, abundances) without assuming any extra condition a priori for the wind (such as a velocity law or a $\varv_\infty/\varv_\text{esc}$ ratio).
	For that purpose, we simultaneously calculate the line-acceleration $g_\text{line}$ by means of the so-called line-force parameters $(k,\alpha,\delta)$ from m-CAK theory \citep{cak,abbott82,ppk}, whereas we calculate the hydrodynamics by solving the stationary equation of motion for the wind using the code \textsc{HydWind} \citep{michel04}, for different sets of stellar parameters for massive stars.
	This iterative procedure is hereafter named \textit{m-CAK prescription}, it was introduced in \citet{alex19} and it has demonstrated to provide values for $\dot M$ of the same order of magnitude than those obtained with self-consistent NLTE studies in the co-moving frame such as \citet{kk17,kk18} or \citet{alex21}.
	Moreover, the wind hydrodynamics used as input in the NLTE radiative transport code is able to reproduce reliable synthetic spectra, in agreement with the observations \citep{alex22a}.

	However, although the calculation of the line-force parameters has been extended for even cooler temperatures \citep{lattimer21,poniatowski22}, in \citet{alex19} our m-CAK prescription is restricted into the range of stellar parameters where $(k,\alpha,\delta)$ can be assumed as constants.
	This is an important consideration because, even though line-force parameters have been commonly treated as constants \citep{shimada94,noebauer15}, there are studies which suggest that the values for $k$, $\alpha$ and $\delta$ may radially change with the wind \citep{schaerer94,puls00}.
	About this, \citet[][Sec.~2]{alex22a} performed an quantitative analysis on the standard deviation and numerical errors in the calculation of our line-force parameters, and by extension on the error margins for the self-consistent wind parameters $\dot M$ and $\varv_\infty$, finding that uncertainties for $(k,\alpha,\delta)$ can be neglected for stars with $T_\text{eff}\ge30$ kK and $\log g\ge3.2$.
	Therefore, we set these thresholds as the range of validity for our m-CAK prescription.
	
	Thus, self-consistent mass loss rate ($\dot M_\text{sc}$) derived from the m-CAK prescription is parametrised as a function of stellar parameters as shown in Eq.~7 from Paper I:
	\begin{align}\label{mdotformula1}
		\log\dot M_\text{sc,predicted}=&-40.314 + 15.438\,w + 45.838\,x - 8.284\,w\,x \nonumber\\
		&+ 1.0564\,y -  w\,y / 2.36 - 1.1967\,x\,y + 11.6\, z \nonumber\\
		&- 4.223\,w\,z - 16.377\,x\,z + y\,z / 81.735\;,
	\end{align}
	where $w$, $x$, $y,$ and $z$ are defined as:
	\begin{equation}
		w=\log \left(\frac{T_\text{eff}}{\text{kK}}\right)\;,\nonumber\\
		x=\frac{1}{\log g}\;,\\
		y=\frac{R_*}{R_\odot}\;,\\
		z=\log \left(\frac{Z_*}{Z_\odot}\right)\;.
	\end{equation}
	
	This intervariable fitting, besides minimising the separation with respect to the real calculated wind solution by line-force parameters, allows a deeper examination of the dependance of mass loss rate on stellar parameters.
	As an example, we found from Eq.~\ref{mdotformula1} that dependency in metallicity does not follow the classical exponential law $\dot M\propto Z^a$ with a constant exponent, but with $a$ ranging from $\sim0.53$ for $M_*\sim120$ $M_\odot$ to $\sim1.02$ for $M_*\sim25$ $M_\odot$.
	In other words, we can approximate this $Z$-dependence for mass loss rate as an intrinsic relationship of stellar mass, as
	\begin{equation}\label{zm_dependence}
		\dot M_\text{sc}\propto Z^{0.4+\frac{15.75}{(M_*/M_\odot)}}\;.
	\end{equation}

	An analogous result, but for an intrinsic luminosity-dependence, was found by \citet{kk18} for their global wind models; where the exponent for the metallicity was less steep for higher stellar luminosity (and subsequently, stellar mass).
	Explanations of this intrinsic mass-dependence may rely on the fact that the more massive the star, the nearer it is from the Eddington limit.
	In that case, the continuum should become increasingly important in contributing to the radiative acceleration $g_\text{rad}$, in decline of the line-driving \citep[see Eq.~6 from][]{alex21}.
	The effect of the continuum is less dependent on metallicity because it does not involve interaction between the radiative field and absorption lines.
	This condition for the mass loss rate was previously pointed out by \citet{grafener08} for WNL stars, where the $Z$-dependence was weaker closer to the Eddington limit (i.e., more mass).
	All in all we can state that Eq.~\ref{zm_dependence} is an important step to understand the dependence on metallicity for the mass loss rates at the earlier evolutionary stages of massive stars.
	
	Our new recipe for mass loss rate is implemented inside the \textsc{Genec} code, which runs stellar evolutionary models to explain and predict the physical properties of massive stars \citep{eggenberger08}.
	We used the same prescriptions as described in \citet{ekstrom12}; that is, solar abundances from \citet{asplund05,asplund09}, opacities from \citet{iglesias96}, and overshoot parameter $\alpha_\text{ov}\simeq0.10$, except for the mass loss rates.
	The self-consistent m-CAK prescription has been already incorporated inside \textsc{Genec} for stars satisfying $T_\text{eff}\ge30$ kK and $\log g\ge3.2$.
	Below such thresholds, the recipe for mass loss rate reverts to Vink's formula.
	
	Because self-consistent mass loss rates are a factor of $\sim3$ lower with respect to Vink's formula at the beginning of MS \citep[i.e., for $T_\text{eff}\gtrsim40$ kK and $\log g\sim4.0$, see][]{alex22a}, self-consistent evolution models predict that stars will retain more mass during their H-core burning stage and therefore they are larger and more luminous.
	These changes are proportional to the initial mass of the star; from important differences in the final stellar properties at the end of MS for $M_\text{zams}=120\,M_\odot$, to almost negligible differences for $M_\text{zams}=25\,M_\odot$.
	Also, evolution models adopting $\dot M_\text{sc}$ predict a smaller production of the isotope Aluminium-26, which means that massive stars contribute in a lesser proportion to feed the interstellar medium with this isotope, compared to other sources \citep{palacios05,wang09}.
	
	Such results are important, but still represent only a first glance to the implications that the self-consistent wind models have for stellar evolution.
	The next step is the incorporation of rotation.
	Stellar rotation is important for stellar evolution because of two reasons.
	First, because the angular momentum in the core at the time of core collapse may strongly impact the final stages of a massive star \citep{yoon06}.
	Powerful stellar winds not only make stars lose mass, but also angular momentum, subsequently affecting the rotation and the evolution of the star \citep{georgy11,keszthelyi17}.
	And second, because rotational mixing \citep{maeder10} changes the distribution of the abundances of the elements inside the star and in particular at the surface \citep[see, e.g.,][]{brott11,ekstrom12}.
	The changes of the abundances at the surface is particularly interesting, because modification in the chemical composition of the atmosphere induces variations in the line-acceleration and therefore in the self-consistent mass loss rate, something that it has been either not considered or neglected by previous authors \citep{bjorklund22}.
	For that reason, before running new evolution models with rotation adopting $\dot M_\text{sc}$, it is necessary to study the influence of rotation over the m-CAK prescription.

%_____ROTATION INSIDE m-CAK PRESCRIPTION
	\begin{figure*}[t!]
%		\sidecaption
		\centering
		\begin{subfigure}{0.45\linewidth}
			\includegraphics[width=\linewidth]{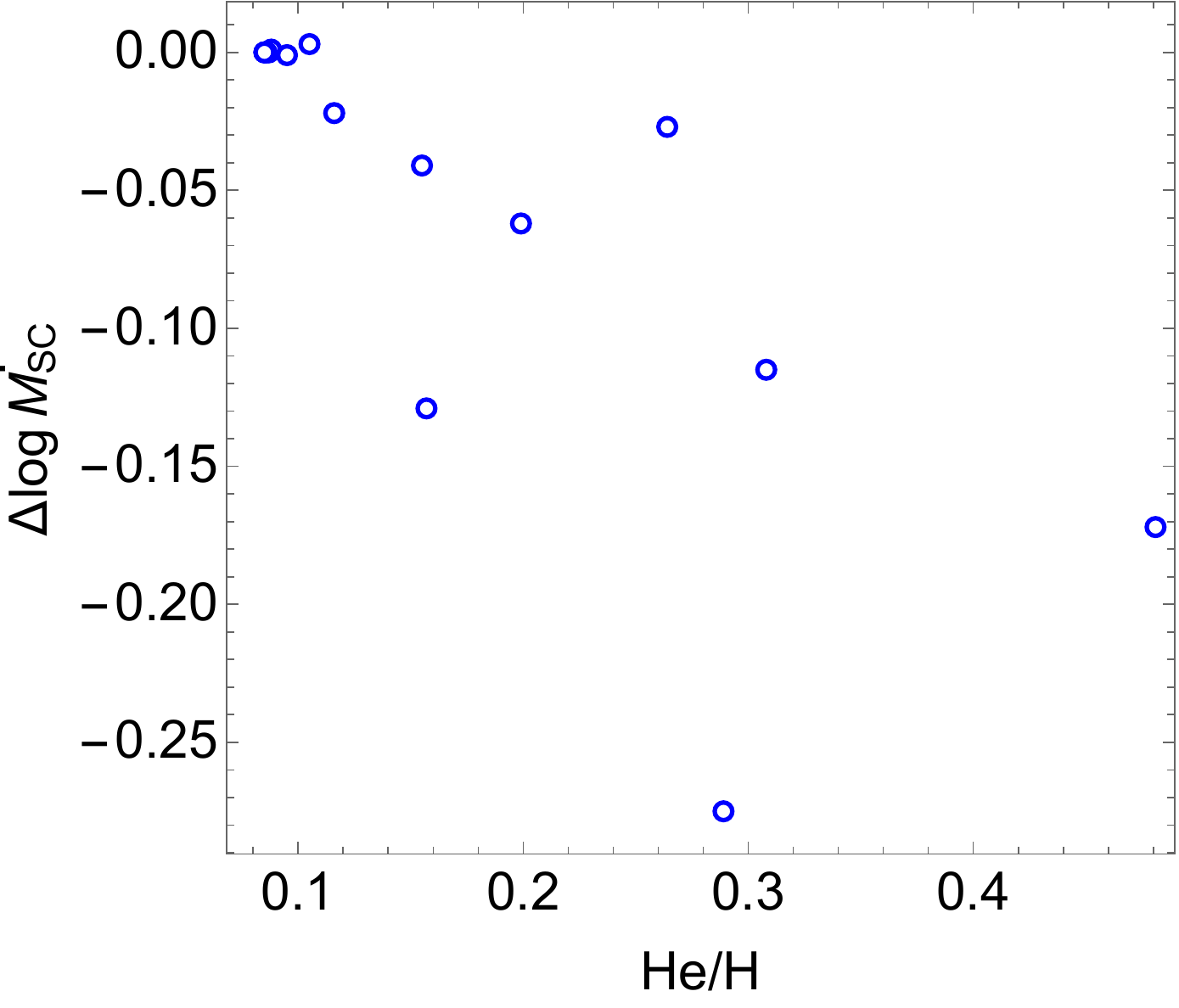}
			\caption{ }
			\label{heh_mdots}
		\end{subfigure}
		\hspace{1cm}
		\begin{subfigure}{0.45\linewidth}
			\includegraphics[width=\linewidth]{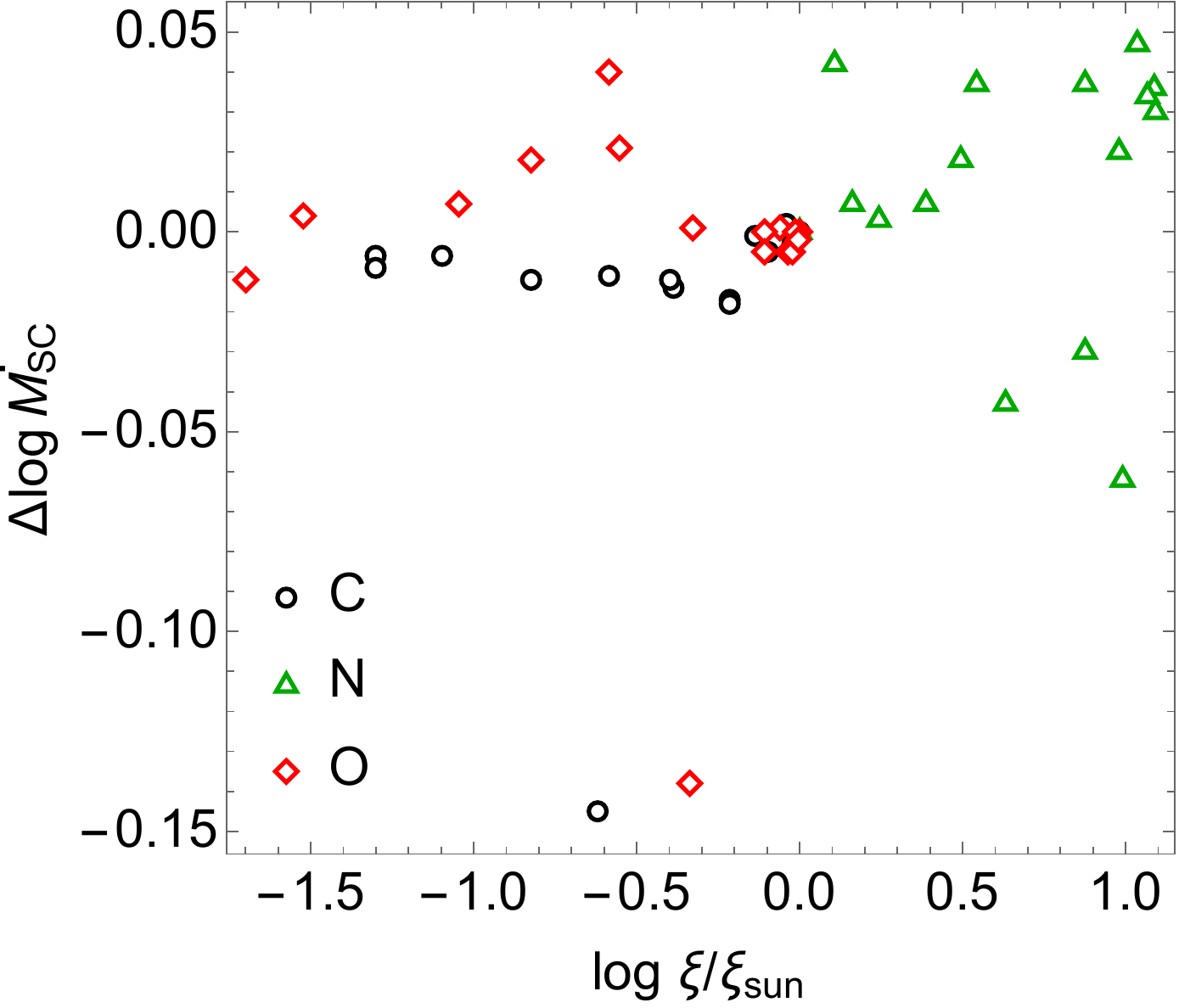}
			\caption{ }
			\label{cno_mdots}
		\end{subfigure}
		\caption{\small{{Differences in the resulting self-consistent mass loss rate $\Delta\log\dot M_\text{sc}$ for the abundances tabulated in Table~\ref{standardtable}, (a) due to the modification of the He/H ratio with respect to the default He/H$=0.085$, (b) due to the modification of any of the CNO elements with respect to $\xi/\xi_\odot=1.0$ (with $\xi$ being carbon, nitrogen or oxygen depending on the respective symbol).}}}
		\label{abund_mdots}
	\end{figure*}

\section{Rotation at the m-CAK prescription}\label{rotationmdot}
\subsection{Enhancing mass loss due to rotational effects}
	Models by \textsc{Genec} are computed from the formula by Vink \citep{vink01} with the correcting factor induced by rotation as given by \citet[][hereafter MM00]{maeder00}, that is
	\begin{eqnarray}\label{rotationmm00}
		\dot M(\omega)&=&F_\omega\dot M(0)\nonumber\\
		&\text{ with } & F_\omega=\frac{(1-\Gamma_\text{Edd})^{\frac{1}{\alpha}-1}}{\left[1-\frac{\omega^2}{2\pi G\rho_m}-\Gamma_\text{Edd}\right]^{\frac{1}{\alpha}-1}}\;,
	\end{eqnarray}
	where $\dot M(0)$ is the mass loss rate without rotation, $\alpha$ is the line-force parameter from CAK theory, $\omega$ is the angular velocity at the stellar surface, and $\Gamma_\text{Edd}$ is the Eddington factor
	\begin{equation}\label{eddington}
		\Gamma_\text{Edd}=\frac{L}{L_\text{Edd}}=\frac{\kappa L}{4\pi cGM}\;,
	\end{equation}
	with $\kappa$ being the total opacity.
	Notice that  Eq.~\ref{rotationmm00} introduces a correction factor independent of the adopted recipe for mass loss rate without rotation; in other words, $\dot M(0)$ can be computed either from Vink's formula or from Eq.~\ref{mdotformula1}.
	Actually, $\dot M(\omega)$ results from an integration over the surface and the integration accounts for the change of the local mass loss rate with the latitude.
	
	Equation~\ref{rotationmm00} incorporates the impacts on mass loss rate produced not only by rotation but also the radiative acceleration due to the continuum (hence the presence of the Eddington factor), through the so-called $\Omega\Gamma$-limit.
	This formulation implies that the stellar break-up or $\Omega$-limit (i.e., when the centrifugal forces compensate gravity in a rotating star) is reached for reduced rotation velocities if $\Gamma$ is large enough.
	The velocity associated to this break-up is the so called critical rotational speed,
	\begin{equation}\label{vcrit}
		\varv_\text{crit}=\sqrt{\frac{GM_*(1-\Gamma_\text{Edd})}{R_*}}\,,
	\end{equation}
	and hence, the rotation of the star is commonly expressed as a dimensionless ratio between the rotational velocity and this critical value as
	\begin{equation}\label{omega_def}
		\Omega=\frac{\varv_\text{rot}}{\varv_\text{crit}}\,.
	\end{equation}
	
	On the other hand, the m-CAK prescription incorporates rotational effects in the hydrodynamic solution calculated by \textsc{HydWind} \citep{michel04}, also using $\Omega$ as an input.
	\textsc{Hydwind} solves the equation of motion for the stationary standard line-driven theory,
	\begin{equation}\label{eommichel}
		\varv\frac{d\varv}{dr}=-\frac{1}{\rho}\frac{dP}{dr}-\frac{GM(1-\Gamma)}{r^2}+\frac{\varv_\phi^2(r)}{r}+g_\text{line}\;,
	\end{equation}
	by finding the eigenfunction that satisfies the velocity profile, $\varv=\varv\,(r)$, and the eigenvalue, which is proportional to the mass loss rate, $\dot M$ \citep{michel07}.
	The rotation factor $\Omega$ is implicitly included in Eq.~\ref{eommichel} by means of $\varv_\phi=\varv_\text{rot}R_*/r$, whereas the additional terms are the same as described by \citet{araya17,araya21} or \citet{alex22a}.
	In this work, we use moderate to intermediate values of $\Omega$ ($=0.4$), therefore all our wind solutions are in the \textit{fast} wind regime and not in the \textit{$\Omega$-slow} wind regime which starts approximately from $\Omega \gtrsim 0.75$ \citep{araya18}.
		
	In order to evaluate the effect of the rotation over the self-consistent wind solutions, we repeat the methodology of Paper I and we run evolutionary rotating models in \textsc{Genec} for initial masses of $120$, $70$, $40$, and $25$ $M_\odot$ adopting Vink's formula for these tracks, hereafter \textit{original tracks}, as shown in Fig.~\ref{hrd_initial}.
	Then we select a set of representative points over these original tracks, and based on their stellar parameters ($T_\text{eff}$, $\log g$, $R_*/R_\odot$, $\Omega$) we calculate the line-force parameters $(k,\alpha,\delta)$ and their respective self-consistent wind parameters $\dot M_\text{sc}$ and $\varv_{\infty,\text{sc}}$.
	Initial rotational velocity for all evolution models is $\Omega=0.4$, as in \citet{ekstrom12}.
	The results are tabulated in Table~\ref{standardtable} where, besides the former columns already included in the Tables~2, 3 and 4 from Paper I, we add the rotation factor $\Omega$ as an extra stellar parameter for the input.
	For each one of the points, we calculate an independent solution for $(k,\alpha,\delta)$ for the tabulated $\Omega$ but also for $\Omega=0.0$ (i.e., without rotation), with their respective self-consistent solution for terminal velocity and mass loss rate.
	The difference in the self-consistent mass loss due to the incorporation of rotation is expressed as the ratio $\dot M(\Omega)/\dot M(0)$, which is compared with the correction factor predicted by MM00 (i.e., Eq.~\ref{rotationmm00}).
	Here, we found that both approaches for the rotational effects over $\dot M$ produce similar results, which is remarkable considering the error bars associated with the self-consistent mass loss rate.
	Based on this outcome, we decide to perform our new evolutionary tracks keeping Eq.~\ref{rotationmm00} as correction factor for $\dot M_\text{sc}$.

	The other physical ingredients related to rotation are the same as in \citet{ekstrom12}, \citet{georgy13} and \citet{eggenberger21}.
	Angular momentum is transported by an advective equation as described in \citet{zahn92}.
	The horizontal and vertical shear diffusion coefficients are from respectively \citet{zahn92} and \citet{maeder97}.

%_____VARIATION IN ABUNDANCES____________________________________________________________________
	\begin{figure*}[t!]
		\centering
		\includegraphics[width=0.455\linewidth]{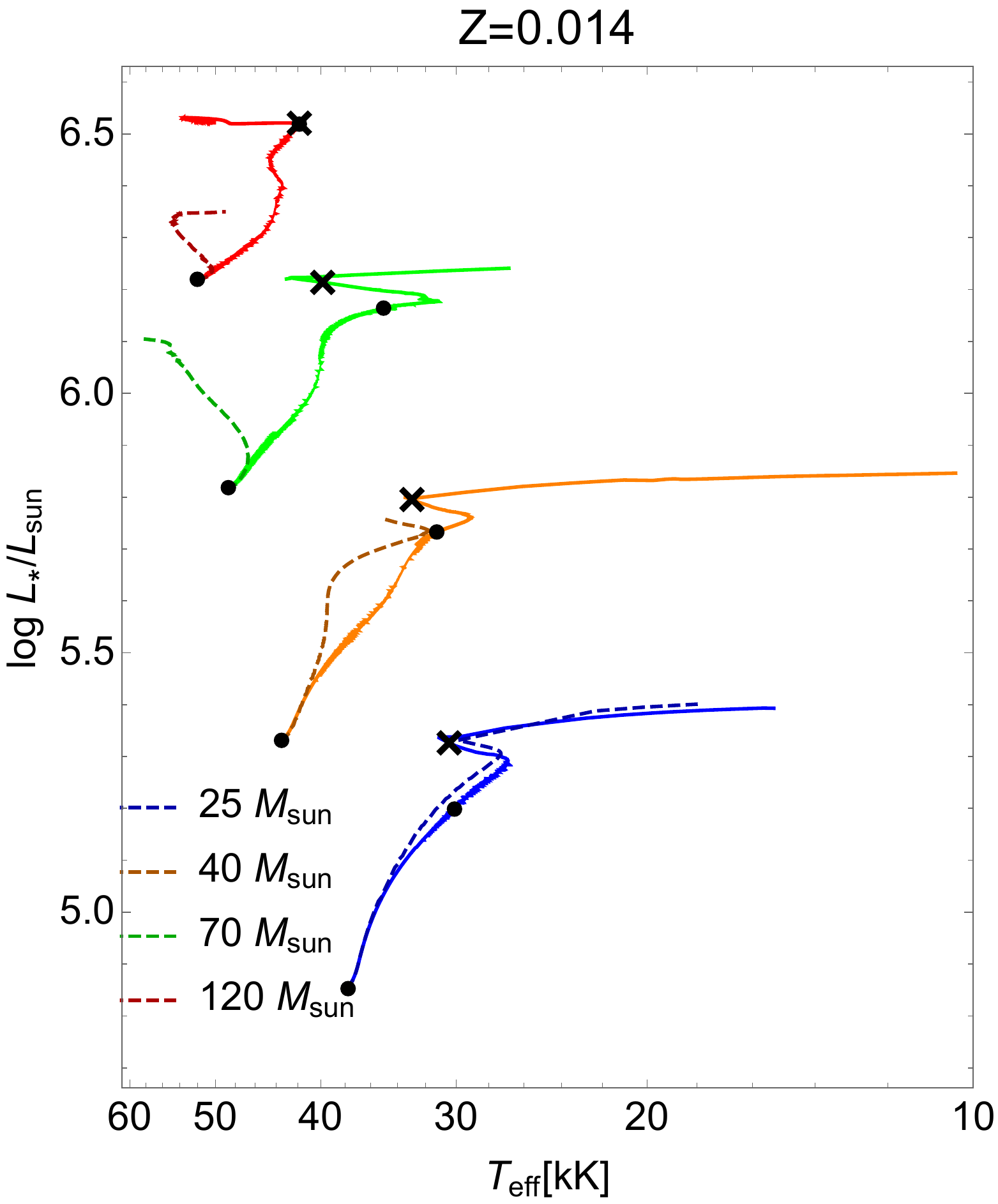}
		\hspace{1cm}
		\includegraphics[width=0.45\linewidth]{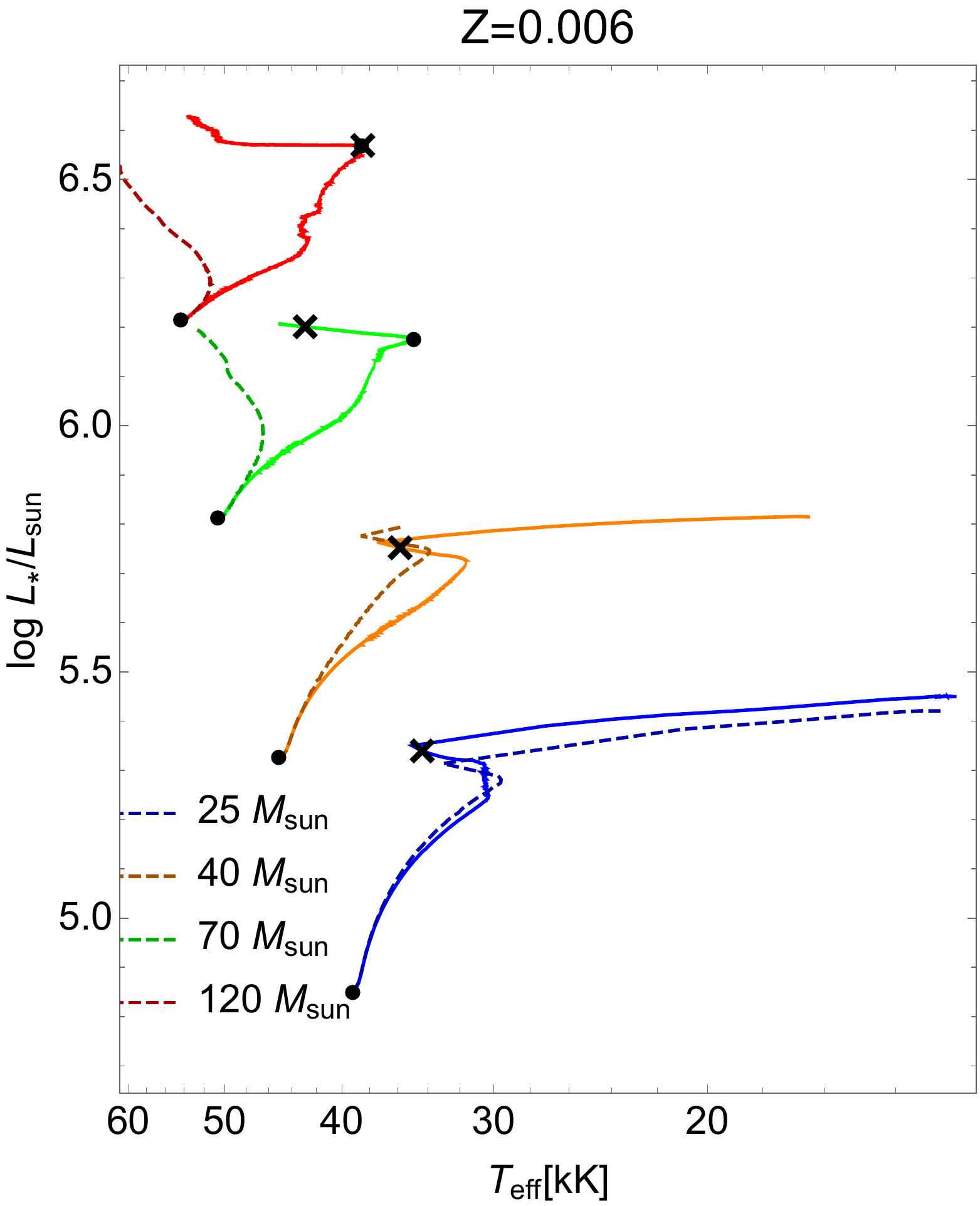}
		\caption{\small{HR diagram for the self-consistent evolution models (solid lines), compared with classical evolutionary tracks (darker dashed lines) from \citet[][$Z=0.014$]{ekstrom12} and \citet[][$Z=0.006$]{eggenberger21}.
		All plots cover the main sequence and part of the Hertzsprung gap.
		Self-consistent tracks are marked with two black dots which represent the zero-age main sequence and the switch of the mass loss recipe from self-consistent to Vink's formula (when either $T_\text{eff}=30$ kK or $\log g=3.2$), whereas the end of the main sequence (terminal-age main-sequence, TAMS) is represented with a black cross.}}
		\label{hrd_final}
	\end{figure*}
	\begin{figure*}[t!]
		\centering
		\includegraphics[width=0.42\linewidth]{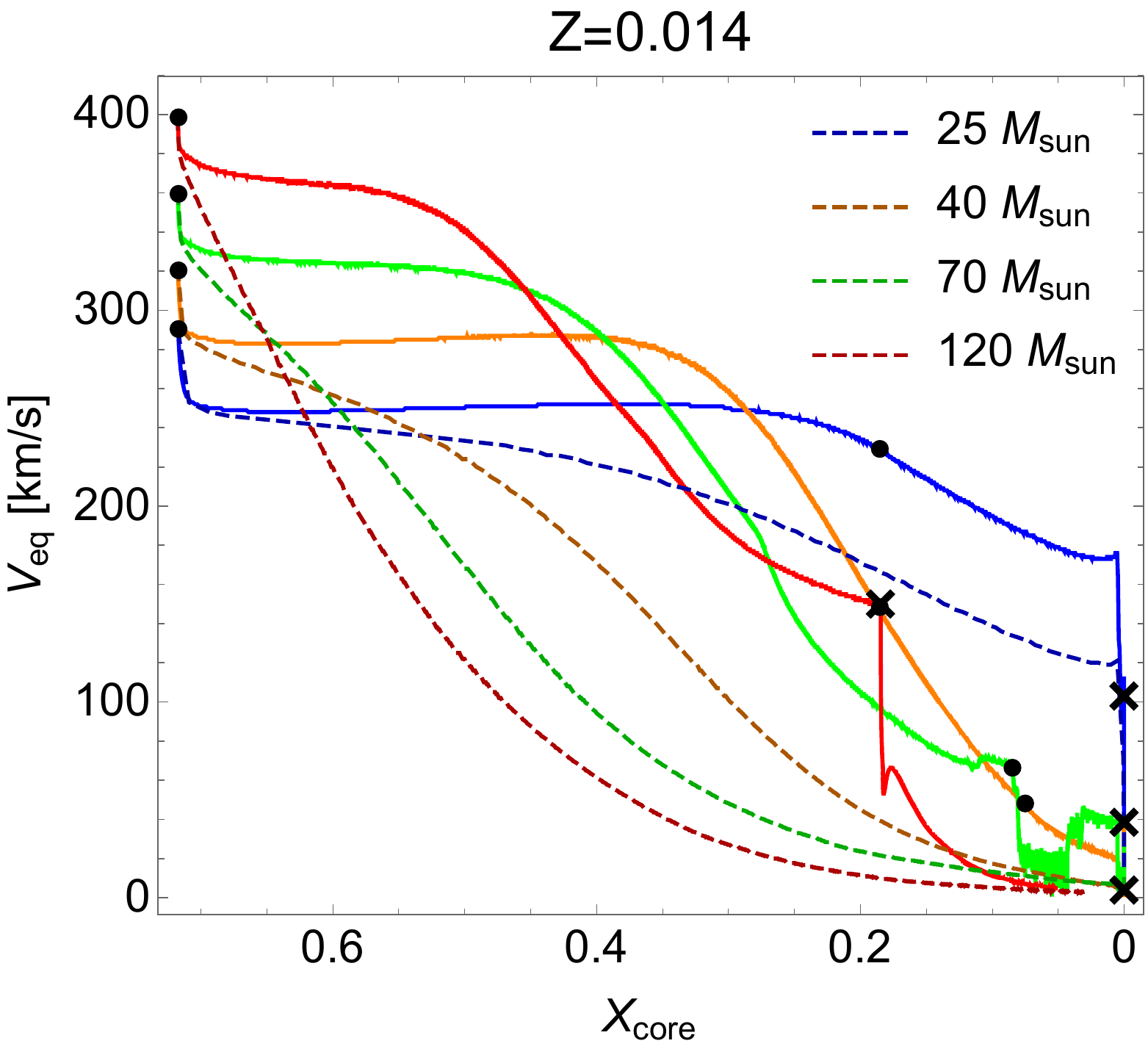}
		\hspace{1cm}
		\includegraphics[width=0.42\linewidth]{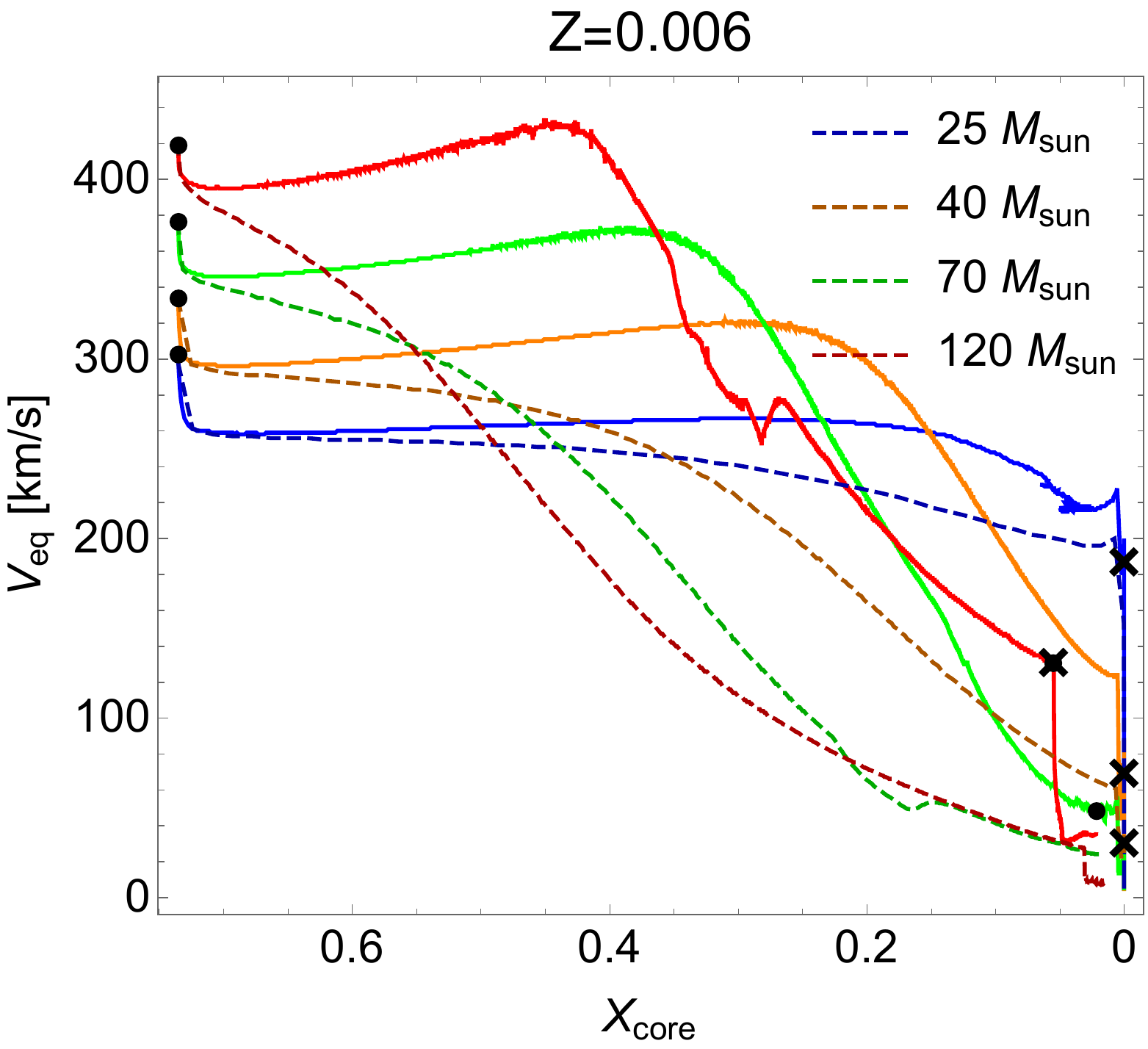}\\
		\includegraphics[width=0.42\linewidth]{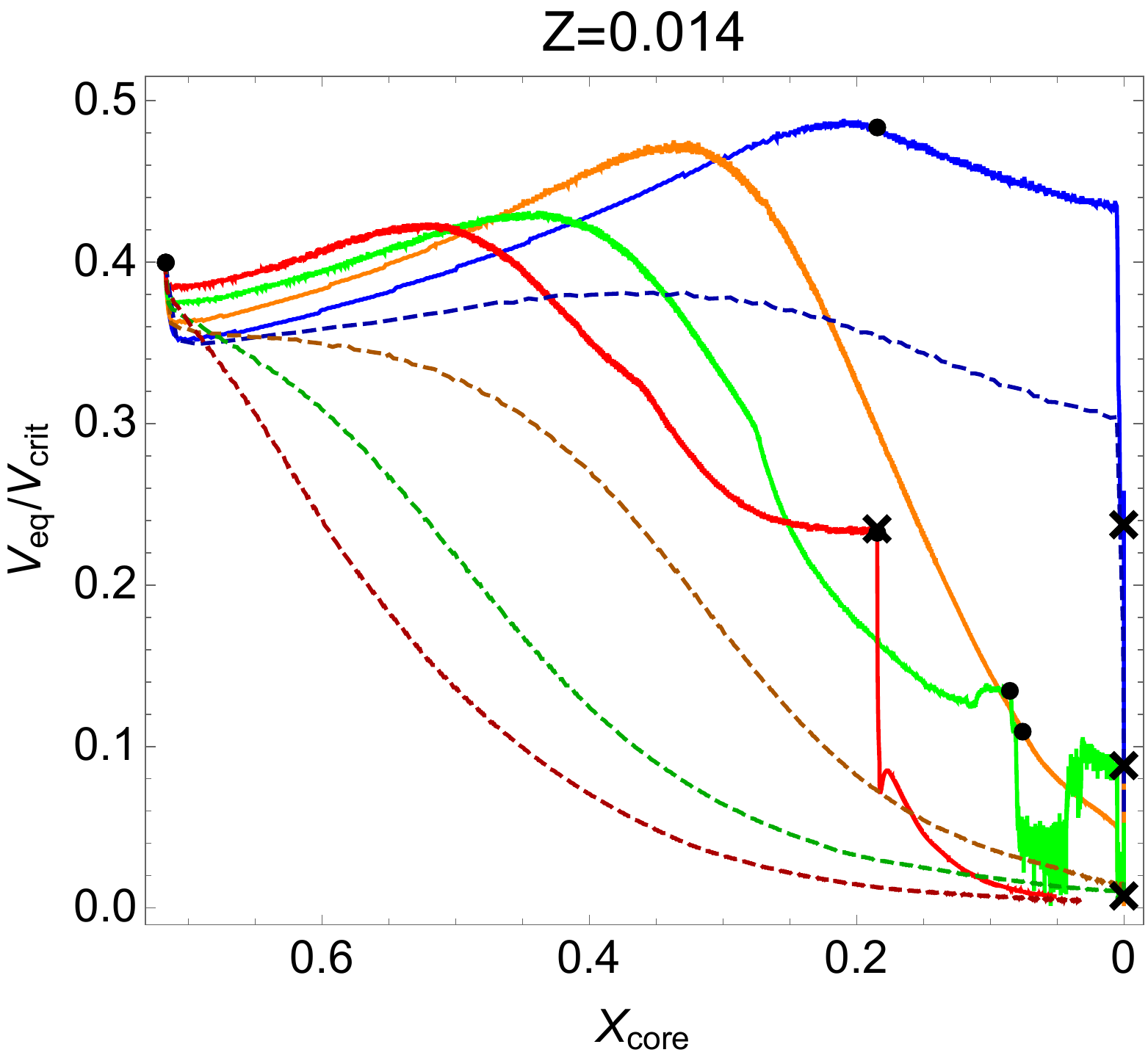}
		\hspace{1cm}
		\includegraphics[width=0.42\linewidth]{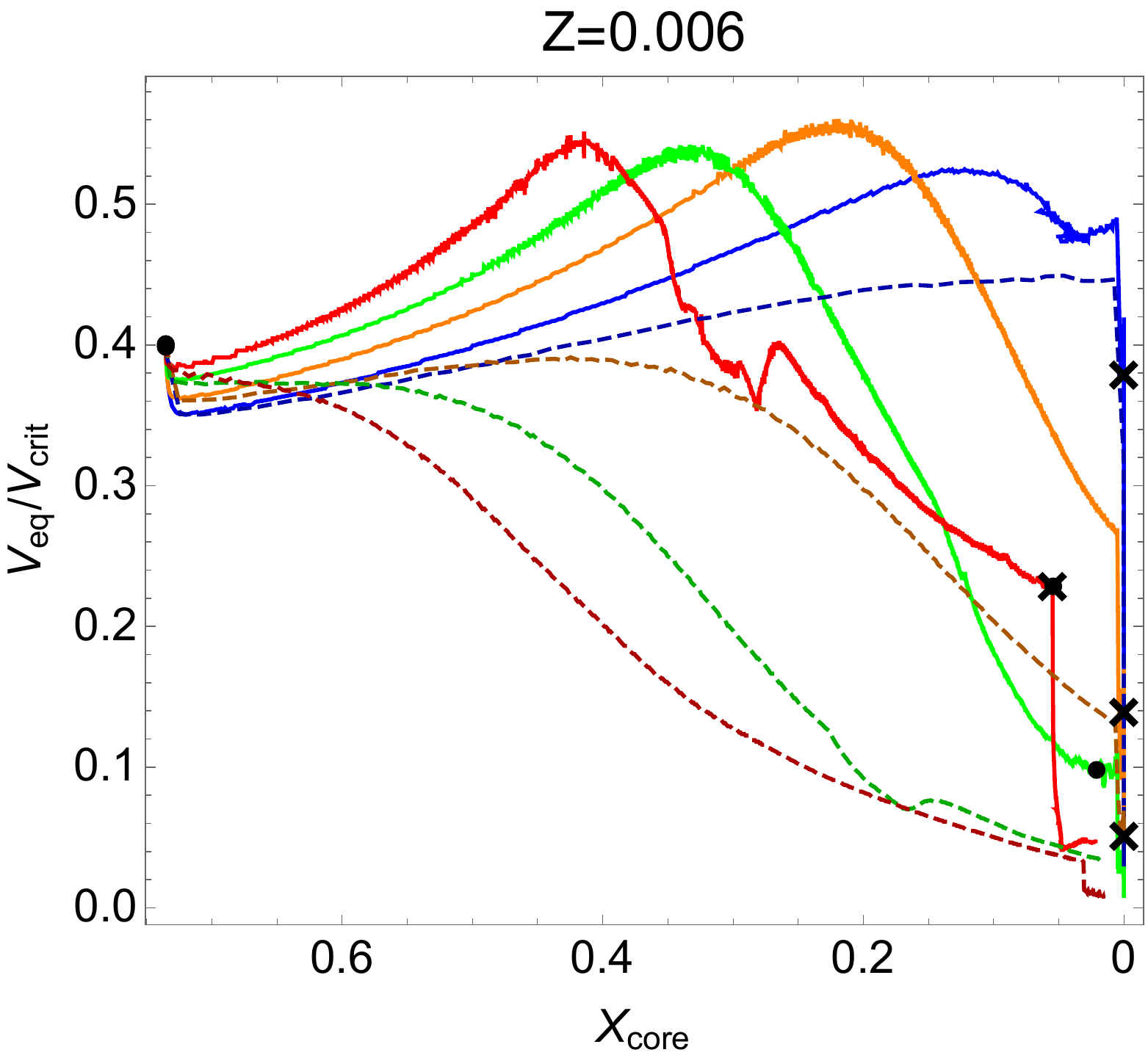}
		\caption{\small{Evolution of surface equatorial velocities and rotation parameter $\Omega$ as a function of the mass fraction of hydrogen at the core, which decreases from $X_\text{core}\sim0.72$ at ZAMS to $X_\text{core}\sim0.05$ at the end of the H-core burning stage.
		The symbols (dots and crosses) have the same meaning as indicated in the caption of Fig.~\ref{hrd_final}.}}
		\label{veq_final}
	\end{figure*}

%_____Variation in abundances
	\begin{table}[t!]
		\centering
		\caption{\small{Summary of the formulae implemented for old and new evolutionary tracks.}}
		\begin{tabular}{ccc}
			\hline\hline
			& Old winds & New winds\\
			\hline
			$\dot M$ without rotation & \citet{vink01} & Eq.~\ref{mdotformula1} (Paper I)\\
			Corr. by rotation & Eq.~\ref{rotationmm00} (MM00) & Eq.~\ref{rotationmm00} (MM00)\\
			Corr. by He/H & none & Eq.~\ref{mdotformula2}\\
			\hline
		\end{tabular}
		\label{tablerecipes}
	\end{table}
\subsection{Variation in abundances}
	Besides the stellar parameters such as temperature, surface gravity, radius and rotation, we included in Table~\ref{standardtable} the values for the abundances of helium (expressed as a fraction of the hydrogen abundance by number) and the individual abundances of carbon, nitrogen and oxygen (expressed as a fraction of the solar abundances).
	Due to the nature of the CNO-cycle, the abundances of helium and nitrogen increase over time, whereas carbon and oxygen decrease \citep{przybilla10}.
	The chemical composition of the surface is enriched in material processed in the core by rotational mixing in the outer radiative envelope.
	Since we assume that chemical composition in the stellar wind is \textit{exactly the same} as in the photosphere, the chemical composition of the stellar wind is altered.
	The question that arises here is, do these alterations in the wind chemical abundances affect the line-acceleration and afterwards the self-consistent mass loss rate?

	To solve that question, we artificially alter the individual abundances for the He and the CNO elements from our stellar models of Table~\ref{standardtable}, and then calculate the respective line-force parameters and self-consistent mass loss rate for each one of the modifications.
	For example, we compare the $\dot M_\text{sc}$ for the stellar model \texttt{P070z10-04} (with N/N$_\odot~=10.86$) with the expected $\dot M_\text{sc}$ if N/N$_\odot~=~1$, whereas the other individual abundances and the rest of stellar parameters ($T_\text{eff}$, $\log g$, etc.) are kept fixed.
	Same for the helium: we compare the self-consistent mass loss rate for, e.g., He/H$~=0.20$ with the expected value for $\dot M_\text{sc}$ if He/H$~=0.085$.
	These differences are expressed as
	\begin{equation}\label{deltamdot}
		\Delta\log\dot M_\text{sc}=\log\dot M_\text{sc}-\log\dot M_{\text{sc},\odot}\;,
	\end{equation}
	with $\dot M_\text{sc}$ the mass losses tabulated in Table~\ref{standardtable}, and $\dot M_{\text{sc},\odot}$ the self-consistent mass loss rate calculated if the abundance of the evaluated element were solar.
	
	Results of these modifications over the $\dot M_\text{sc}$ are displayed in Fig.~\ref{heh_mdots} for changes in the He/H ratio, and in Fig.~\ref{cno_mdots} for modifications of the CNO elements.	
	We see that the variation of each one of the CNO elements even by a factor of $\sim10$ barely affects the $\dot M_\text{sc}$, keeping the difference $\Delta\log\dot M_\text{sc}$ close to zero.
	Indeed, differences in mass loss rate are just of the order of $\pm0.05$ in log scale ($\sim12\%$ of $\dot M_\text{sc}$), smaller than the error bars shown in Table~\ref{standardtable}.
	On the contrary, the increase in the He/H ratio makes $\dot M_\text{sc}$ decrease in larger magnitudes.

	Results from Fig.~\ref{abund_mdots} are quickly explained due to their absolute numbers of the evaluated abundances.
	Even though the changes in CNO abundances are by a factor of $\sim10$ (compared with the initial values at ZAMS), they still represent a small percentage of the total composition of the star.
	On the contrary, the abundance of helium plays a major role because it grows (in mass fraction) from $\sim26\%$ up to $\sim70\%$, thus significantly changing the mean molecular weight of the wind.
	Also, helium abundance is involved in the computation of the ratio $N_e/\rho$ and in the radiative acceleration due to Thomson scattering\footnote{Here, $\Gamma$ is the Eddington factor when only free electron scattering opacity is accounted for (the so called classical Eddington factor).}
	\begin{equation}\label{densitymichel}
		\frac{N_\text{e}}{\rho}=\frac{1}{m_\text{p}}\frac{1+2\,X_\text{He}}{1+4\,X_\text{He}}\;,
	\end{equation}
	\begin{equation}\label{radiationthomson}
		\Gamma=\frac{N_\text{e}}{\rho}\left(\frac{\sigma_\text{Th}L_*}{4\pi cGM_*}\right)\;,
	\end{equation}
	where $X_\text{He}$ is the helium abundance by number, $N_\text{e}$ is the density of free electrons, and $m_\text{p}$ is the proton mass \citep[the meaning of the other symbols are detailed in][Sec.~2]{michel04}.
	Hence, when the helium abundance increases, the number of free electrons is reduced and then $\Gamma$ decreases.
	As a consequence, the effective mass $M_\text{eff}=M_*(1-\Gamma)$ becomes larger in Eq.~\ref{eommichel}, and accordingly effective gravity, meaning that the line-acceleration now is able to remove less material as stellar wind (i.e., mass loss rate decreases).
	However, it is important to mention that this theoretical analysis is made upon the basis of the isolated alteration of the He/H ratio, and therefore this decreasing in the resulting $\dot M$ does not imply any modifications in temperature or in any other stellar parameter.
	For this reason, it differs from the $\dot M\propto X_\text{He}$ relationship found for \citet{hamann95}, which was the result of a global analysis over the wind regime of WR stars.
	
	Thus we decide to introduce an extra variable to the recipe of Eq.~\ref{mdotformula1}, based on a simple fit of Fig.~\ref{heh_mdots}
	\begin{equation}\label{mdotformula2}
		\log\dot M_\text{sc}=\log\dot M_\text{sc,0}-0.559\times\left[\left(\text{He}/\text{H}\right)_*-\left(\text{He}/\text{H}\right)_0\right]\;,
	\end{equation}
	where (He/H)$_0$ is the He/H ratio at ZAMS \citep[0.085 from][]{asplund09}.
	We see from Table~\ref{standardtable} that (He/H)$_*$ may grow up to $\sim0.3-0.45$ at the end of the self-consistent regime, thus leading to a reduction of the mass loss rate of around $\sim30-40\%$.
	However, same as in Paper I, this modification applies only when the self-consistent mass loss rate is adopted in the evolutionary tracks, for $T_\text{eff}\ge30$ kK and $\log g\ge3.2$, and hence it is still unknown the magnitude of the decreasing factor in $\dot M$ prior to the end of the MS, when hydrogen abundance is $X=0.3$ at the surface.
	
	Finally, details of the formulae used for both old and new evolution models are summarised in Table~\ref{tablerecipes}.

%_____RESULTS____________________________________________________________________________________
% Starts with Z=0.014: 24,22,21,19
% End of H-burning with Z=0.014: 845, 3782, 9395, 4096
% Starts with Z=0.006: 24, 22, 19, 18
% End of H-burning with Z=0.006: 615, 1399, 2889, 4228
	\begin{table*}[t!]
		\centering
		\caption{\small{Properties of the stellar models at the end of the Main Sequence.}}
		\begin{tabular}{cccc|ccccccc}
			\hline
			\hline
			$M_\text{zams}$ & $Z$ & $\varv_\text{rot,ini}$ & mass loss & \multicolumn{7}{c}{End of Main Sequence}\\
			& & & & $t_\text{MS}$ & $M_*$ & $R_*$ & $\varv_\text{rot,surf}$ & $Y_\text{surf}$ & N/C & N/O\\
			\hline
			% Mass=120
			120.0 & 0.014 & 398.0 & $\dot M_\text{Vink}$ & 2.361 & 87.121 & 16.3 & 20.6 & 0.686 & 79.03 & 65.04 \\ % 67261
			& & & $\dot M_\text{sc}$ & 2.603 & 110.15 & 34.6 & 149.0 & 0.687 & 55.86 & 26.31 \\ % 40971
			120.0 & 0.006 & 419.0 & $\dot M_\text{Vink}$ & 2.513 & 100.504 & 16.5 & 79.7 & 0.694 & 71.32 & 57.79 \\ % 34891
			& & & $\dot M_\text{sc}$ & 3.070 & 111.196 & 43.4 & 131.0 & 0.694 & 28.94 & 11.14 \\ % 42281
			\hline
			% Mass=70
			70.0 & 0.014 & 360.0 & $\dot M_\text{Vink}$ & 4.199 & 42.679 & 18.7 & 7.1 & 0.894 & 73.30 & 69.46\\ % 118521
			& & & $\dot M_\text{sc}$ & 3.935 & 60.363 & 26.9 & 3.9 & 0.628 & 18.20 & 5.64 \\ % 93931
			70.0 & 0.006 & 376.0 & $\dot M_\text{Vink}$ & 3.882 & 57.250 & 15.5 & 37.0 & 0.694 & 34.00 & 13.38 \\ % 40911
			& & & $\dot M_\text{sc}$ & 3.956 & 64.60 & 21.7 & 26.2 & 0.509 & 8.04 & 2.44 \\ % 28891
			\hline
			% Mass=40
			40.0 & 0.014 & 321.0 & $\dot M_\text{Vink}$ & 5.794 & 31.86 & 19.2 & 9.9 & 0.553 & 13.86 & 3.90\\ % 67331
			& & & $\dot M_\text{sc}$ & 5.535 & 36.428 & 23.6 & 37.4 & 0.460 & 5.46 & 1.50\\ % 37821
			40.0 & 0.006 & 334.0 & $\dot M_\text{Vink}$ & 5.622 & 36.196 & 16.2 & 30.6 & 0.440 & 6.101 & 1.647 \\ % 19931
			& & & $\dot M_\text{sc}$ & 5.382 & 38.180 & 18.5 & 71.7 & 0.370 & 3.664 & 0.953 \\ % 13991
			\hline
			% Mass=25
			25.0 & 0.014 & 291.0 & $\dot M_\text{Vink}$ & 7.977 & 23.591 & 15.9 & 68.4 & 0.343 & 3.316 & 0.756\\ % 10831
			& & & $\dot M_\text{sc}$ & 7.623 & 24.274 & 16.0 & 113.0 & 0.320 & 2.618 & 0.587 \\ % 8441
			25.0 & 0.006 & 302.0 & $\dot M_\text{Vink}$ & 7.708 & 24.272 & 14.1 & 159.0 & 0.309 & 3.389 & 0.687 \\ % 4111
			& & & $\dot M_\text{sc}$ & 8.105 & 24.567 & 16.7 & 185.0 & 0.331 & 4.267 & 0.786 \\ % 4111
			\hline
		\end{tabular}
		\label{timescalesXY}
	\end{table*}

\section{Results}\label{rotationtracks}
	Evolutionary models with rotation, both original and self-consistent, are shown in Fig.~\ref{hrd_final} in the HR diagram, whereas the rotation and equatorial velocities for tracks adopting both mass loss recipes are shown in Fig.~\ref{veq_final}.
	Additional plots such as the comparison with self-consistent non-rotating models from Paper I, the evolutionary tracks in the $\log g$ vs. $T_\text{eff}$ plane, evolution of the mass loss rate, stellar mass and radii, He/H ratio, Eddington factor, and  convective core mass are shown in Appendix~\ref{extraplots}.
	Also, hereafter all plots comparing new and old recipes of mass loss rate will use solid lines for models with $\dot M_\text{sc}$ and dashed lines for models with $\dot M_\text{Vink}$.
	Major features of the stellar models at the end of the MS, comparing self-consistent tracks with the tracks adopting Vink's formula \citep{ekstrom12,eggenberger21}, are tabulated in Table~\ref{timescalesXY}.
	
	Overall, compared with Paper I \citep[see Table~8 from][]{alex22b} we find that the lifetime during the main sequence is $\sim18-22\%$ longer than for the non-rotating models, regardless of the adopted mass loss recipe.
	The difference with non-rotating models is also appreciated in Fig.~\ref{rot_vs_norot}, where the evolutionary tracks with and without rotation are compared.
	The rotating tracks remain in bluer regions of the HR diagram compared to the non-rotating ones.
	Similar to Paper I, self-consistent mass loss rates are of a factor about $\sim3$ lower than the mass loss rates from Vink's formula (Fig.~\ref{mdots_final}), which makes the stars retain more mass (Fig.~\ref{mass_final}) and have larger radii (Fig.~\ref{radii_final}) when $\dot M_\text{sc}$ is adopted.
	Moreover, besides the straightforward differences in the final masses and the final radii, we also observe that the less strong winds from the self-consistent prescription predict in general weaker surface enrichments of helium and nitrogen at the surface, as seen in the last columns of Table~\ref{timescalesXY} and in the evolution of the He/H ratio (Fig.~\ref{heh_final}).
	The only noticeable exception is the model with $M_\text{zams}=25\,M_\odot$ and $Z=0.006$, described in detail  in the subsection below.

	Concerning rotation, self-consistent evolutionary tracks have higher velocities for all models as appreciated in Fig.~\ref{veq_final}, where we compute equatorial velocities and $\Omega=\varv_\text{eq}/\varv_\text{crit}$ as a function of the core hydrogen abundance from ZAMS (starting with $X_\text{core}\sim0.72$) until the end of the main sequence.
	Same as in Paper I, the largest impact of the new self-consistent mass loss rates are for the higher initial mass stars: the weaker $\dot M_\text{sc}$ produces stars that rotate faster.
	Even though the rotational velocity drops down at TAMS for all models, this braking is slower than the one expected for models with the old mass loss recipe.
	This weaker braking is more prominent for the low metallicity, where the stars adopting $\dot M_\text{sc}$ keep their rotational velocity almost constant for a longer time before the final deceleration.
	This implies that, in our regime of metallicities (between $Z=0.006$ and $Z=0.014$) mass loss is still a prominent process that impacts the evolution of the rotational velocity.
	Evolution of rotational velocities is explained by the balance between the transport of angular momentum from the convective core (which is contracting during the main sequence stage) to the stellar surface, and the removal of outer layers due to mass loss \citep{ekstrom08,demink13}.
	At extreme low metallicities ($Z=0.0002=1/35\,Z_\odot$), massive stars increase their $\varv_\text{rot}$ because their mass losses are not strong enough to counteract the transport of angular momentum from the contracting core \citep{szecsi15}.
	Also for $Z=0.006$, we see that $\Omega$, which is set initially as $=0.4$, increases up to a constant peak of $\sim0.55$ before the final decrease, as a direct consequence of the initial constancy in the equatorial velocity and the decreasing in the evolution of the critical velocity (Eq.~\ref{vcrit}).
	Weaker winds predicts a nearly exponential increasing in the stellar radii for all initial masses and both metallicities (Fig.~\ref{radii_final}), whereas both the stellar mass and Eddington factor show a linear and metallicity dependent difference with respect to the models computed with the Vink's formula (Fig.~\ref{mass_final} and Fig.~\ref{edd_final}).
	Therefore, because they are weaker, self-consistent winds makes our evolution models to approach closer to the break-up limit for rotational velocity in agreement with \citet{ekstrom08}, prior to the final drop in $\varv_\text{rot}$ at the end of main sequence in agreement with \citet{szecsi15}.

	Now, let us discuss some more detailed effects of the use of the new mass loss rate for the different initial mass models considered in the present work.

%_____Case 120 M_sun______________________________________________________________________________
	\begin{figure}
		\centering
		\includegraphics[width=0.95\linewidth]{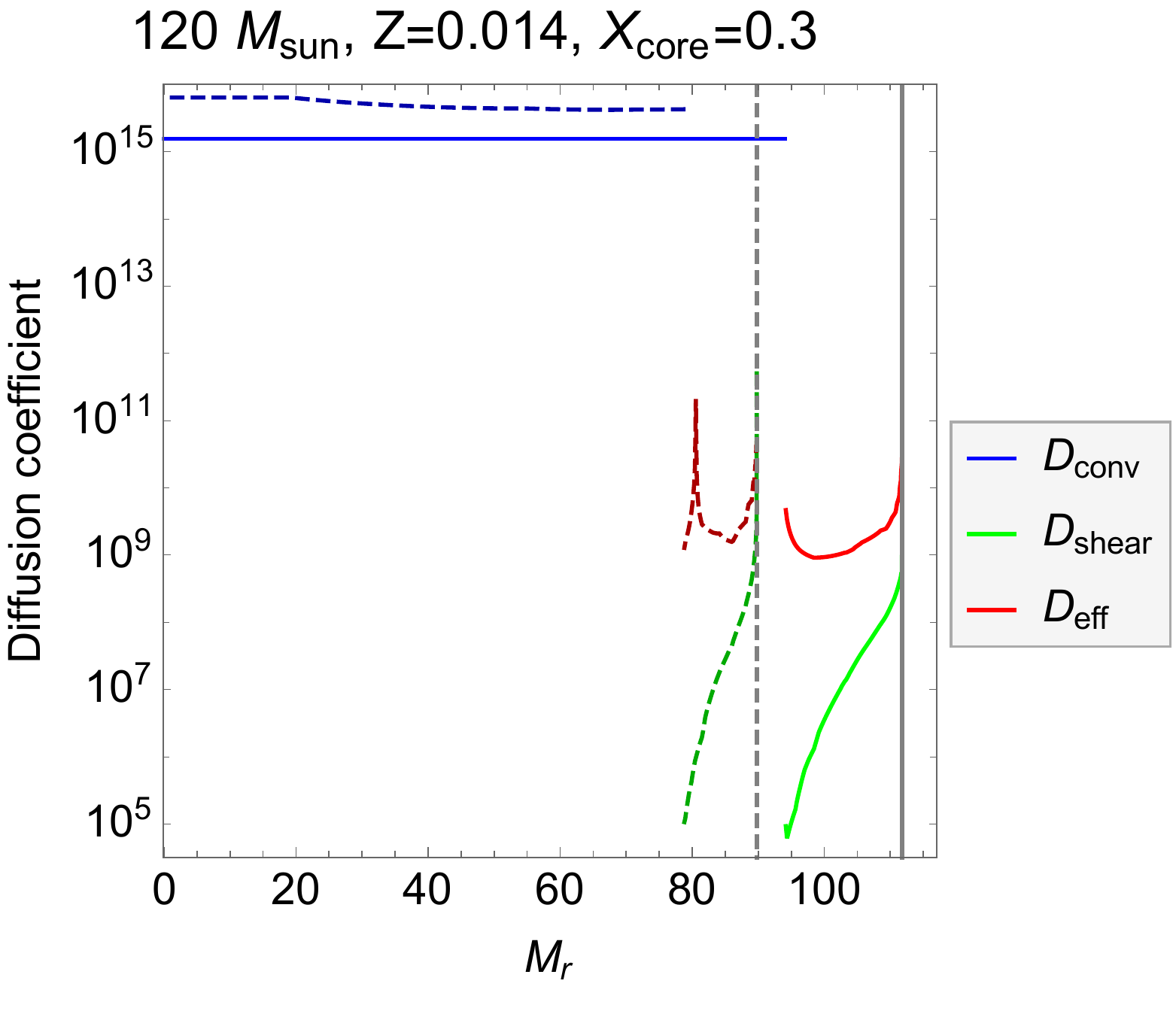}
		\caption{\small{Diffusion factors for convective turbulence ($D_\text{conv}$), shear turbulence ($D_\text{sh}$) and effective turbulence ($D_\text{eff}$) as a function of the Lagrangian mass coordinate, for our evolution models with $M_\text{zams}=120\,M_\odot$ and $Z=0.014$, adopting old and new winds at an intermediate point in its main sequence stage ($X_\text{core}=0.3$).
		Vertical grey lines represent the total mass of both models.
		Symbology for solid and dashed lines, is the same as in Fig.~\ref{hrd_final}.}}
		\label{diffusions}
	\end{figure}
	\begin{figure*}[t!]
%		\sidecaption
		\centering
		\begin{subfigure}{0.45\linewidth}
			\includegraphics[width=\linewidth]{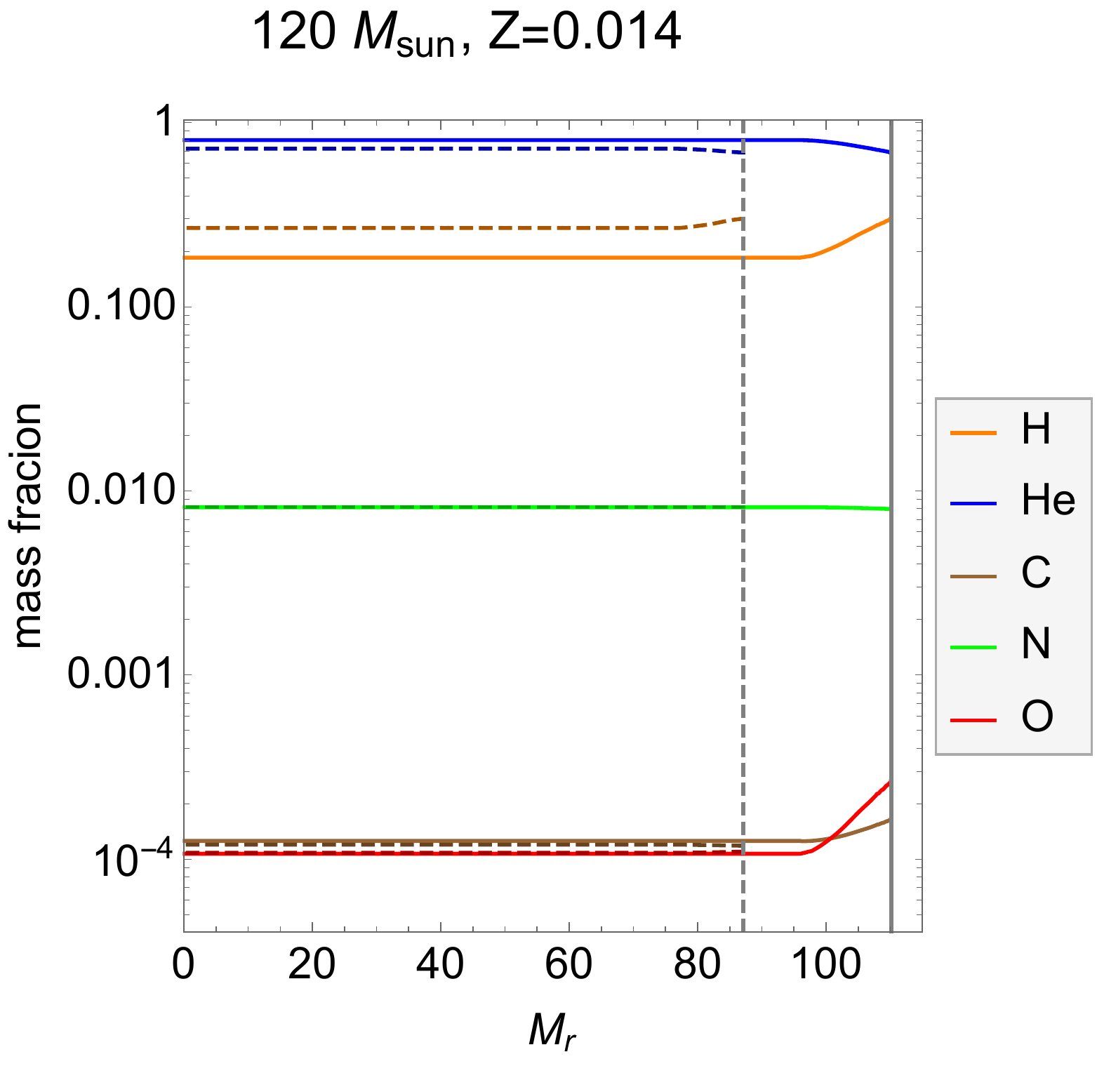}
			\caption{ }
			\label{P120z14_inner}
		\end{subfigure}
		\hspace{1cm}
		\begin{subfigure}{0.45\linewidth}
			\includegraphics[width=\linewidth]{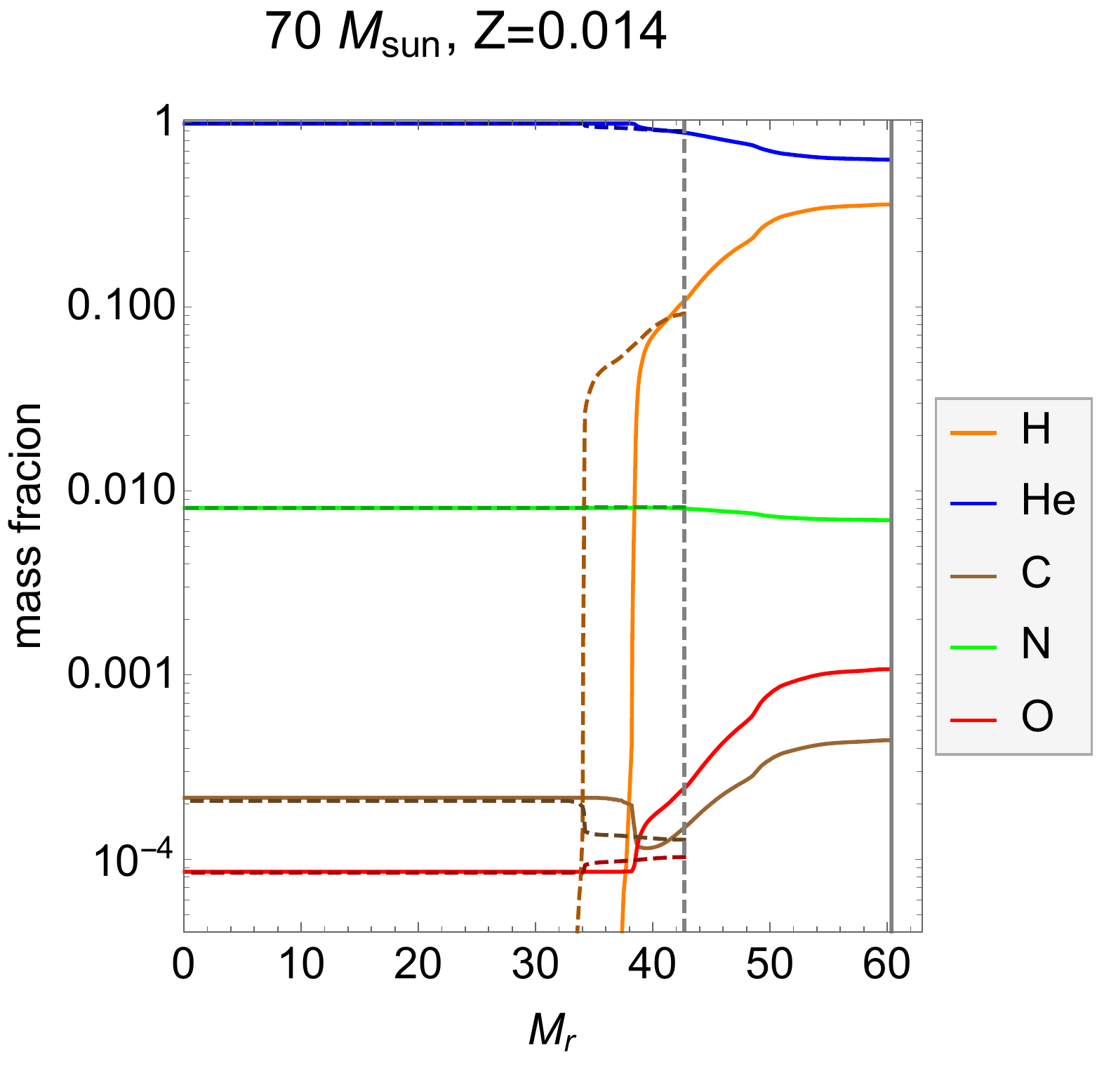}
			\caption{ }
			\label{P070z14_inner}
		\end{subfigure}
		\caption{\small{Variation in a logarithmic scale of the mass fraction of hydrogen, helium and CNO elements, as a function of the Lagrangian mass coordinate in solar units inside the (a) $120\,M_\odot$ model and (b) the $70\,M_\odot$ model, for the two prescription of the mass losses studied in the present work at the end of the main sequence stage.
		The meaning of the different line style is the same as in Fig.~\ref{hrd_final}.
		Vertical grey lines represent the total mass of both models, which correspond to the final masses at the end of MS tabulated in Table~\ref{timescalesXY}.}}
		\label{struct_P120070}
	\end{figure*}

\subsection{Case $M_\text{zams}=120$ $M_\odot$}
	This is the range where we find the \textit{very massive stars} \citep[VMS][]{yusof13}, whose properties and evolution have aroused great interest in recent years \citep{higgins22,sabhahit22}.
	Here, the tracks with $Z=0.006$ have a special relevance because the most studied cluster containing VMSs is 30 Dor, located in the LMC \citep{crowther10,bestenlehner14}.
	
	Same as in \citet{yusof13}, we consider that the star becomes a Wolf-Rayet (i.e., it changes from optically thin to optically thick wind) when its surface hydrogen abundance goes below 0.3 and $T_\text{eff}>10$ kK.
	After that point, mass loss rate is assumed to follow the recipe from \citet{grafener08} for WR winds, generating the discontinuity in $\dot M$ after the black cross as observed at Fig.~\ref{mdots_final}.
	Notice that prior to the change in the recipe, evolution of the self-consistent mass loss rate proceeds as expected due to the large increment in the He/H ratio (Fig.~\ref{heh_final}) as a consequence of Eq.~\ref{mdotformula2}.
	Due to the extreme strength of the winds, the VMS is depleted from the $70\%$ of its surface hydrogen even before H is completely exhausted in its core, as seen from Fig.~\ref{P120z14_inner}, generating the stars with spectral type WNh \citep{hamann06,martins22}.
	This switch not only in the $\dot M$ prescription but also in the wind regime (from optically thin to optically thick wind) is also evidenced by Fig.~\ref{radii_final}, where the radius needs to the recalculated after the black crosses based on the wind opacity \citep[see, e.g,][]{meynet05} due to the absence of a well-defined photosphere \citep{schaller92}.
	
	The most remarkable difference between classical and self-consistent tracks in Fig.~\ref{hrd_final}, is the drift `redwards' (i.e., towards the cool range of temperatures in the HRD) that the self-consistent models do since the beginning of the main sequence.
	For both metallicities, stars with $\dot M_\text{sc}$ end their MS lifetime as O-type at $T_\text{eff}\sim40$ kK, before suddenly increasing their temperature and becoming WNh stars.
	We note that the model using the weaker self-consistent $\dot M$ shows in general a slightly stronger surface enrichment than the original model.
	This explains why these models evolve to redder regions of the HRD during the MS phase.
	It is a well known effect that a star showing a nearly homogeneous internal chemical composition keep blue colours in the HRD diagram \citep{maeder87c}.
	A homogeneous chemical composition can result from a very strong internal mixing, from very strong mass losses or from both processes.
	In contrast an internal non-homogeneous composition will produce redwards evolution on the HRD.
	The extension to the red is favoured when a larger convective core is present and the radiative envelope is not too efficiently mixed, otherwise we would have a situation approaching the one of a homogeneous internal composition and the track will remain in the blue region of the HRD \citep{martinet21}.
	Here we see that reducing the mass loss favours redwards tracks.
	In Fig.~\ref{diffusions}, we show the diffusion coefficients in the new and old mass loss rate models in the middle of the MS phase.
	For the case of $120\,M_\odot$ and $Z=0.014$, the model with a lower mass loss rate has a more massive convective core at that stage, even though it represents a slightly minor fraction of the total mass: $\sim84\%$ for the self-consistent model, against $\sim87\%$ for the old one.
	This is a result of the higher difference in the mass retained and also to the possible higher value of the effective diffusion coefficient, $D_\text{eff}$, just above the convective core.
	In the rest of the radiative envelope, there is not much difference between the values of $D_{\rm eff}$.
	Since the diffusion timescale varies as $R^2/D_{\rm eff}$, and since the radius of the self-consistent model is larger, we see that the self-consistent model will take a longer time to reach a given level of surface enrichment than the original one.
	This imposes redder evolutionary tracks for the model with $\dot M_\text{sc}$. 
	
\begin{figure*}[t!]
%		\sidecaption
		\centering
		\begin{subfigure}{0.415\linewidth}
			\includegraphics[width=\linewidth]{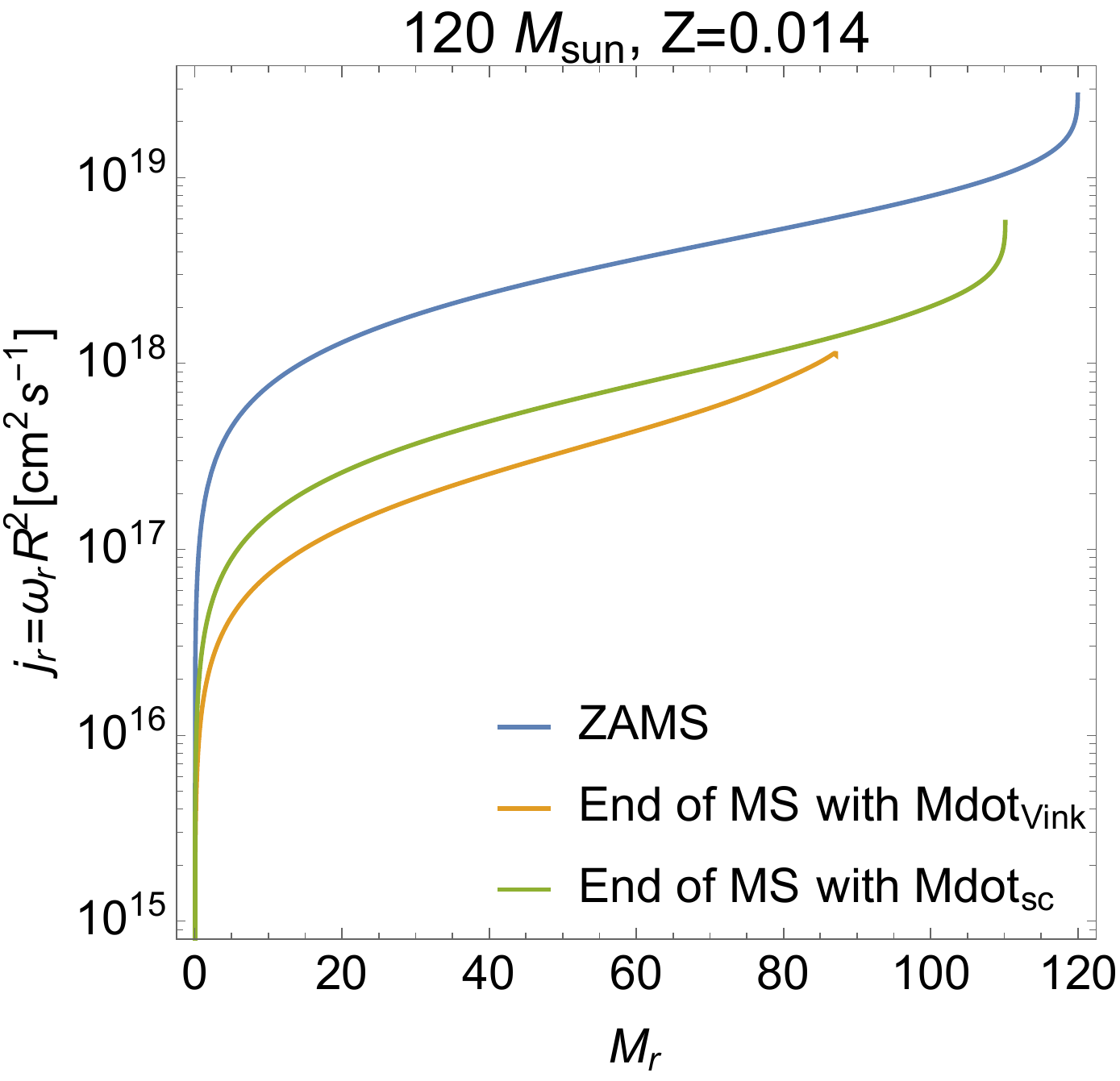}
			\caption{ }
			\label{angmom_P120z14}
		\end{subfigure}
		\hspace{1cm}
		\begin{subfigure}{0.45\linewidth}
			\includegraphics[width=\linewidth]{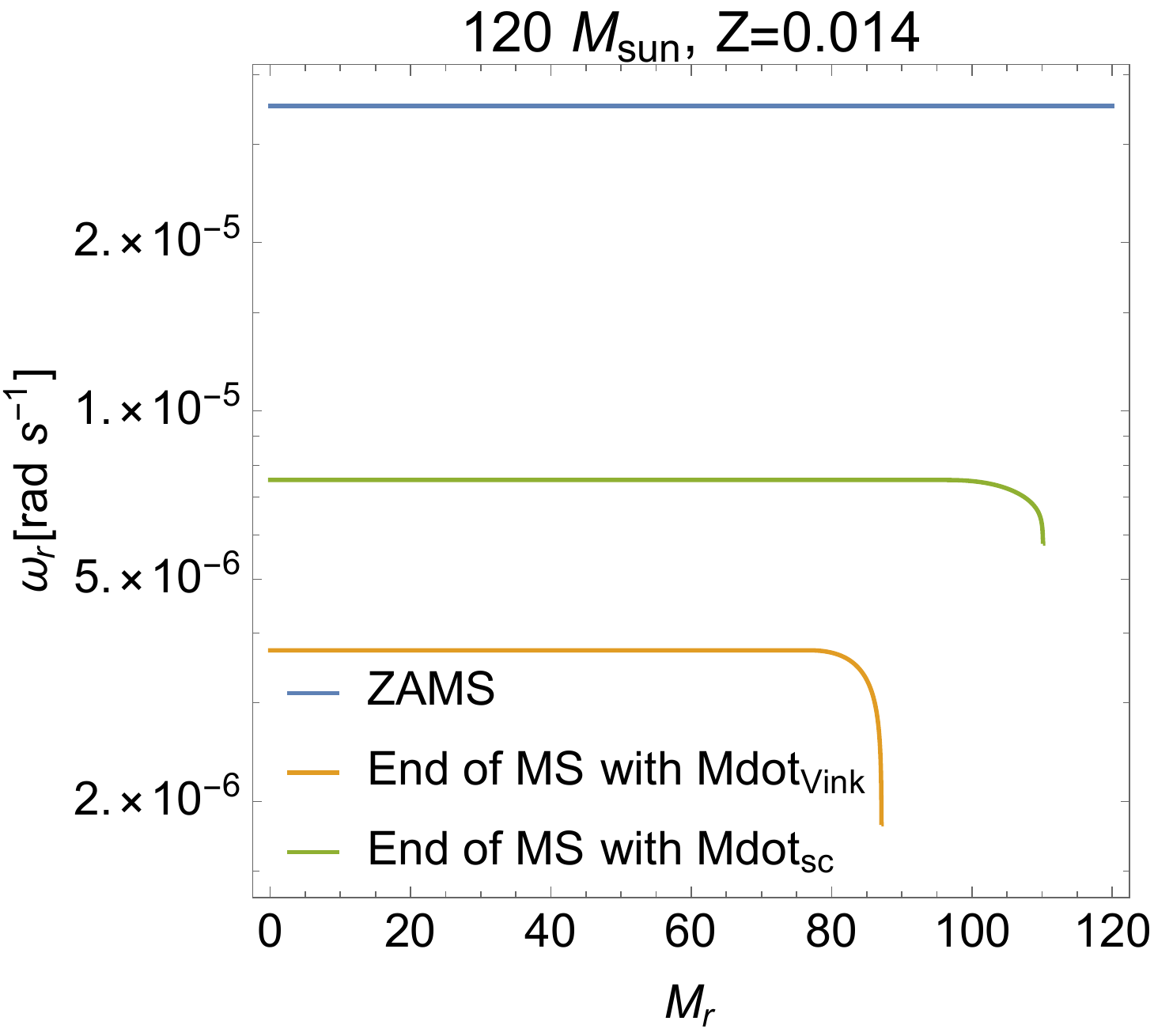}
			\caption{ }
			\label{vangr_P120z14}
		\end{subfigure}
		\caption{\small{Angular properties for the inner structure of our $120\,M_\odot$ model on the ZAMS and at the end of the MS phase for the two prescriptions of the mass losses studied in the present work.
		(a) Variation in a logarithmic scale of the specific angular momentum as a function of the Lagrangian mass coordinate in solar units.
		(b) Angular velocity, in radians per second, also as a function of the Lagrangian mass coordinate.}}
		\label{momang_P120z14}
	\end{figure*}

	Figure~\ref{P120z14_inner} compares the chemical composition of the self-consistent and original models at the end of the MS phase.
	As expected the model with the lower mass loss rate is showing the greatest contrasts between the surface composition and that of the core.
	Figure~\ref{angmom_P120z14} shows the variation of the specific angular momentum $j_\text{r}=r^2\omega$ inside the $120\,M_\odot$ at the beginning and end of the MS phase for the two different mass loss prescriptions.
	Let us first remind that in the absence of any angular momentum transport, $j_\text{r}$ remains constant.
	The evolution that we see between the beginning and end of the core H-burning phase (a global decrease of $j_\text{r}$ in the whole star) is due to the fact that the transport processes will in general transport angular momentum from the inner to the outer layers of the star.
	At the surface the angular momentum is removed by stellar winds, thus we see that when the mass loss rates are smaller, $j_\text{r}$ is larger.
	Indeed, by decreasing the mass loss rate by about a factor three during the main sequence (compared to Vink's formula), we obtain that the specific angular momentum is shifted by $~0.3$ dex upwards when $\dot M_\text{sc}$ is used, whereas the value of the angular velocity of the core is increased by a factor of $\sim2$ (see Fig.~\ref{vangr_P120z14}).

	Finally, we notice that for both metallicities our rotating models reach the WNh stage at $T_\text{eff}\gtrsim40$ kK.
	Since we know that stars will increase their temperature through the WR stage, this means that rotating models with $M_\text{zams}=120\,M_\odot$ do not cross the temperature regime around $\sim25$ kK during their MS phase, where the so-called bi-stability jump \citep{vink99} takes place.
	Because of this reason, and contrary to the non-rotating models from Paper I, we do not observe LBV behaviour such as eruptive mass losses at this mass range.
	The fact that only initially slow rotating models can go through an eruptive mass loss regime at the end of the main sequence supports the idea that the LBV phenomenon is very rare for very massive stars \citep{grafener21}.

%_____Case 70 M_sun_______________________________________________________________________________

\subsection{Case $M_\text{zams}=70$ $M_\odot$}
	Same as before, besides the differences in the stellar mass (Fig.~\ref{mass_final}) we also observe a divergence bluewards-redwards for the evolution across the HR diagram (Fig.~\ref{hrd_final}).
	However, unlike the previous case, this time the beginning of the WR winds stage coincides with the end of the H-core burning, which can be appreciated in Fig.~\ref{P070z14_inner} where hydrogen is depleted in the core at the end of both (standard and self-consistent) tracks.
	Again we see the change in the slope of the evolution in mass loss rate (Fig.~\ref{mdots_final}), this time with a flattening in $\dot M$ prior to reaching the second black dot, associated with the increasing of the He/H ratio (Fig.~\ref{heh_final}).
	
	Concerning rotation, the braking in the equatorial velocity only becomes significant after the star has burnt almost half of the hydrogen in its core: $X_\text{c}\sim0.38$ for $Z=0.014$ and $X_\text{c}\sim0.36$ for $Z=0.006$, which also represents the peak of $\Omega\sim0.42$ for Galactic metallicities and $\Omega\sim0.54$ for SMC metallicities (Fig.~\ref{veq_final}).
	This is interconnected one more time with the retaining of more specific angular momentum for the models adopting $\dot M_\text{sc}$, and with the larger stellar radii (Fig.~\ref{radii_final}).
	Indeed, the switch in the wind regime from $\dot M_\text{sc}$ to $\dot M_\text{Vink}$ (marked with an abrupt increase in mass loss rate after the second black dot of Fig.~\ref{mdots_final}) carries an important drop in the rotational velocity which creates some numerical instability, but does not affect the general trend in the evolution of observed for all the rotation models.
	
	From Fig.~\ref{P070z14_inner}, we see that the variation of the element abundances across the stellar structure from core to surface is more prominent for the model with $\dot M_\text{sc}$, which is evidence for the less efficient rotational mixing for the weaker wind.
	However, on the contrary of the case with $M_\text{zams}=120\,M_\odot$, where the inner structure was almost fully homogeneous, now there are remarkable distinctions between core and surface abundances because the mass loss has been less efficient in reducing the mass of the envelope and revealing the inner layers. 
	
	The redward drift of the evolutionary tracks for the reduced mass loss rate models is in line with the results found by \citet{bjorklund22}, who implemented a mass loss recipe derived from the wind theoretical prescription from \citet{bjorklund21} where $\dot M$ is also $\sim3$ times below the values coming from Vink's formula.
	In their study, they used \textsc{Mesa} to trace the evolutionary track of a star with $M_\text{zams}=60$ $M_\odot$, $\varv_\text{rot,ini}=350$ km s$^{-1}$, and solar metallicity; where they found the drift bluewards at $T_\text{eff}\sim26$ kK (see their Fig.~6).
	On our side, by making an eye inspection on Fig.~\ref{hrd_final} and Fig.~\ref{rot_vs_norot}, we conjecture that the point of turning bluewards for one of our self-consistent models with $60\,M_\odot$ and $Z=0.014$ must be between $\sim29-31$ kK, i.e., around $\sim5$ kK hotter than found by \citet{bjorklund22}.
	Such a discrepancy is relatively minor considering the differences between \textsc{Genec} and \textsc{Mesa} \citep[check the comparison for the models of these both codes, as done by][]{agrawal22}.
	Therefore, we can conclude that the self-consistent m-CAK prescription for stellar winds predicts evolutionary tracks in fair agreement with the recent study of \citet{bjorklund22}, as expected given that both studies are based on mass loss rates of the same order of magnitude.

%_____Cases 25 and 40 M_sun__________________________________________________________________________
	\begin{figure*}[t!]
%		\sidecaption
		\centering
		\begin{subfigure}{0.46\linewidth}
			\includegraphics[width=\linewidth]{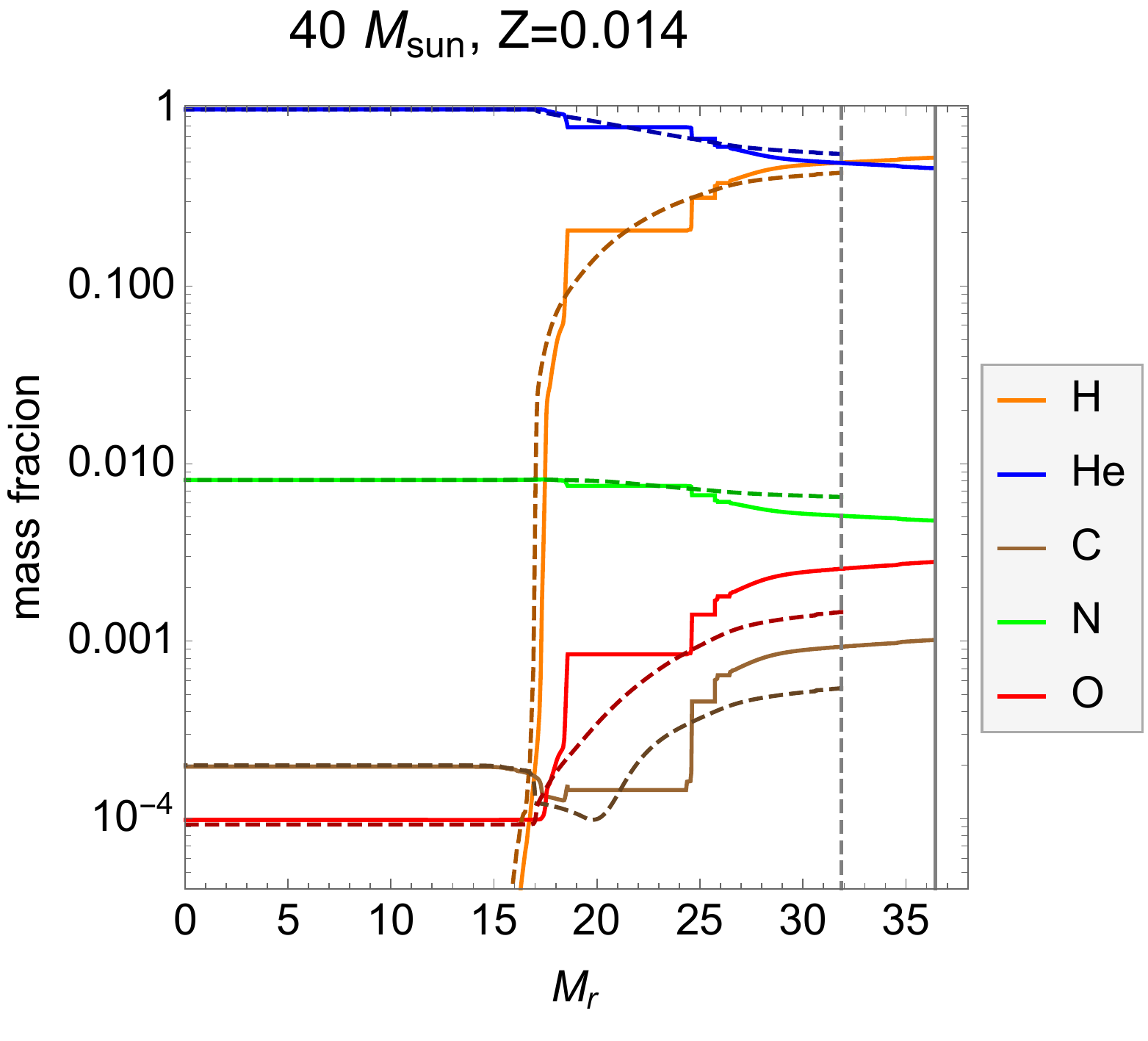}
			\caption{ }
			\label{P040z14_inner}
		\end{subfigure}
		\hspace{1cm}
		\begin{subfigure}{0.46\linewidth}
			\includegraphics[width=\linewidth]{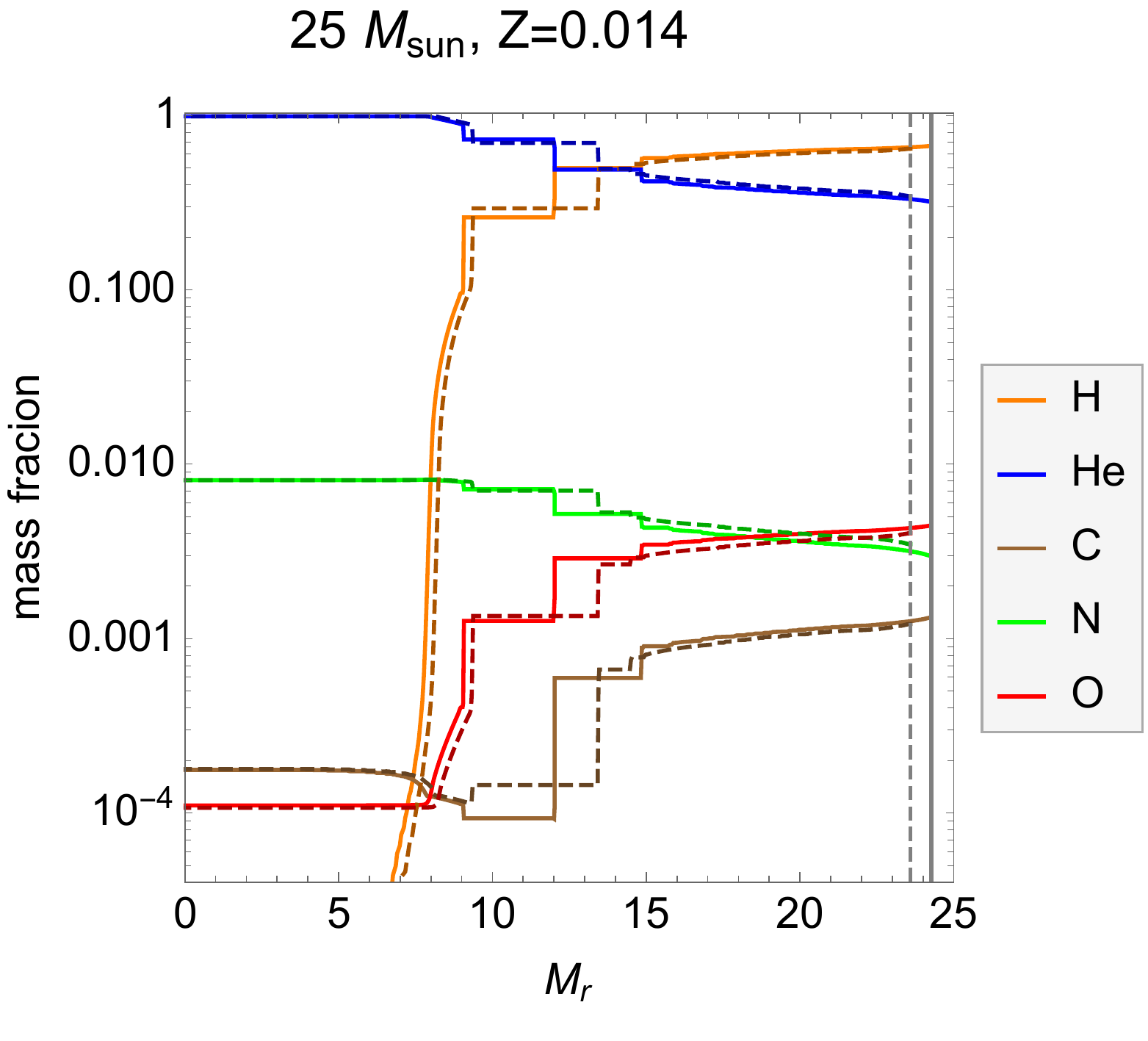}
			\caption{ }
			\label{P025z14_inner}
		\end{subfigure}
		\caption{\small{Same as Fig.~\ref{struct_P120070}, but for models with $40$ (a) and $25\,M_\odot$ (b).}}
		\label{struct_P040025}
	\end{figure*}
	\begin{figure*}[t!]
%		\sidecaption
		\centering
		\begin{subfigure}{0.415\linewidth}
			\includegraphics[width=\linewidth]{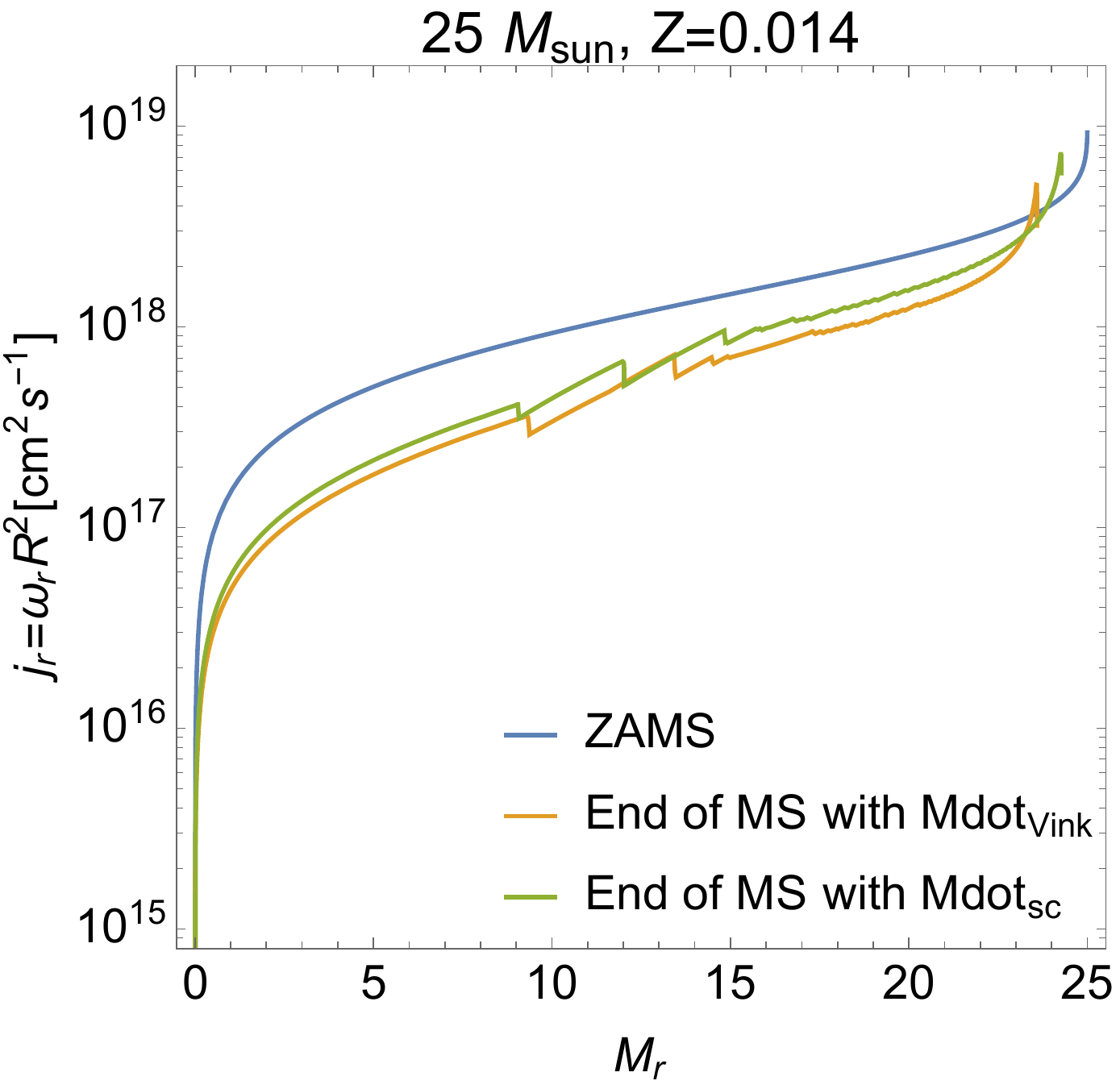}
			\caption{ }
			\label{angmom_P025z14}
		\end{subfigure}
		\hspace{1cm}
		\begin{subfigure}{0.45\linewidth}
			\includegraphics[width=\linewidth]{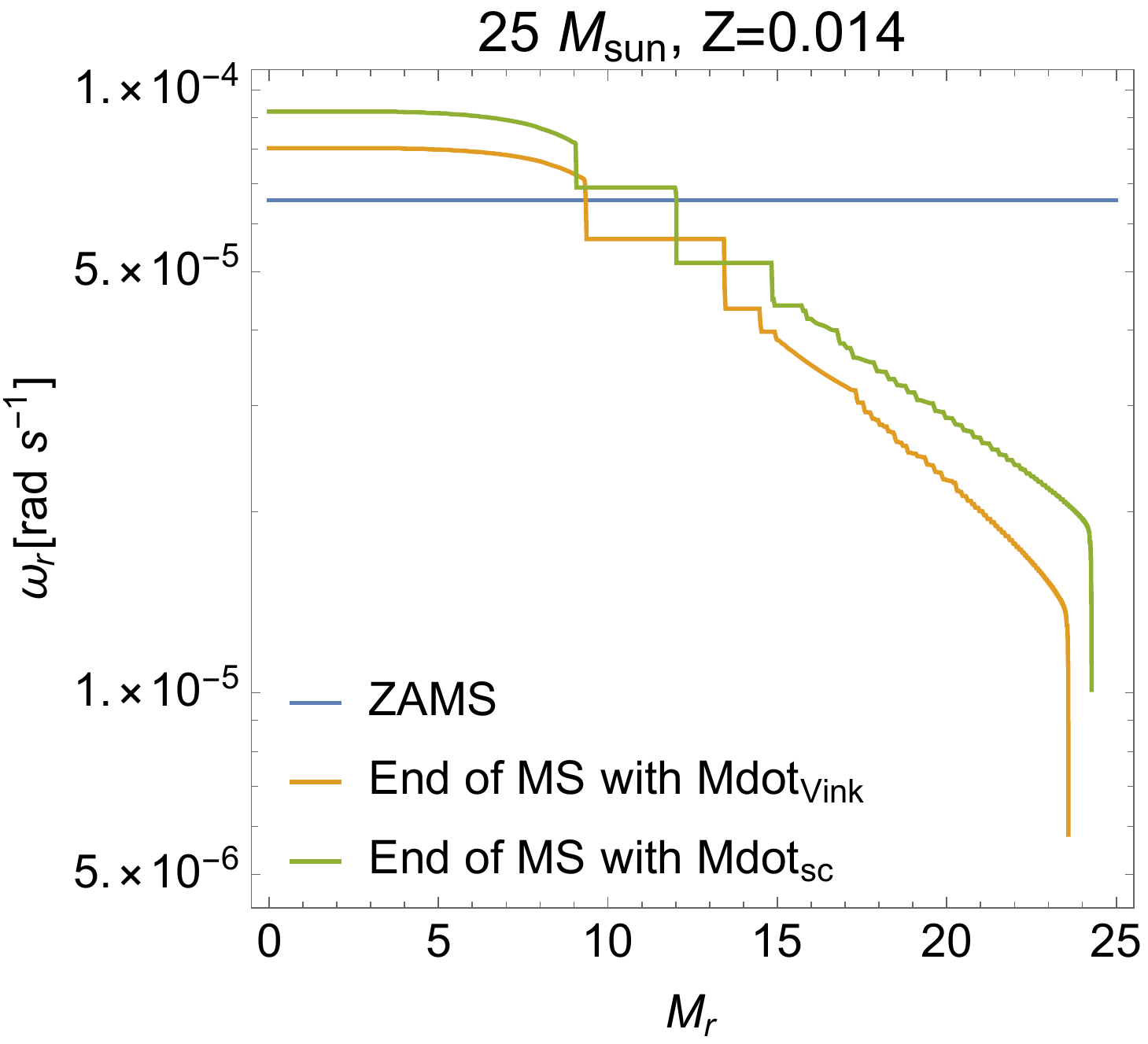}
			\caption{ }
			\label{vangr_P025z14}
		\end{subfigure}
		\caption{\small{Same as Fig.~\ref{momang_P120z14}, but for our model with $M_\text{zams}=25\,M_\odot$ and $Z=0.014$.}}
		\label{momang_P025z14}
	\end{figure*}

\subsection{Cases $M_\text{zams}=25$ and $40$ $M_\odot$}
	Effects of self-consistent mass loss rates over the evolution of stars within these mass ranges are less pronounced, as it is seen in Fig.~\ref{hrd_final} and in the resulting final masses and radii from Table~\ref{timescalesXY}.
	We also observe a `redder' evolution for the models adopting $\dot M_\text{sc}$, but still `bluer' compared with non rotating cases (Fig.~\ref{rot_vs_norot}).
	Differences in the rotational velocities are also less remarkable, though they still follow the same trend of the more massive models.
	For the case of $Z=0.006$ (LMC), we observe that the end of H-burning is reached inside the region of validity of the self-consistent tracks (Fig.~\ref{shrd_final}), and therefore the second dot overlaps again with the cross.
	Because we observe this situation only with $Z=0.002$ (SMC) in the non-rotating cases from the Paper I, it implies that the evolution models adopting m-CAK wind prescription cover a broader range of the main sequence when rotation is incorporated.
	
	The impact of the different mass losses on the chemical structures of the $25$ and $40\,M_\odot$ stellar models can be seen in Fig~\ref{struct_P040025}.
	The convective cores are $8.5$ and $18.5\,M_\odot$ for the $25$ and $40\,M_\odot$ stars, respectively, when the high mass loss rates are used (Fig.~\ref{masscc_final}).
	The composition in the radiative envelope shows some significant differences between the two models.
	This is particularly visible for the $40\,M_\odot$ star, where in the low mass loss rate model a convective region is associated to the H-burning shell (between about $17$ and $25\,M_\odot$ coordinates), while no convective zone is associated to that shell in the high mass loss rate model.
	We see therefore, that reducing the mass of the envelope disfavours the formation of an intermediate convective shell associated to the H-burning shell.
	
	The specific angular momentum inside our model of $25\,M_\odot$ is shown in Fig.~\ref{angmom_P025z14}.
	We see along the curves small abrupt variations (as for instance around $9\,M_\odot$ at the end of the MS phase).
	These come from regions where there is a transition between a convective and a radiative zone, where the chemical composition changes, also observed as an abrupt jump in the inner angular velocity (Fig.~\ref{vangr_P025z14}).
	Like for the $120\,M_\odot$ star, the model computed with the self-consistent mass loss rates ends the MS phase with slightly larger values of the specific angular momentum.
	The increase is less strong since the mass loss rates are anyway weaker when the initial mass decreases.
	The changes in the interior angular velocity distribution are also much weaker than for the $120\,M_\odot$ model.
	We note as expected a slightly faster rotating core in the self-consistent stellar model compared to the original one.
	
%_____DISCUSSION________________________________________________________________________________
%\section{Discussion}
\section{Comparison with rotational surveys}\label{discussion}
	\begin{figure}[t!]
		\centering
		\includegraphics[width=0.9\linewidth]{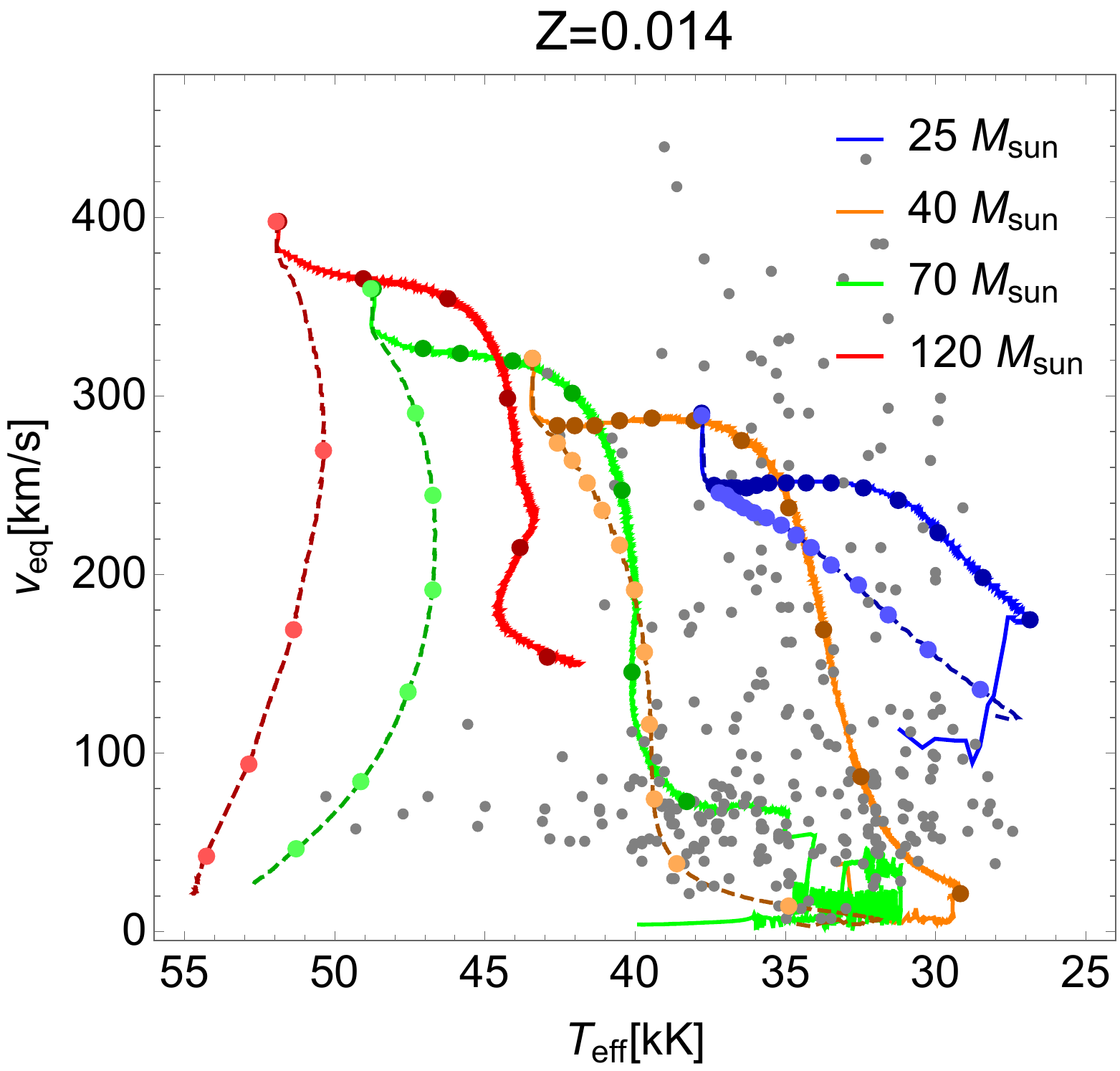}
		\caption{\small{Rotational velocities as a function of the effective temperature, for evolutionary tracks adopting old (dashed) and new (solid) winds.
		The coloured dots represent the intervals of age with a step of $0.5$ Myr.
		Grey dots represent the sample of O-type stars taken from the survey of \citet{holgado22}.}}
		\label{teffvrot_h22}
	\end{figure}
	\begin{figure}[t!]
		\centering
		\includegraphics[width=0.9\linewidth]{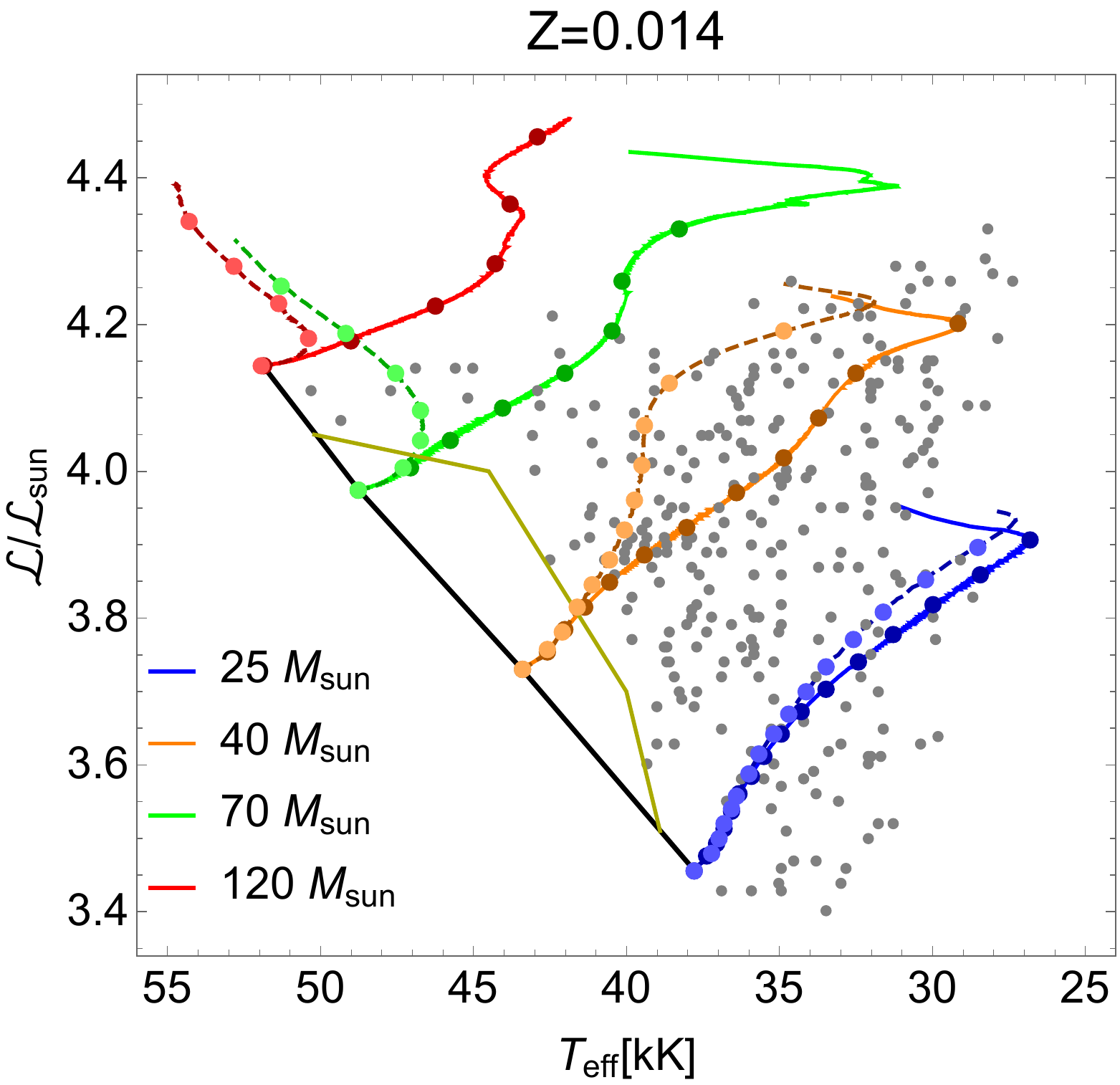}
		\caption{\small{Spectroscopic Hertzsprung-Russell diagram \citep[][sHRD, with $\mathcal L:=T_\text{eff}^4/g$]{langer14}, for evolutionary rotating tracks adopting old (dashed) and new (solid) winds.
		Grey dots represent the sample of O-type stars taken from the surveys of \citet{holgado22}.
		The black line represents the ZAMS for all models, whereas dark yellow line represents the region close to the ZAMS where no stars are found \citet{holgado20}.}}
		\label{shrd_h22}
	\end{figure}
	Because the self-consistent m-CAK prescription predicts less strong winds than previous studies, stellar models computed with such prescription evolve at higher luminosities and reach larger radii during the MS phase than the models of \citet{ekstrom12} or \citet{eggenberger21}, as already detailed in Section~\ref{rotationtracks}.
	Besides, evolutionary tracks adopting $\dot M_\text{sc}$ also predict that the rotational velocity during the main sequence will be higher as seen in Fig.~\ref{veq_final}, implying a weaker braking in the equatorial velocity of massive stars.
	This weaker braking is also represented in Fig.~\ref{teffvrot_h22}, where we plot rotational velocities as a function of the effective temperatures for evolutionary tracks adopting old and new winds, and where the timesteps of $0.5$ Myr are illustrated with the respective coloured circles.
	Velocities from evolution models with $\dot M_\text{Vink}$ quickly decreases after the first $\sim1-2$ Myr, passing from little variations in temperature for $M_\text{zams}=120\,M_\odot$ to moderate variations for $M_\text{zams}=25\,M_\odot$.
	On the contrary, the $\varv_\text{rot}$ from self-consistent evolutionary tracks keep relatively constant at the beginning of the main sequence for all tracks, at timescales inversely proportional to their initial masses (from $\sim1.0$ Myr for $M_\text{zams}=120\,M_\odot$ to $\sim6.0$ Myr for $M_\text{zams}=25\,M_\odot$), whereas $T_\text{eff}$ decreases $\sim8$ kK per each model prior to the final braking.

	In parallel, we plot in Fig.~\ref{teffvrot_h22} the sample of 285 O-type stars taken from the recent survey of \citet{holgado22}, whose catalog is available online\footnote{\url{https://cdsarc.cds.unistra.fr/viz-bin/cat/J/A+A/665/A150}}.
	In their study, they revisited the rotational properties of 285 Galactic massive O-type as part of the IACOB Project \citep{simondiaz15,holgado18,holgado20}.
	This survey covers a range of stellar masses from $15$ to $80$ solar masses, and a mixture of ages within the main sequence stage.
	Besides, they compared the rotation of these stars with the state-of-the-art evolution models \citet{brott11} and \citet{ekstrom12}, finding that neither of the two sets of rotating evolutionary tracks was able to reproduce such features of the survey.
	From one side, models from \citet{brott11} cannot reproduce the existence of stars with $\varv_\text{rot}\lesssim150$ km s$^{-1}$ across the entire domain of O-type stars \citep[see Fig.~9 of][]{holgado22}.
	From the other side, models from \citet{ekstrom12} cannot adequately reproduce the scarcity of stars with $\varv_\text{rot}\gtrsim75$ km s$^{-1}$ for the left side of Fig.~\ref{teffvrot_h22}, which is also appreciated by the lack of stars matching the old tracks with $70$ and $120$ $M_\odot$.
	
	In contrast, in Fig.~\ref{teffvrot_h22} we observe that new self-consistent evolutionary tracks can adequately explain both issues.
	Tracks from $25$ to $70$ $M_\odot$ match to the set of stars with rotational velocities below 150 km s$^{-1}$ for the range of temperatures from $27$ to $40$ kK.
	Also, the empty region of stars with $T_\text{eff}\ge42.5$ kK and $\varv_\text{rot}\ge75$ km s$^{-1}$ would only be populated by our $120$ $M_\odot$ model, which is out of the mass range of the sample.
	Therefore, evolution models adopting weaker winds better interpret the rotational properties of the survey of \citet{holgado22}, keeping with initial $\Omega=0.4$ and without the need of decreasing the initial equatorial velocity for our models.
	
	Nonetheless, despite this encouraging result, there are still important aspects of the rotational properties of Galactic O-type stars that need to be taken into account.
	For instance, although our track of $M_\text{init}=25\,M_\odot$ can encircle a larger fraction of stars in the top-right side of Fig.~\ref{teffvrot_h22}, there are still fast rotators ($\varv_\text{rot}>300$ km s$^{-1}$) outside the scope of any track, which only could be explained by binary interaction effects \citep{demink13,wangchen20}.
	Moreover, the slow braking observed for self-consistent evolutionary tracks indicates that stars born with masses between $25$ and $70$ $M_\odot$ should spend the $\sim75\%$ of the lifetime in the main sequence stage with $\varv_\text{rot}$ between $\sim250$ km s$^{-1}$ and $\sim330$ km s$^{-1}$.
	As a consequence, we should expect to find a larger fraction of O-type stars in this range of rotational velocities.
	However, because of its proximity to the ZAMS, this range also covers an important portion of the region where there is a lack of empirically detected O-type stars, as shown in the Fig.~\ref{shrd_h22} and described by \citet{holgado20}.
	For that reason, it is not surprising to find a moderate number of stars in the range from $250$ to $330$ km s$^{-1}$ and temperatures between $30$ and $42.5$ kK.
	Therefore, even though our evolution models adopting self-consistent winds represent a relevant upgrade, there are still important challenges in the study of O-type stars to deal with, such as the expansion of current samples to more hidden places by the use of infrared observations.

%_____IMPLICATIONS________________________________________________________________________________
\section{Implications of evolution models with self-consistent winds}\label{implications}
	The evolutionary tracks across the HRD from our rotating models adopting self-consistent winds exhibit considerable contrasts with the old models adopting Vink's formula, leading into a better description of the rotational properties of the most recent observational diagnostics.
	In this Section we move one step beyond, and we explore some of the implications that the incorporation of these self-consistent models have at Galactic scale.
	We remark however that these implications are here discussed at an introductory level, and their respective conclusions are part of forthcoming studies.

\subsection{Enrichment of the $^{26}$Al isotope}
	\begin{figure}[t!]
		\centering
		\includegraphics[width=.95\linewidth]{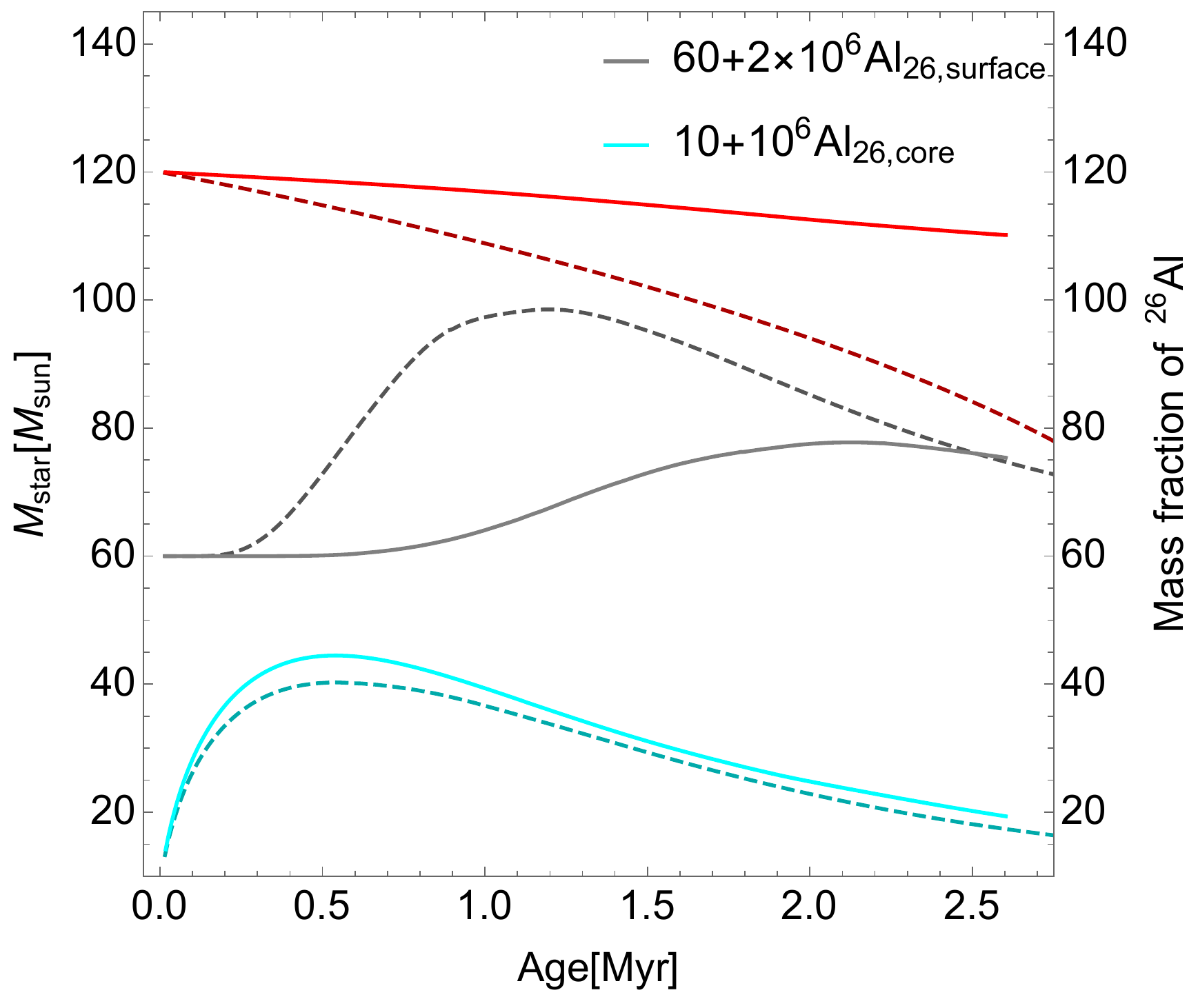}
		\caption{\small{ Variation as a function of time of the total mass (upper red curves), of the mass fraction of $^{26}$Al at the centre (bottom cyan curves) and at the surface (grey curves beginning at an ordinate equal to 60) in the $120\,M_\odot$ rotating models with the old (upper panel) and new (lower panel) mass loss rates.}}
		\label{al26_rot}
	\end{figure}
	
	Figure~\ref{al26_rot} shows how the new mass loss rates impact the quantity of $^{26}$Al ejected by the stellar winds of our $120\,M_\odot$ models computed with the two mass loss rate prescriptions.
	Compared to the models discussed in Paper I for the non rotating case, the current rotating models allow $^{26}$Al to appear at the surface well before layers enriched by nuclear burning in $^{26}$Al are uncovered by the stellar winds.
	This is the effect of the rotational mixing that allows diffusion of the $^{26}$Al produced in the core to reach the surface. 
	Thus we see that, like for the case in Paper I, reducing the mass loss rate reduces the maximum value of the abundance of $^{26}$Al reached at the surface.
	This maximum also occurs at a more advanced age for the model with self-consistent wind, despite the peak of production of $^{26}$Al in the core is found at almost the same age for both (old and new) wind prescriptions.
	Such time difference between the peak abundance at the centre and at the surface is easily explained by the reduction on mass loss rate, which makes the star to take more time to remove their outer layers and then expose its inner composition, and for the less intense mixing as previously evidenced in Fig.~\ref{P120z14_inner}.
	\begin{table}[t!]
		\centering
		\caption{\small{Values for the integration of the mass fractions $^{26}$Al$_\text{surface}$ for our evolution model of $M_\text{zams}=120\,M_\odot$ and $Z=0.014$ (see Fig.~\ref{al26_rot}), compared with the results from Paper I (non-rotating tracks).}}
		\begin{tabular}{cccc}
			\hline\hline
			Track & \multicolumn{2}{c}{$\int \dot M\;\text{$^{26}$Al}_\text{surf}(t)\,dt$}\\
			& \multicolumn{2}{c}{$[M_\odot]$}\\
			& non-rotating & rotating\\
			\hline
			Classic & $3.47\times10^{-4}$ & $4.15\times10^{-4}$\\
			Self-consistent & $1.82\times10^{-4}$ & $4.9\times10^{-5}$\\
			\hline
		\end{tabular}
		\label{integratedal26}
	\end{table}
	
	The total amount of $^{26}$Al released from the stellar surface to the interstellar medium during the MS for each one of our wind prescriptions, together with previous results from Paper I, are tabulated in Table~\ref{integratedal26}.
	Although the total amount of the aluminium-26 ejected by old winds is almost the same regardless if rotation is considered, for self-consistent winds the rotating model predicts the ejection of $\sim8$ times less fraction of the isotope during the MS, compared to the rotating evolution model adopting Vink's formula.
	Such a difference is related not only to the mass loss due to the line-driven mechanism.
	Whereas non-rotating models predict eruptive processes associated to LBVs (Fig.~7 from Paper I), rotating models for $M_\text{zams}=120\,M_\odot$ never reach those magnitudes for $\dot M$ (Fig.~\ref{mdots_final}).
	Therefore, the contribution of $^{26}$Al to the interstellar medium predicted by self-consistent winds is even weaker (compared with models adopting Vink's formula) for rotating models than for non-rotating models.
	
	However, it is important to remark that Fig.~\ref{al26_rot} covers only the lifetime of the star when the wind is optically thick, then excluding optically thick winds from WNh (H-core burning) and WR (He-core burning) phases.
	In contrast, the study of \citet{martinet22} calculated the evolution of $^{26}$Al for evolution models with mass ranges from $12$ to $300$ $M_\odot$ and metallicities from $Z=0.002$ to $Z=0.020$, but without making any distinction between thin and thick wind regimes.
	Instead, they made the distinction between H-core and He-core burning stages, showing that the production of $^{26}$Al abruptly decreases when the star enters into the He-burning.
	As a consequence, even though here we are selecting only one mass and one metallicity to do a quick comparison with the more complete analysis from \citet{martinet22}, we infer that the larger contribution of $^{26}$Al from very massive stars to the ISM must stem from the later stages exhibiting optically thick winds, such as H-core WNh stars, to explain the current estimations of $^{26}$Al total abundance in the Milky Way \citep{knodlseder99,diehl06,wang09,pleintinger19}.
	Certainly, we need a extensive study implementing self-consistent evolution models for wider ranges of masses and metallicities, together with updates in the convection criteria \citep{georgy14,kayser20} and overshooting values \citep{martinet21} in order to have more complete analysis.

\subsection{Massive stars at the Galactic Centre}
	The implications of the new evolution models for massive stars with self-consistent mass loss rate are large, not only across the main sequence but also over the subsequent evolutionary stages.
	One of such consequences is how using models computed with the weaker self-consistent mass loss rates may actually provide estimates of evolutionary masses that are lower than when models with $\dot M$ from \citet{vink01} are adopted.
	
	An example of this situation is the study of the Ofpe\footnote{Ofpe is a spectral classification for stars showing intermediate properties between O-type and WR stars \citep{walborn77,walborn82}. For that reason, they are assumed to be a transition between both spectroscopic phases according to the Conti scenario \citep{conti75}.} stars located at the Galactic Centre.
	Massive stars at the Galactic Centre play an important role feeding the supermassive black hole Sgr~A* by means of their stellar winds \citep{cuadra15,ressler18,calderon20}.
	This is a consequence of the remarkable situation of OB and WR stars representing a large fraction of the stars at the Galactic Centre \citep[e.g.,][]{lu13, vonfellenberg22}, even though this is a region thought to be hostile to star formation \citep[e.g.,][]{genzel10}.
	The wind properties of these massive stars have been analysed and constrained by \citet{martins07} for the O-type and WRs, and by \citet{habibi17} for the B-type stars.
	\citet{martins07} calculated the stellar and wind parameters for a set of stars at evolved stages such as Ofpe and WN, which are particularly relevant for feeding Sgr A*.  
	Their analysis was performed by spectral fitting using the CMFGEN code and the evolutionary tracks from \citet{meynet05}.
	Therefore, it is important to check whether the state-of-the-art evolutionary tracks suggest a revision of the properties of the massive stars at the Galactic Centre, and the consequences these modifications imply for the accretion on to Sgr~A*.

	\begin{figure}[t!]
		\centering
		\includegraphics[width=0.9\linewidth]{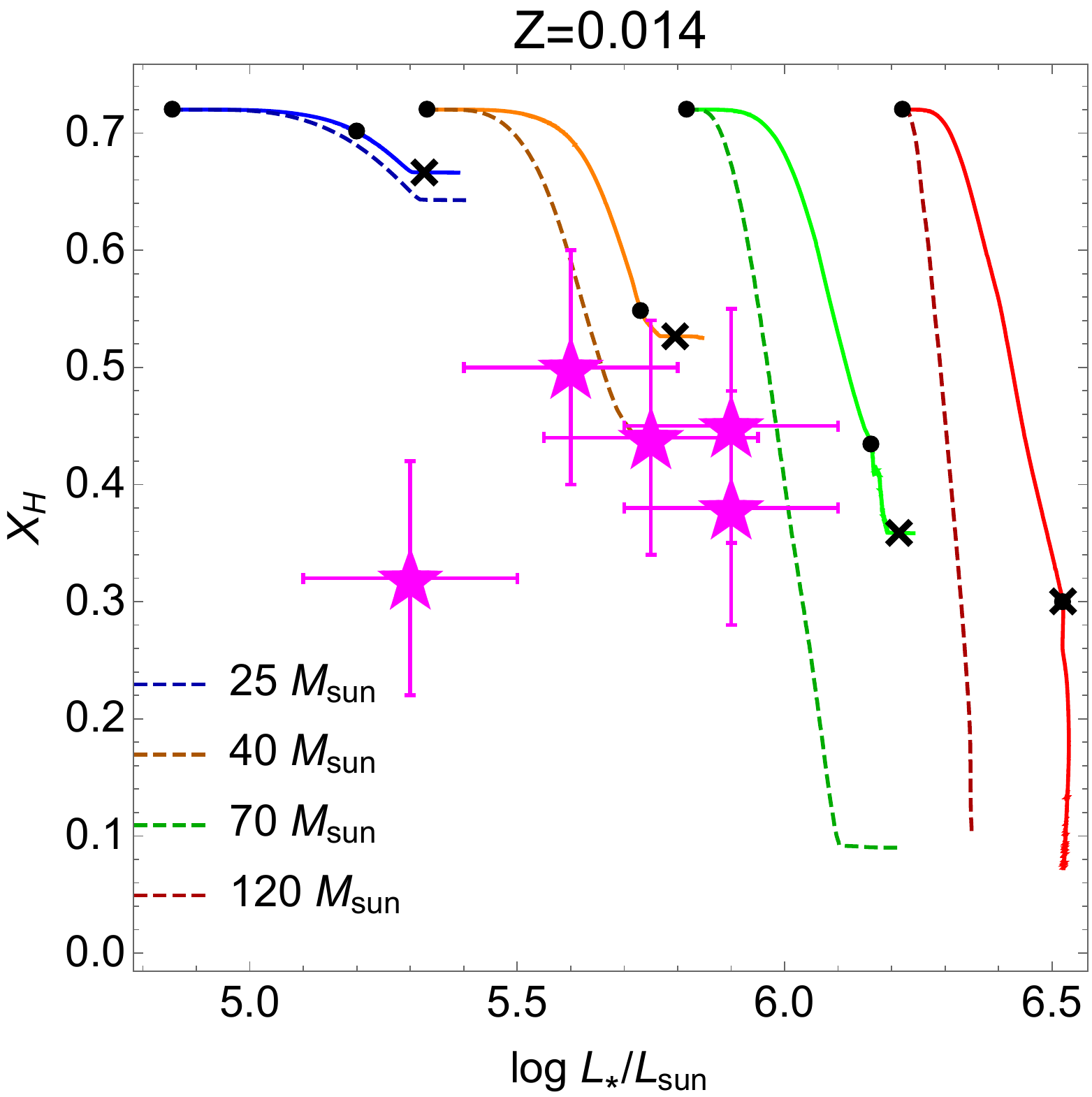}
		\caption{\small{Surface hydrogen abundance (in mass fraction) as a function of the stellar luminosity for different evolutionary tracks.
		Continuous lines are models computed with the self consistent mass-loss rate, while dashed lines are computed with Vink's original recipe.
		Magenta symbols are the Ofpe stars from the Galactic Centre as plotted by \citet[][compare with their Fig.~21]{martins07}.}}
		\label{martins07}
	\end{figure}

	To illustrate the issue, Fig.~\ref{martins07} shows the hydrogen abundance at the stellar surface as a function of luminosity for standard and self-consistent evolutionary tracks, plus five of the Ofpe stars analysed by \citet[][compare with their Fig.~21]{martins07}.
	From Fig.~\ref{martins07} we can see that for a given hydrogen surface abundance and for a given initial mass, the tracks computed with the weaker self-consistent mass-loss rates are overluminous.
	These tracks also end the MS phase with higher surface hydrogen abundances.
	We notice that the lowest luminosity Ofpe star in the sample cannot be reproduced by any of the two families of tracks.
	The other four stars can reasonably be fitted by both types of models.
	On average, although the initial masses that will be deduced from comparison with self-consistent mass-loss rate tracks are smaller than the masses deduced from tracks computed with Vink's mass loss rates.
	
	Given that the observed Ofpe stars in the Galactic center would have lower initial masses, we can speculate that their wind parameters (terminal velocity and mass loss rate) will be lower as well. 
	The terminal velocity is an important parameter in this context, since the Galactic centre is a region with large stellar density, and the winds collide with each other, creating a high-temperature medium.
	However, slower winds ($\lesssim 600$ km/s) produce a plasma of lower temperature ($\lesssim 4\times 10^6$ K) at the collision, which becomes susceptible to hydrodynamical instabilities and ends up forming high-density clumps and streams.
	These clumps can then be captured by the central black hole, increasing its accretion rate \citep{cuadra05, cuadra08, calderon16, calderon20, calderon20bin}.
	
	In a forthcoming paper we will perform a more complete analysis of the Galactic Centre massive star population, extending the tracks to the WR stage and taking into account the non-standard chemical abundance deduced for this region.  
	From the observational side, we will also include data collected after \citet{martins07}, which is expected to afford a reduction in the typical error bars for the stellar luminosities from $0.2$ down to $0.1$ dex (S.\ von Fellenberg, priv.\ comm.).
	The comparison between models and observations will allow us to update the estimates for the initial masses and ages of this population, and also its wind parameters, which have a large influence on the current accretion on to Sgr A*.

%_____CONCLUSION_______________________________________________________________________________
\section{Conclusion}\label{conclusion}
	We have extended the stellar evolution models performed in \citet[][Paper I]{alex22b}, adopting self-consistent m-CAK prescription \citep{alex19,alex22a} for the mass loss rate recipe (Eq.~\ref{mdotformula1}), by including the rotational effects.
	Stellar rotation affects the mass loss of a massive star, not only by changing the balance between gravitational, radiative and centrifugal forces \citep{maeder00}, but also because rotational mixing considerably modifies the internal distribution of the chemical elements and thus impacts the evolutionary tracks in the HRD and hence the mass loss rates.
	The progressive increase of the helium-to-hydrogen ratio in the wind impacts the line-force parameters $(k,\alpha,\delta)$ and henceforth the self-consistent solution of the equation of motion, leading to a decrease of 0.2 dex in the absolute value of the mass loss rate (Eq.~\ref{mdotformula2}).
	
	The updated mass loss rates are implemented for a set of evolutionary models at different initial stellar masses, for metallicities $Z=0.014$ (Galactic) and $Z=0.006$ (LMC).
	New tracks show important differences with respect to the studies of \citet{ekstrom12} and \citet{eggenberger21}, who adopted the formula from \citet{vink01} for the winds of massive stars, but exhibit a fair proximity with the studies of \citet{sabhahit22} for VMS, and with \citet{bjorklund22} and their own self-consistent wind prescription.
	Besides the differences in the tracks across the HRD, already remarked in Paper I, for rotating evolutionary tracks we find as expected that the surface rotation maintains a higher value when the weaker self-consistent mass loss rates are adopted.
	We observe that for the initial rotation considered here, tracks computed with the self consistent mass loss rates extend more to the red part of the HR diagram during the MS phase
	Such an effect is more important for masses above about $40\,M_\odot$.

	New tracks predict evolution of the surface rotational velocities for O-type stars that, at first sight, appear to be in better agreement with the most recent observational diagnostics.
	The slow braking in the evolution of $\varv_\text{rot}$ explains better the rotational properties of the survey of \citet{holgado22} for O-type stars, such as the lack of fast rotators for stars with $T_\text{eff}\gtrsim42.5$ kK and the abundance of stars with $\varv_\text{rot}\le150$ km s$^{-1}$ and temperatures between 27 and 40 kK.
	Besides, the implications of these new evolution models are wide.
	For example, lower mass loss rate predicts a less important stellar wind contribution to the Aluminium-26 enrichment of the ISM during the main sequence phase, at least whereas the stellar wind is optically thin.
	Likewise, the fact that self-consistent models are more luminous in the HRD suggests that the initial mass deduced from evolutionary tracks might be lower when the self-consistent tracks are used.
	This may apply to the mass estimates of Ofpe and WN stars at the Galactic Centre, which leads us to expect that their wind properties (mass loss rate and terminal velocity) might also be overestimated.
	Nevertheless, a more accurate analysis of the stars surrounding Sgr A* and their stellar winds are deferred to a forthcoming paper.

%_____AGRADECIMIENTOS__________________________________________________________________________
\begin{acknowledgements}
We are very grateful for the anonymous referee and their very constructive comments and feedback.
We also acknowledge useful discussions with Sebastiano von Fellenberg and Stefan Gillessen.
This project was partially funded by the Max Planck Society through a ``Partner Group'' grant. 
JC acknowledges financial support from FONDECYT Regular 1211429.
GM has received funding from the European Research Council (ERC) under the European Union's Horizon 2020 research and innovation programme (grant agreement Nº 833925, project STAREX).
MC thanks the support from FONDECYT projects 1190485 and 1230131.
\end{acknowledgements}

%_____BIBLIOGRAFÍA_______________________________________________________________________________
\bibliography{rotationpaper.bib} % your references Yourfile.bib
\bibliographystyle{aa} % style aa.bst
%\end{multicols}

\begin{appendix}
\section{Additional plots}\label{extraplots}
	\begin{figure*}[t!]
		\centering
		\includegraphics[width=0.4\linewidth]{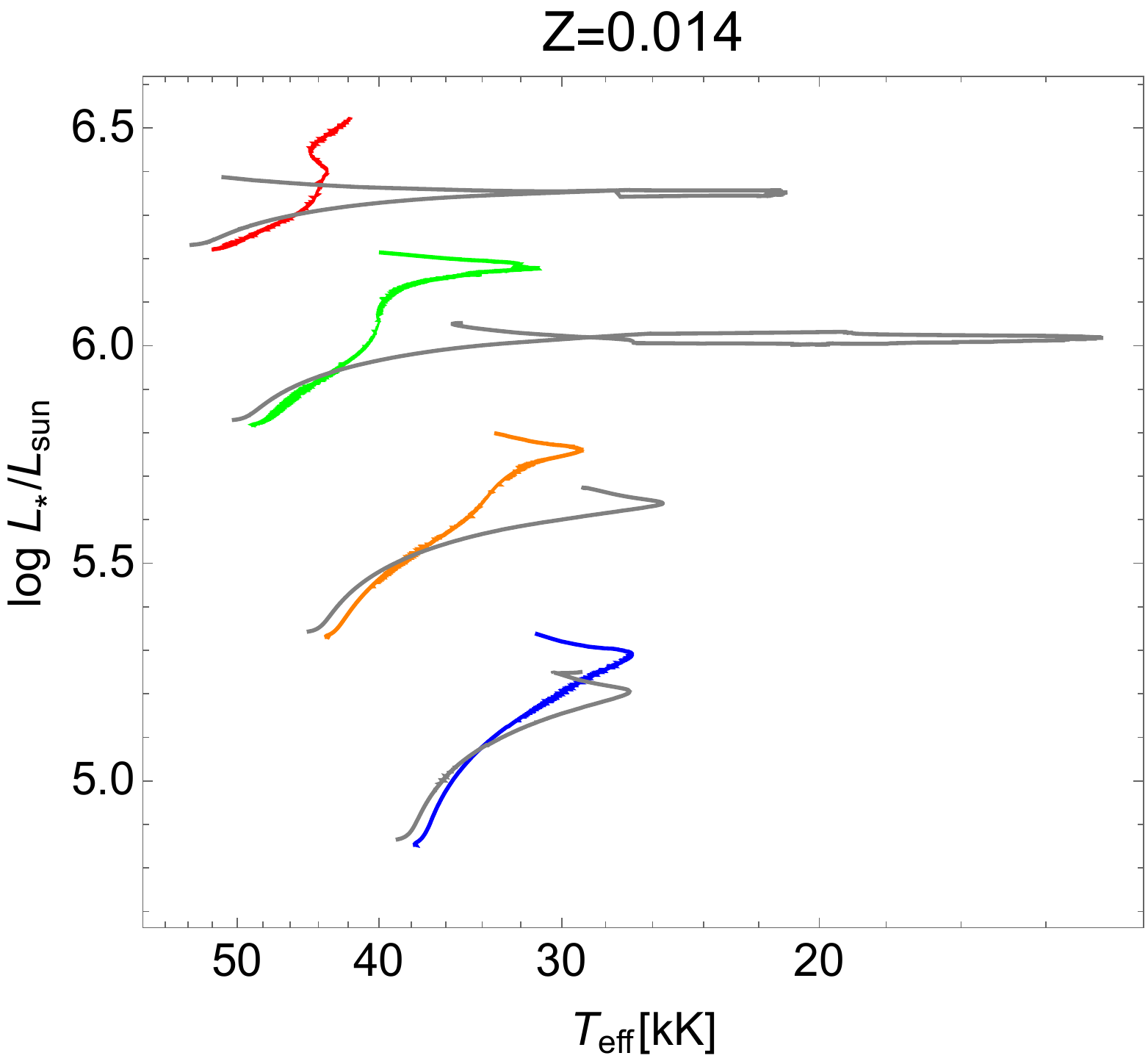}
		\hspace{1cm}
		\includegraphics[width=0.4\linewidth]{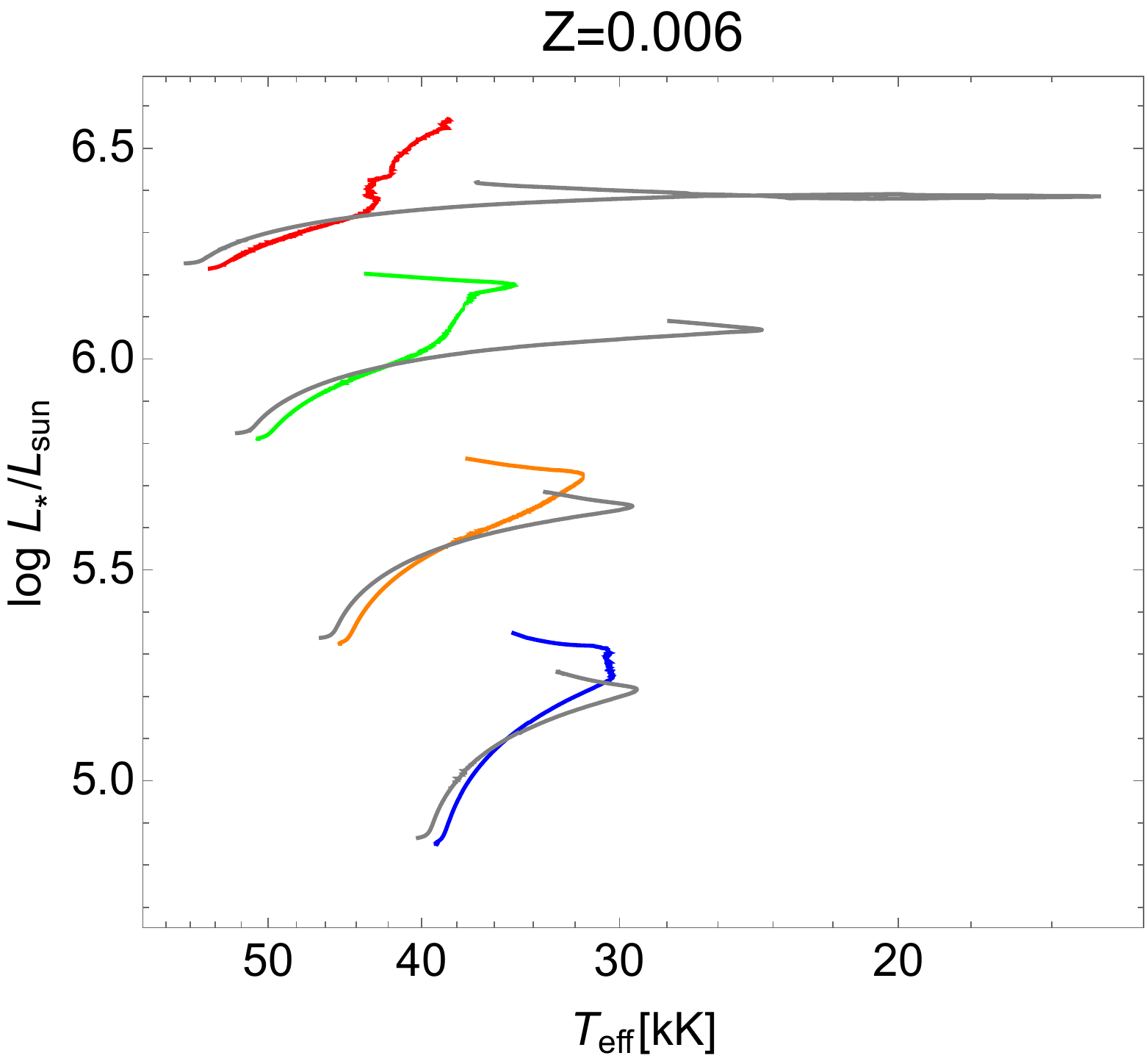}
		\caption{\small{Self-consistent evolution tracks from Fig.~\ref{hrd_final}, compared with the respective self-consistent non-rotating models from Paper I \citep{alex22b}.}}
		\label{rot_vs_norot}
	\end{figure*}
	\begin{figure*}[t!]
		\centering
		\includegraphics[width=0.4\linewidth]{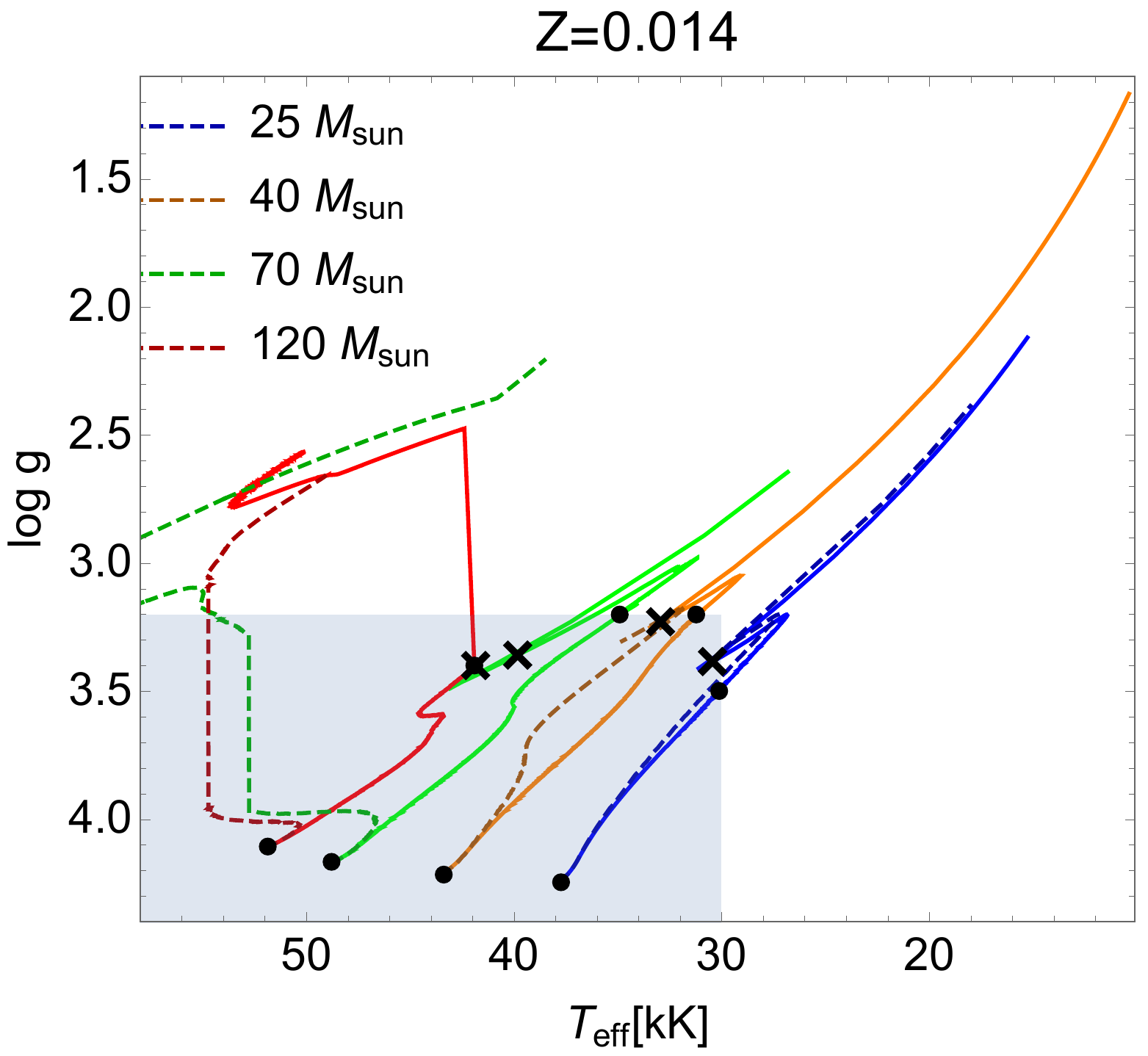}
		\hspace{1cm}
		\includegraphics[width=0.4\linewidth]{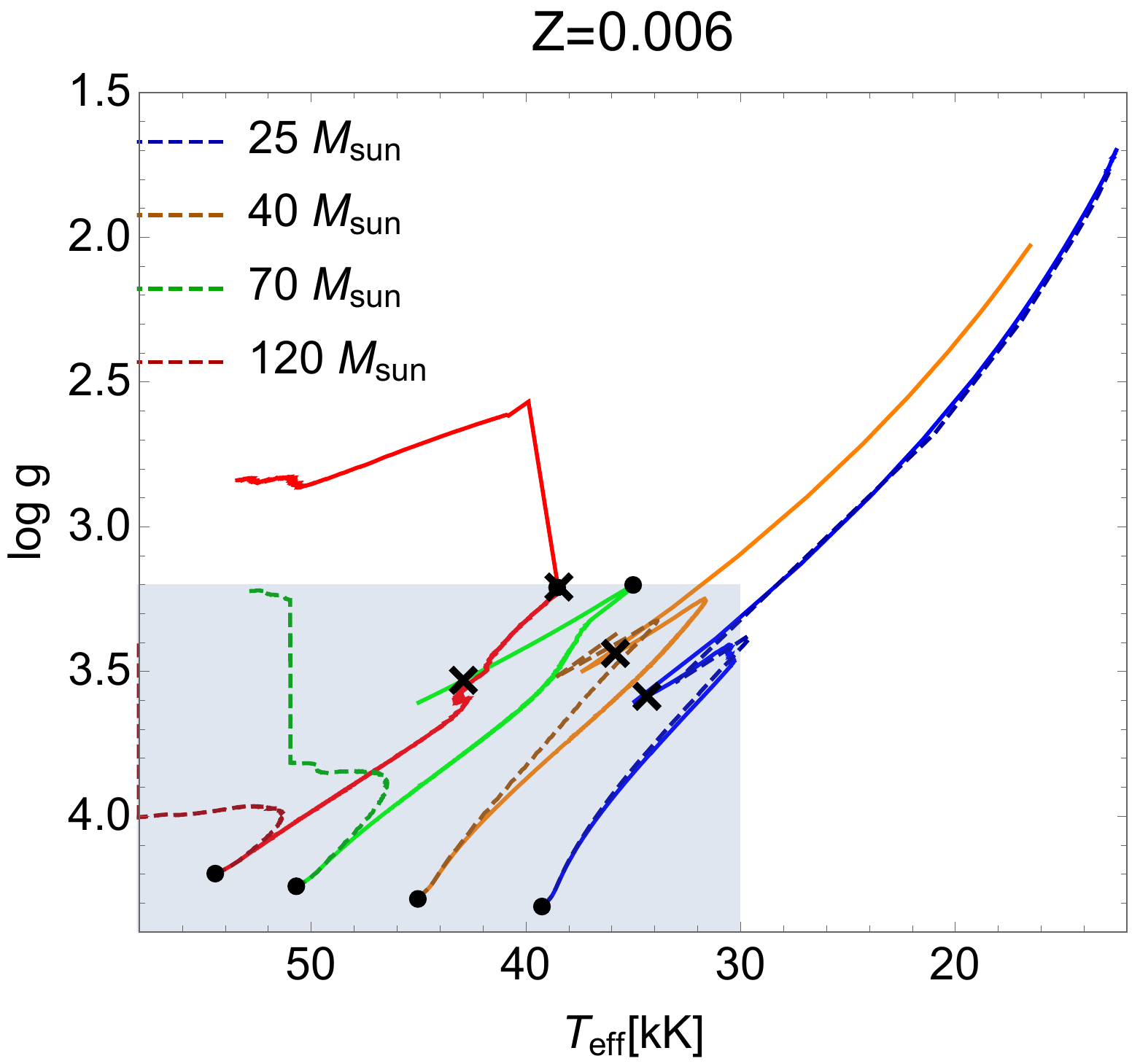}
		\caption{\small{Self-consistent evolution tracks (solid lines), compared with classical evolutionary tracks (darker dashed lines) from \citet[][$Z=0.014$]{ekstrom12} and \citet[][$Z=0.006$]{eggenberger21}, across the plane $(T_\text{eff},\log g)$.
		Shadowed region constrain the range of validity for the m-CAK prescription, as defined in Section~\ref{selfconsistent}.}}
		\label{shrd_final}
	\end{figure*}
	\begin{figure*}[t!]
		\centering
		\includegraphics[width=0.4\linewidth]{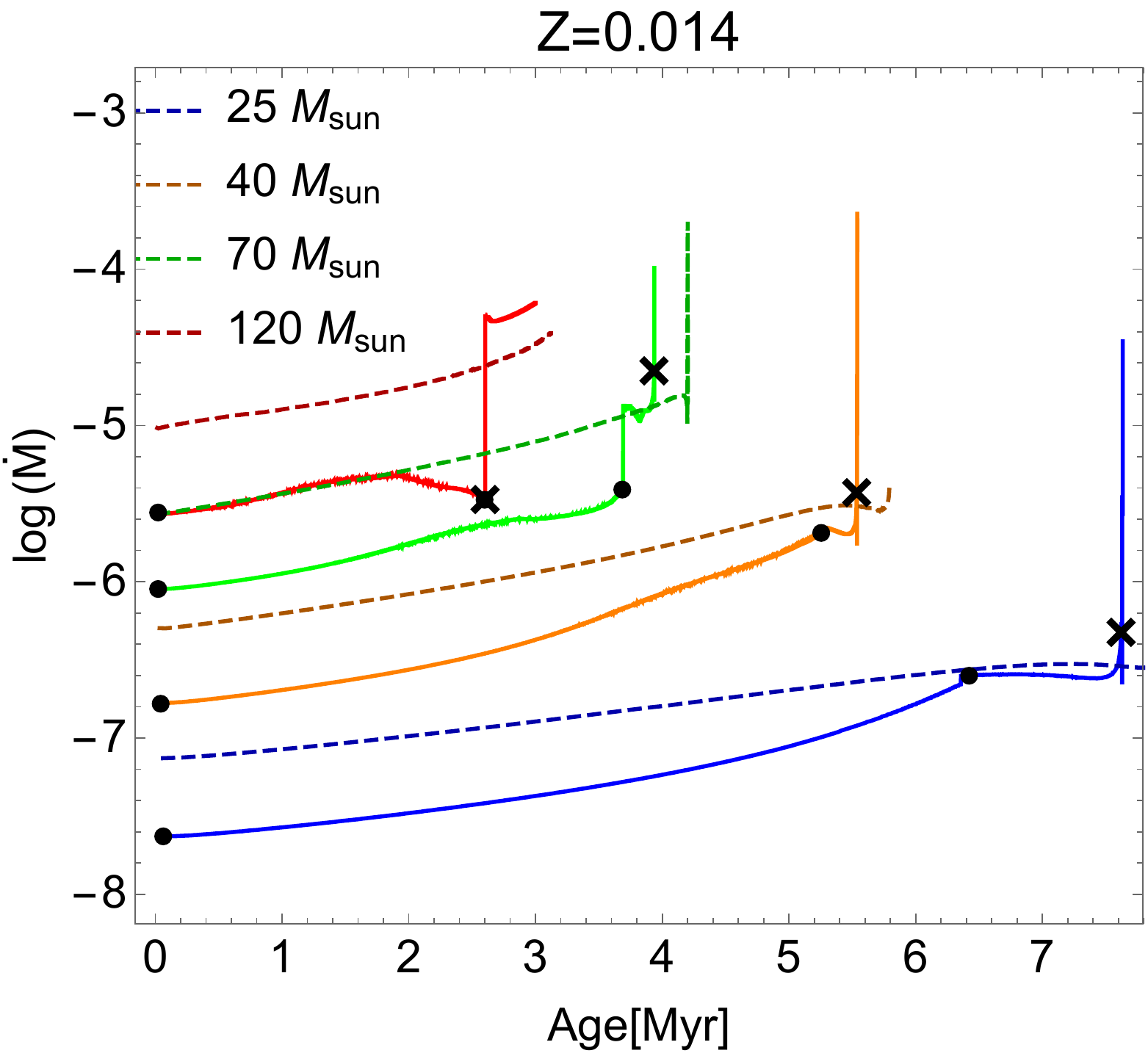}
		\hspace{1cm}
		\includegraphics[width=0.4\linewidth]{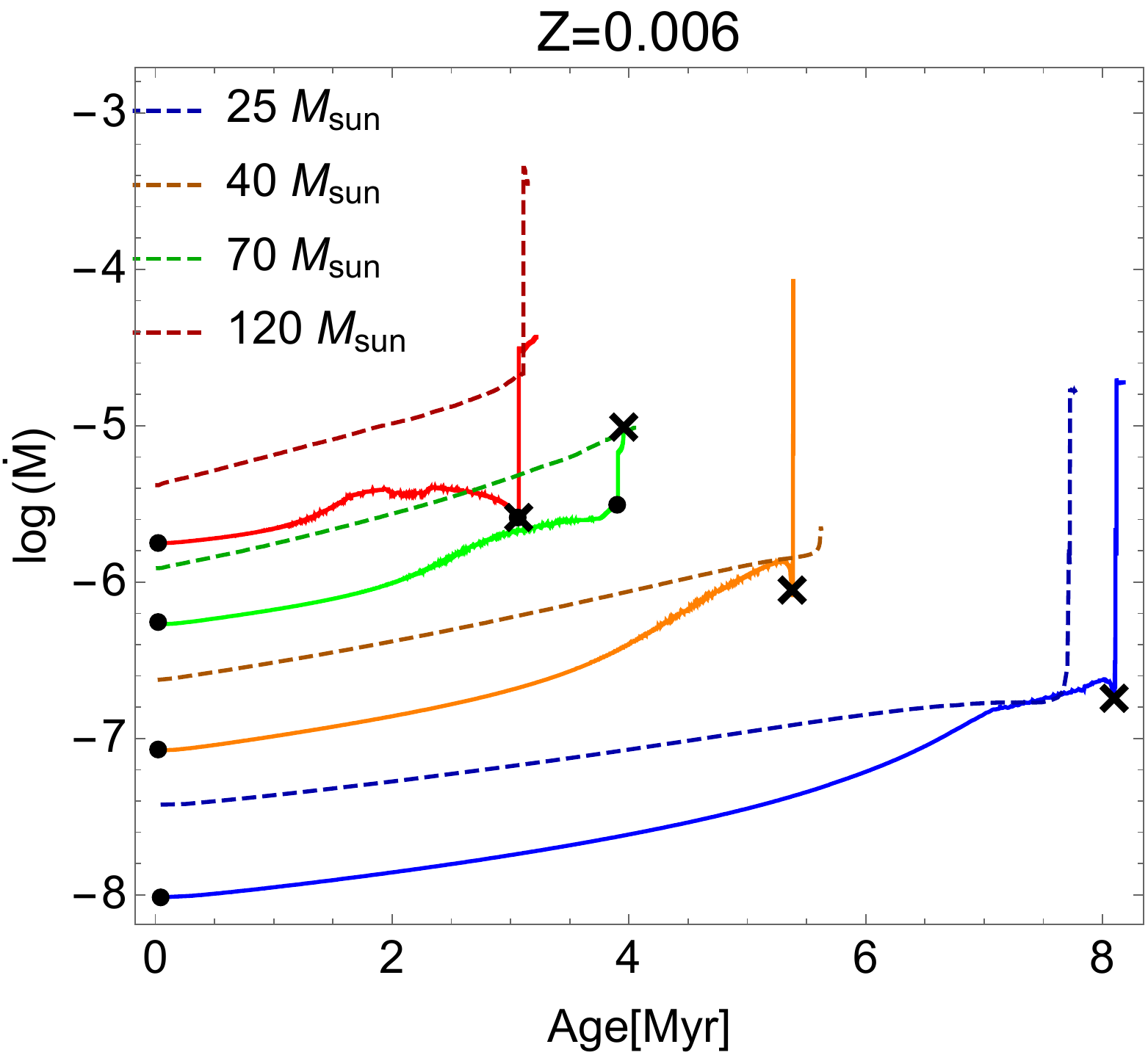}
		\caption{\small{Evolution of the self-consistent mass loss rate, $\dot M_\text{sc}$ (solid lines), compared with the models using Vink's formula (darker dashed lines)}}
		\label{mdots_final}
	\end{figure*}
	\begin{figure*}[t!]
		\centering
		\includegraphics[width=0.4\linewidth]{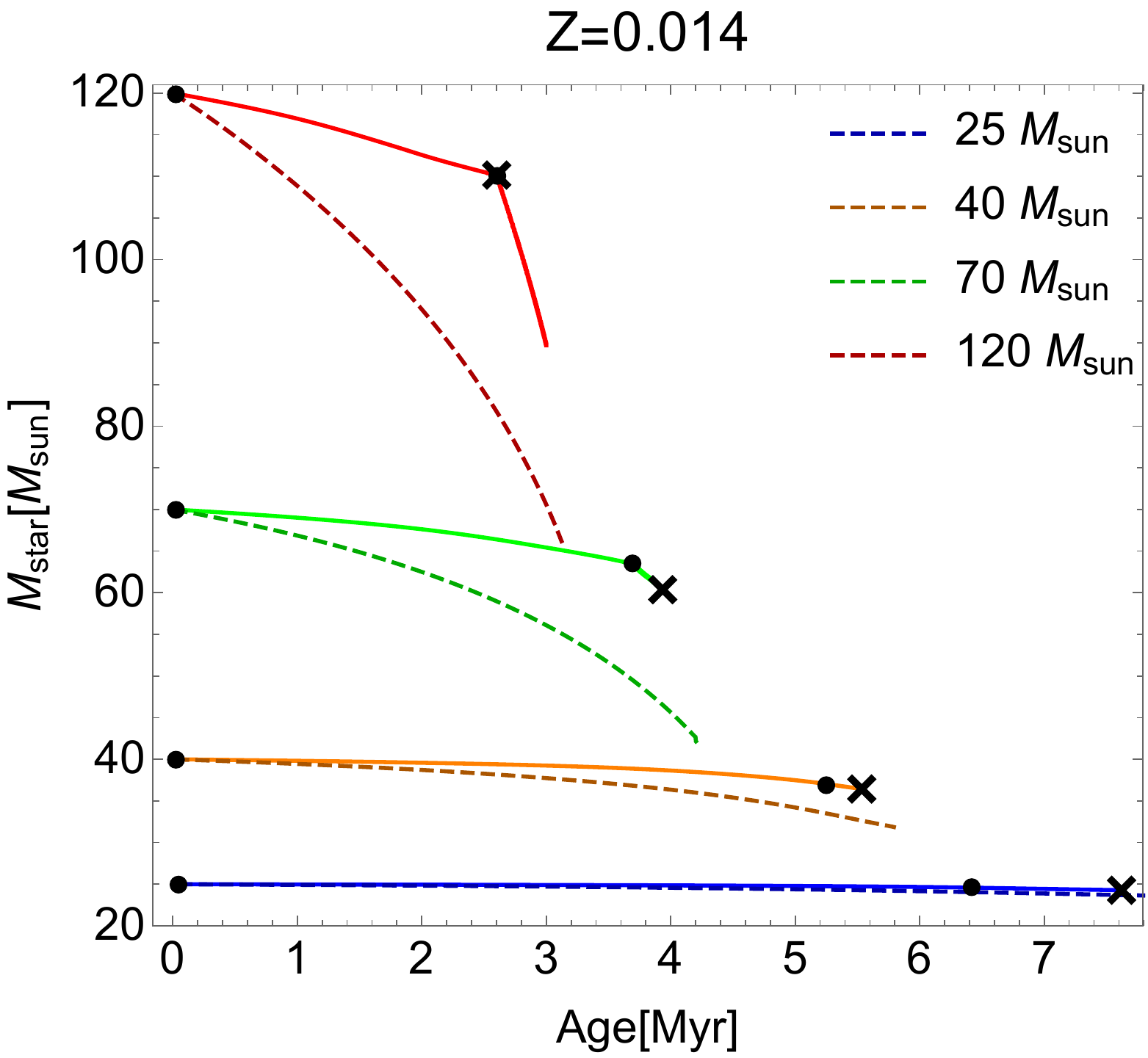}
		\hspace{1cm}
		\includegraphics[width=0.4\linewidth]{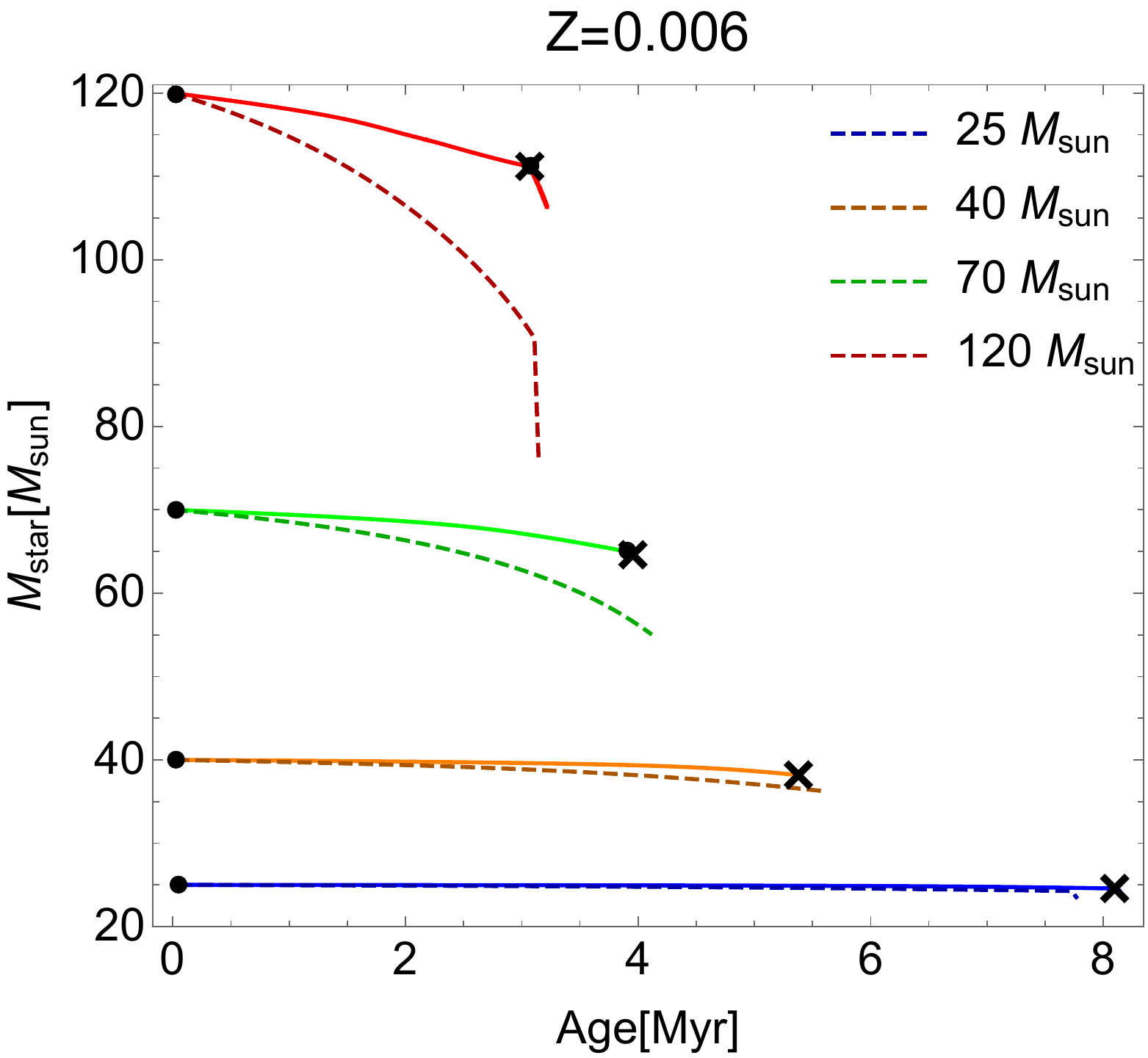}
		\caption{\small{Evolution of stellar mass.}}
		\label{mass_final}
	\end{figure*}
	\begin{figure*}[t!]
		\centering
		\includegraphics[width=0.4\linewidth]{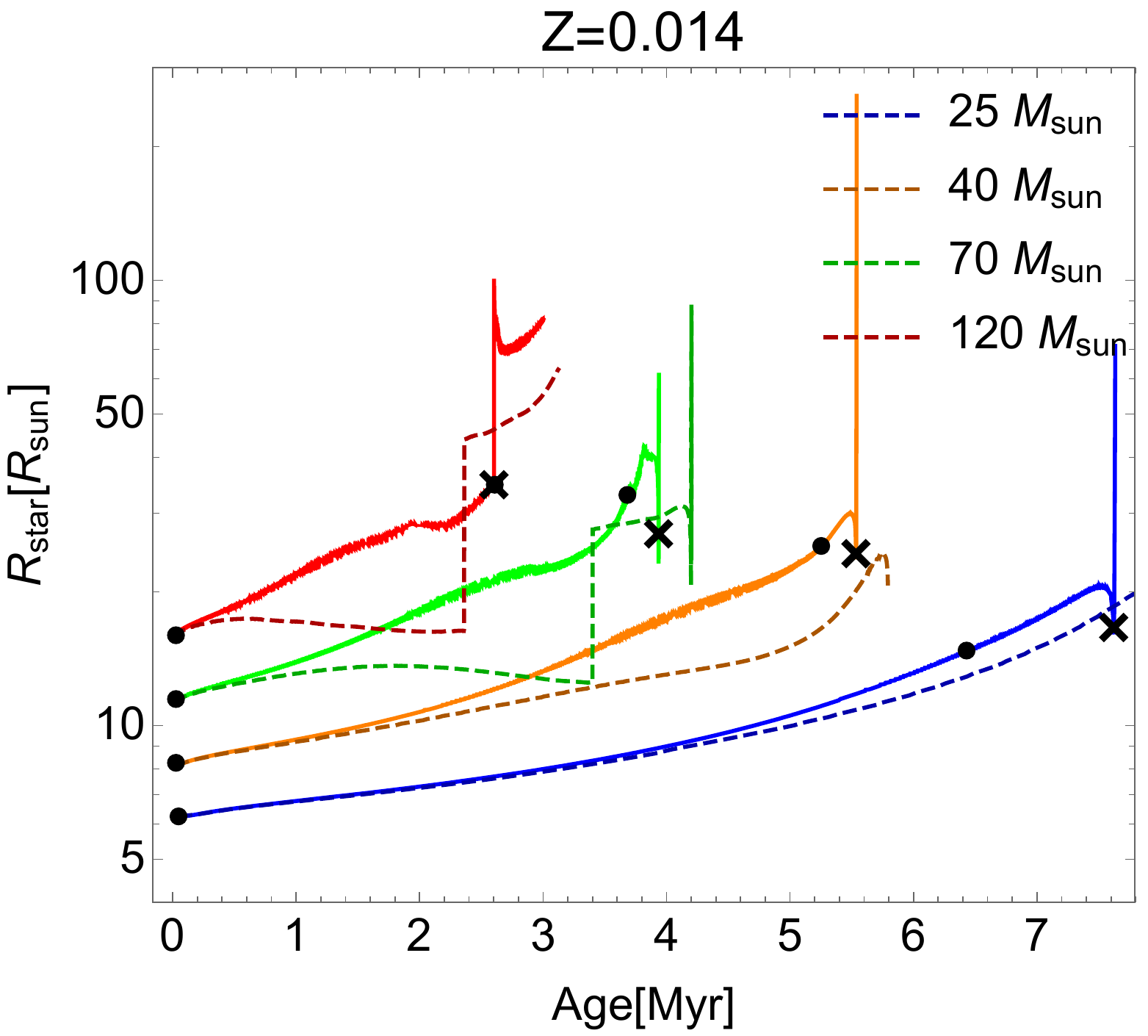}
		\hspace{1cm}
		\includegraphics[width=0.4\linewidth]{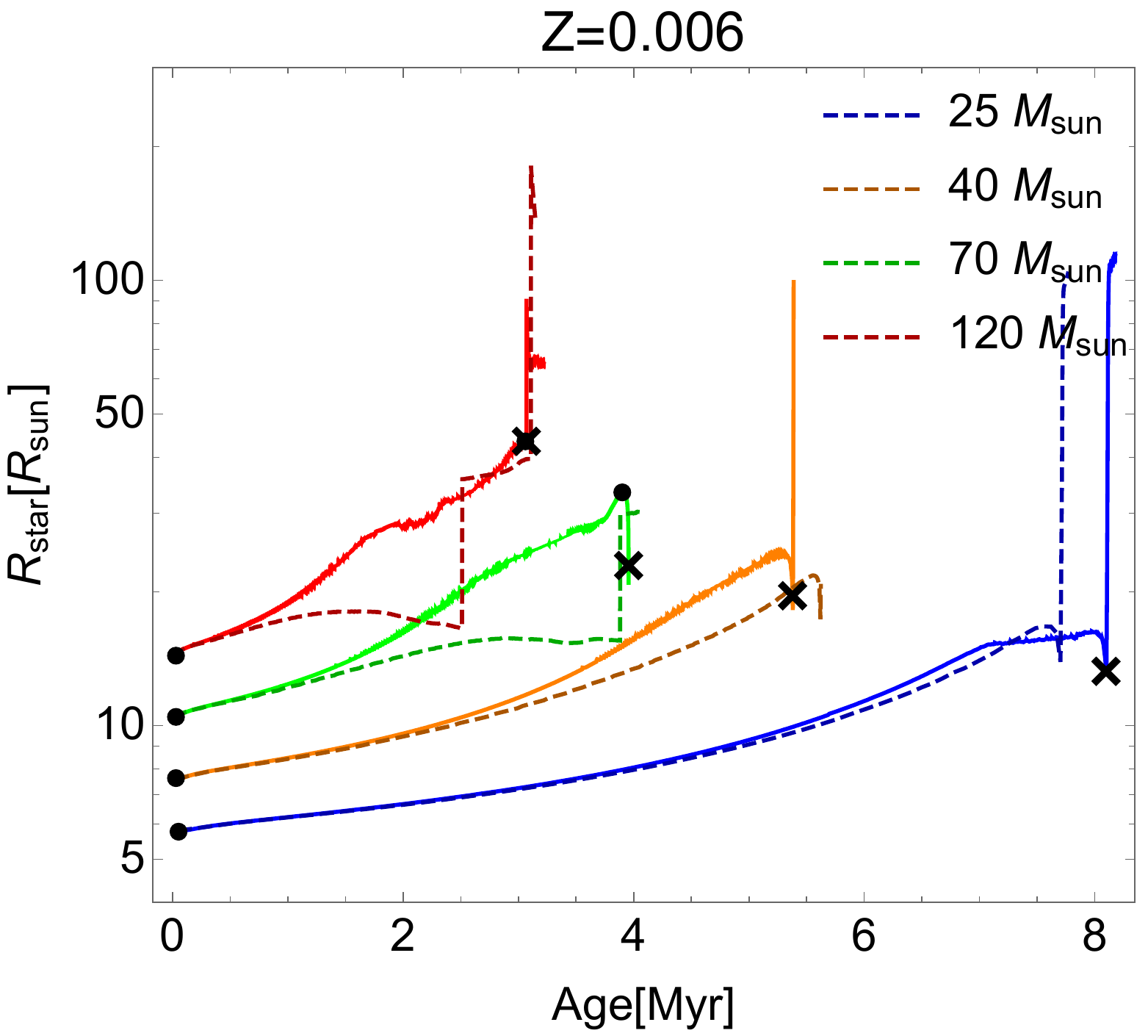}
		\caption{\small{Evolution of stellar radii.}}
		\label{radii_final}
	\end{figure*}
	\begin{figure*}[t!]
		\centering
		\includegraphics[width=0.4\linewidth]{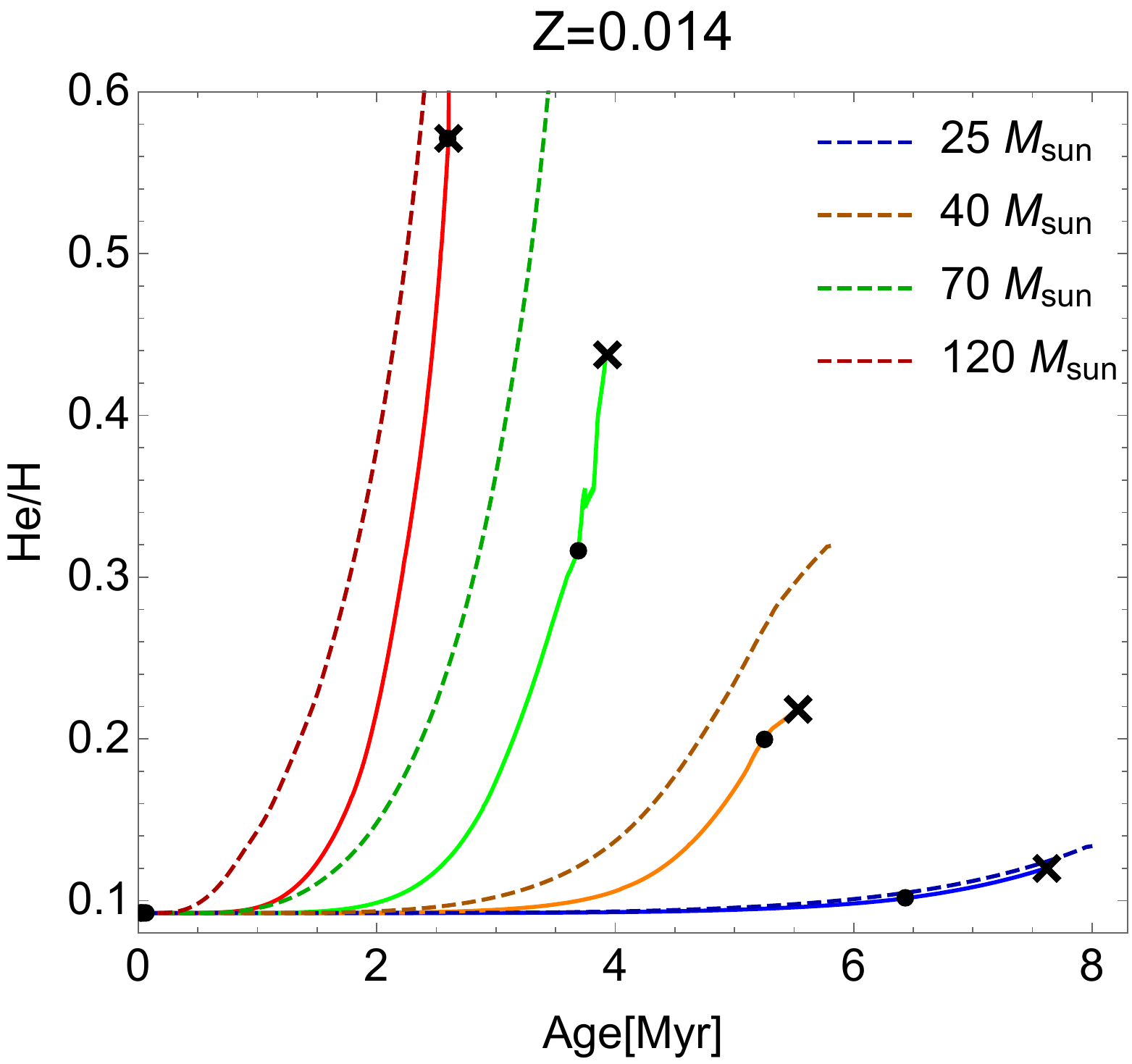}
		\hspace{1cm}
		\includegraphics[width=0.4\linewidth]{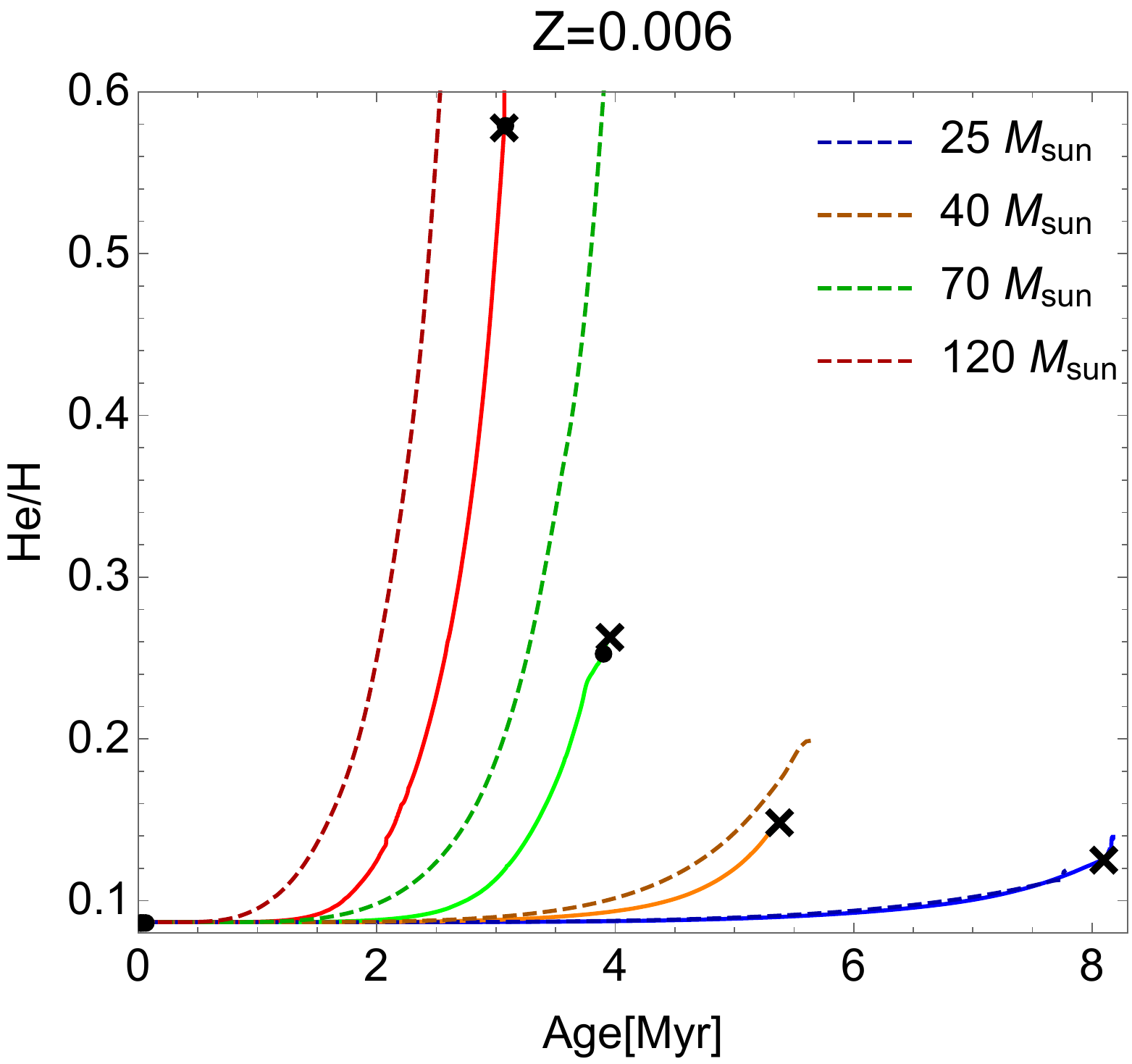}
		\caption{\small{Evolution of He/H ratio.}}
		\label{heh_final}
	\end{figure*}
	\begin{figure*}[t!]
		\centering
		\includegraphics[width=0.4\linewidth]{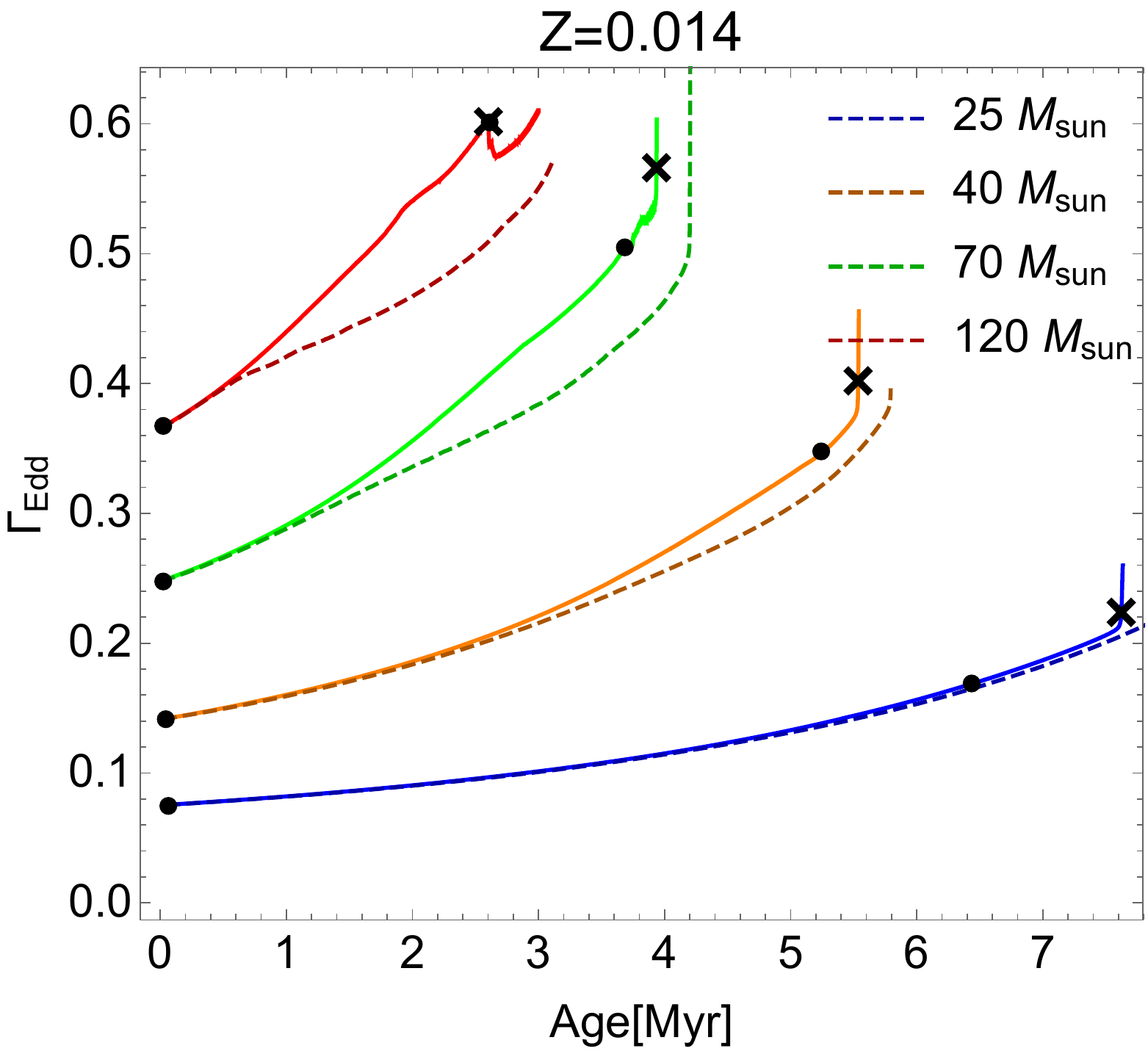}
		\hspace{1cm}
		\includegraphics[width=0.4\linewidth]{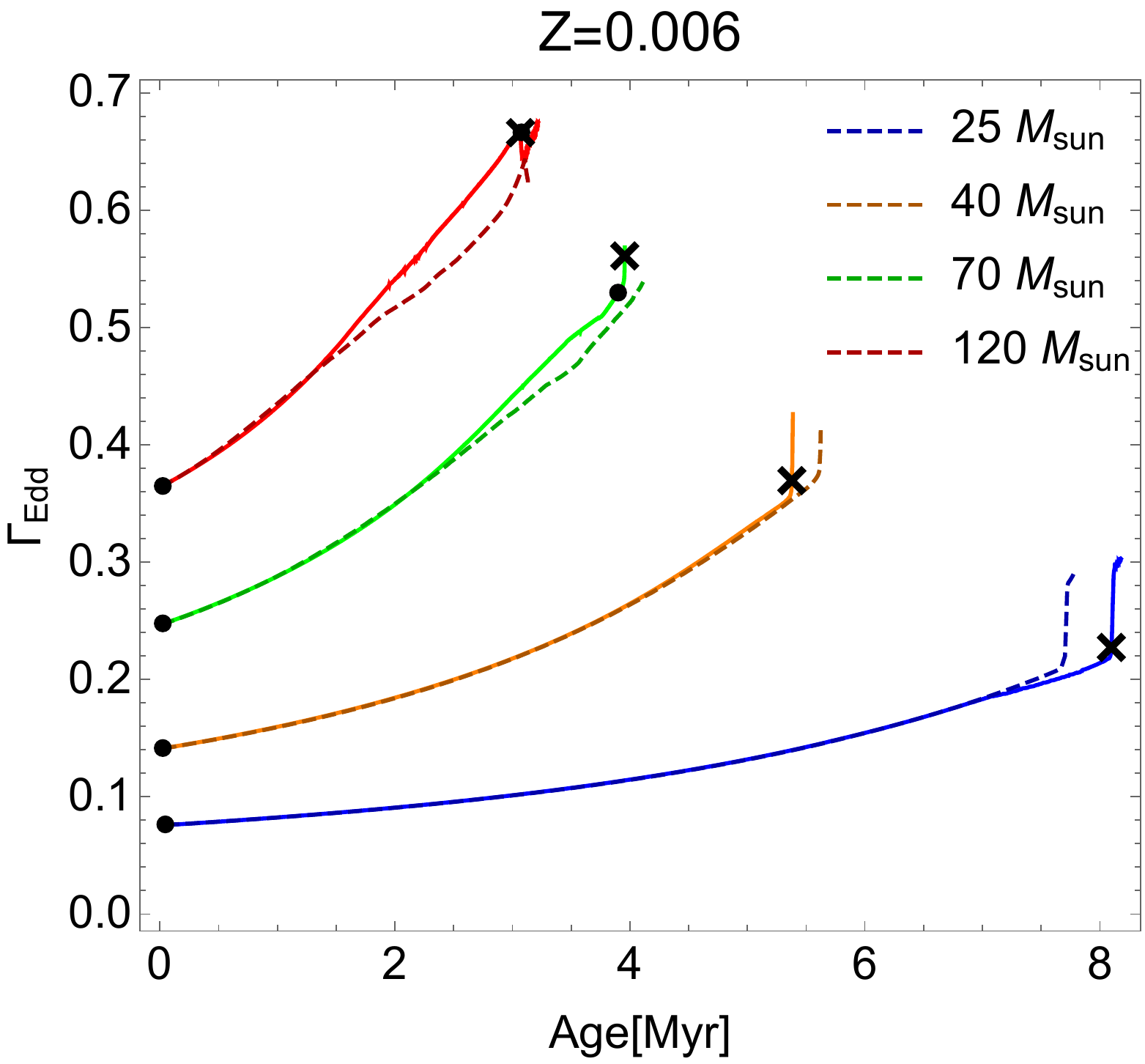}
		\caption{\small{Evolution of Eddington factor, $\Gamma_\text{Edd}$.}}
		\label{edd_final}
	\end{figure*}
	\begin{figure*}[t!]
		\centering
		\includegraphics[width=0.4\linewidth]{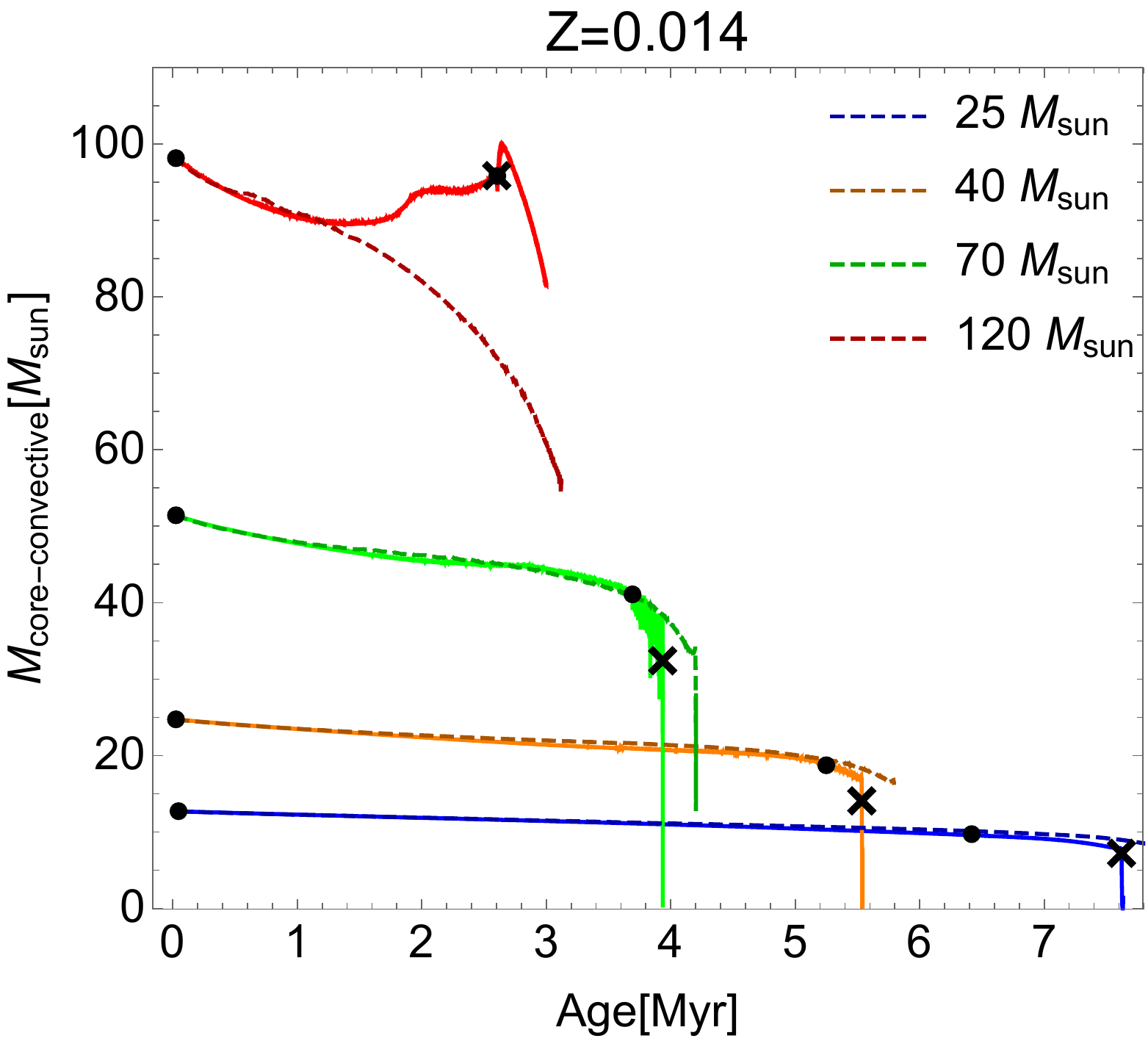}
		\hspace{1cm}
		\includegraphics[width=0.4\linewidth]{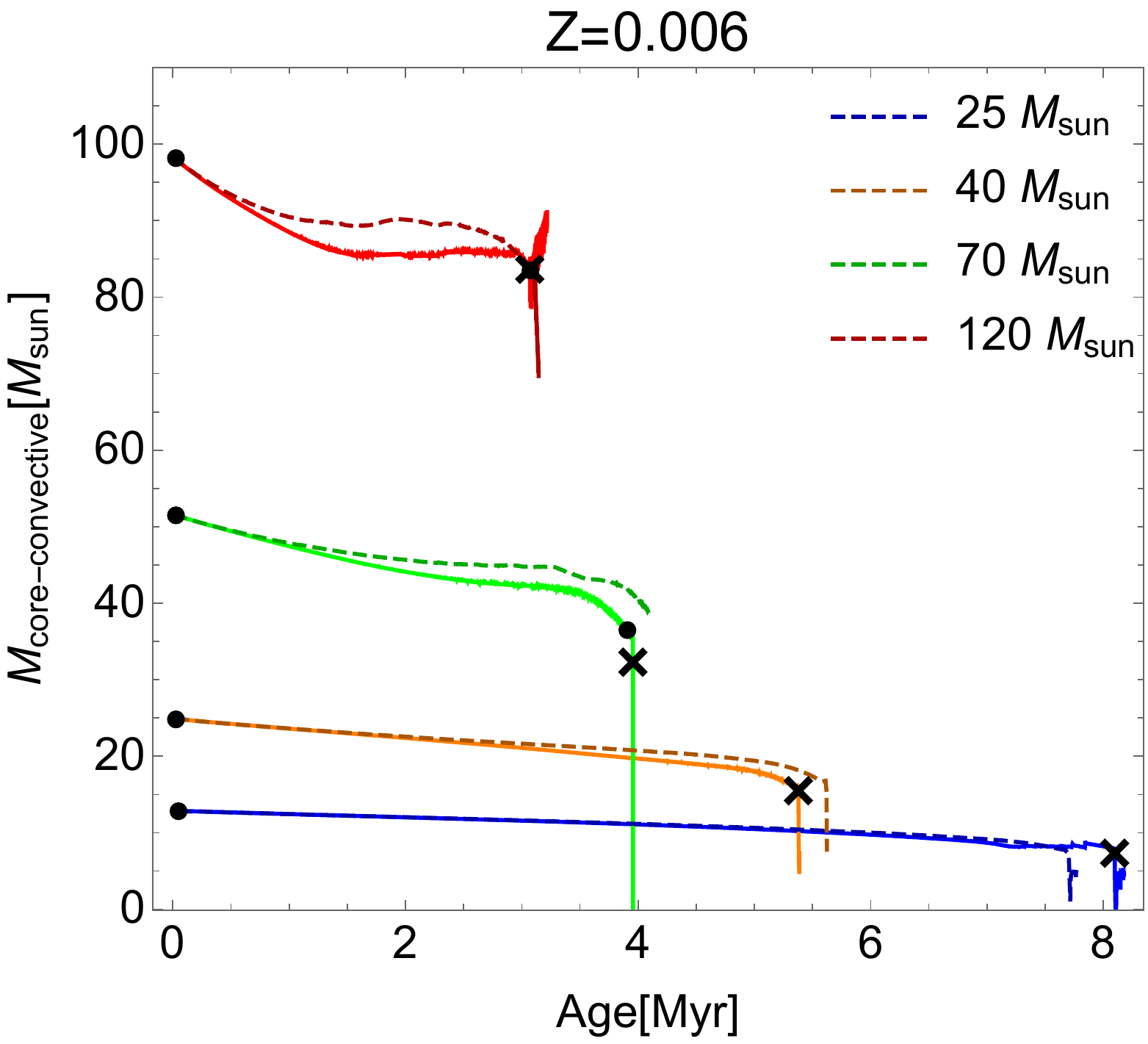}
		\caption{\small{Evolution of the mass of the stellar convective-core.}}
		\label{masscc_final}
	\end{figure*}

\end{appendix}

\end{document}